\documentclass[a4paper,11pt]{article}
\pdfoutput=1 % if your are submitting a pdflatex (i.e. if you have
\usepackage{jheppub} % for details on the use of the package, please
\usepackage{todonotes}
\usepackage{subfigure}
\usepackage{graphicx}
\usepackage{dsfont}
\usepackage{amssymb}
\usepackage{amsmath}
\usepackage{tikz}
\usepackage{tikz-3dplot}
\usetikzlibrary{snakes}
\usetikzlibrary{shapes.misc}
\usetikzlibrary{cd}

\usepackage{ytableau}

\usetikzlibrary{decorations.markings}
\tikzset{line/.style={line width=0.25mm},
curve/.style={line,smooth,tension=1},
->-/.style={decoration={
  markings,
  mark=at position #1 with {\arrow[>=stealth]{>}}},postaction={decorate}},
-<-/.style={decoration={
  markings,
  mark=at position #1 with {\arrow[>=stealth]{<}}},postaction={decorate}},
}

\usepackage{amsthm}
\newtheorem{thm}{Theorem}%[section]

\def\beq{\begin{align}}
\def\eeq{\end{align}}
\def\beqn{\begin{eqnarray}}
\def\eeqn{\end{eqnarray}}

\def\TY{${\rm TY}_{\mathbb Z_N}^{t,\kappa}$\,}
\def\pfTY{pf-${\rm TY}_{\mathbb Z_N}^{t,\kappa,\beta}$\,}

\def\Tr{{\rm Tr}}

\def\={\!\!&=&\!}
\def\+{\!\!&+&\!}
\def\-{\!\!&-&\!}

\def\and{\ \ \ {\rm and} \ \ }

\newcommand{\Z}{\mathbb{Z}}
\newcommand{\C}{\mathcal{C}}
\newcommand{\B}{\mathcal{B}}

\newcommand{\F}{\mathcal{F}}
\newcommand{\om}{\omega}
\newcommand{\Hom}{\mathrm{Hom}}
\newcommand{\End}{\mathrm{End}}
\newcommand{\bC}{\mathbb{C}}

\title{\boldmath Para-fusion Category and Topological Defect Lines in $\mathbb Z_N$-parafermionic CFTs}

\author[a]{Jin Chen,}
\author[b,c]{Babak Haghighat,}
\author[b]{Qing-Rui Wang}
\affiliation[a]{Department of Physics, Xiamen University, Xiamen, 361005, China}
\affiliation[b]{Yau Mathematical Sciences Center, Tsinghua University, Beijing, 100084, China}
\affiliation[c]{Yanqi Lake Beijing Institute of Mathematical Sciences and Applications (BIMSA), Huairou District, Beijing 101408, P. R. China}

 \emailAdd{zenofox@gmail.com}
 \emailAdd{babakhaghighat@tsinghua.edu.cn}
 \emailAdd{wangqr@tsinghua.edu.cn}

\abstract{We study topological defect lines (TDLs) in two-dimensional $\mathbb Z_N$-parafermionic CFTs. Unlike the bosonic case, parafermionic CFTs contain parafermionic defect operators that can live on TDLs and obey fractional statistics. We propose a categorical description of these TDLs, called a \emph{para-fusion category}. Its distinguishing features include $\mathbb Z_M$ $q$-type objects for $M\vert N$ and parafermionic defect operators as specialized 1-morphisms of the TDLs. Para-fusion categories generalize the super-fusion categories that describe TDLs in two-dimensional fermionic CFTs. We study these features from both a two-dimensional CFT perspective and a three-dimensional anyon-condensation perspective. In the latter approach, we introduce generalized parafermionic anyon condensation and use it to establish a functor from the parent fusion category of TDLs in a bosonic CFT to the para-fusion category of TDLs in the parafermionized theory. Finally, we provide examples that illustrate the properties of para-fusion categories and give a full classification of a universal para-fusion category obtained by parafermionic condensation of the Tambara--Yamagami $\mathbb Z_N$ fusion category.

}

\begin{document} 
\maketitle
\flushbottom

\newpage
\section{Introduction}
Global symmetries have played a central role in quantum field theory since its inception. They guide the construction of QFTs and constrain their dynamics. Our understanding of symmetry has evolved significantly in recent years \cite{Gaiotto:2014kfa}. In modern language, ordinary global symmetries are represented by invertible topological defects supported on codimension-one surfaces. Correlation functions containing these defects are invariant under continuous deformations of their support. Consequently, the defects commute with the stress tensor and with the Hamiltonian. Group multiplication and inversion are represented by fusion and orientation reversal of the corresponding topological defects. This viewpoint extends the notion of symmetry to higher-form, higher-group, and non-invertible symmetries, collectively known as generalized global symmetries. A $p$-form symmetry is generated by topological defects of codimension $p+1$, with an ordinary global symmetry corresponding to $p=0$. Nontrivial 't Hooft anomalies among symmetries of different form degrees can lead to higher-group or higher-categorical structures \cite{
Cordova:2018cvg, Benini:2018reh, Antinucci:2022eat,
Bartsch:2022mpm, Decoppet:2022dnz,Gaiotto:2017yup,Gaiotto:2017tne, Heckman:2022branes,Heckman:2022topdown}. Non-invertible symmetry defects are topological defects that are closed under fusion, but a simple defect need not possess a fusion inverse, and the fusion of simple defects may decompose into a direct sum of simple defects \cite{Bhardwaj:2017xup, Chang:2018iay, Choi:2021kmx, Kaidi:2021xfk, Roumpedakis:2022aik, Bhardwaj:2022yxj, Choi:2022zal, Cordova:2022ieu, Choi:2022jqy,
Kaidi:2022uux,Bashmakov:2022jtl,Bashmakov:2022uek, Choi:2022rfe, Kaidi:2023maf, Apruzzi:2022rei,Chen:2023qnv, Bashmakov:2023kwo}. Their fusion and junction data are described by categorical symmetries, which generalize group symmetries. 

In two dimensions, non-invertible symmetries can be realized by one-dimensional topological defect lines (TDLs), which occur widely in 2d conformal field theories. Historically, these defect lines were studied in connection with boundary CFTs, twisted partition functions, and orbifolds \cite{Frohlich:2004ef,Fuchs:2007tx,Frohlich:2009gb,Petkova:2000ip}. From a modern viewpoint, TDLs also provide tools for studying the dynamics and mathematical structures of quantum field theories. For example, generalized 't Hooft anomalies of TDLs can constrain RG flows between conformal and gapped phases \cite{Chang:2018iay, Thorngren:2018bhj, Thorngren:2019iar, Komargodski:2020mxz,Zhang:2023wlu}. A chosen semisimple system of TDLs with finite-dimensional junction spaces and finitely many isomorphism classes of simple lines forms a fusion category when it contains the identity line and is closed under fusion and duals \cite{Fuchs:2002cm,
Carqueville:2012dk,
Bhardwaj:2017xup,Chang:2018iay,etingof_tensor_2015}. Such a fusion category need not contain every topological line of the theory. Within this finite categorical setting, one can also study generalized orbifolds and dualities associated with categorical symmetries. In the Turaev--Viro description associated with a fusion category $\C$, the boundary TDLs are objects of $\C$, whereas the bulk anyons are objects of the Drinfeld center $Z(\C)$ \cite{Gaiotto:2020iye,
Kaidi:2021gbs, Bhardwaj:2023wzd,
Bhardwaj:2023ayw,JIWen2019,ji2022unified,Wen2023}. The canonical bulk-to-boundary tensor functor from $Z(\C)$ to $\C$ need not be injective on isomorphism classes of simple objects and need not preserve simplicity: a nontrivial bulk anyon may restrict to the boundary vacuum or decompose into a sum of boundary lines. Thus boundary TDLs should not be identified directly with bulk anyons. This relation connects the study of TDLs to the classification of topological orders and to anyon condensation~\cite{bais2009condensate,eliens2014diagrammatics,kong2014anyon,burnell2018anyon}.

Along this line, the study of categorical symmetries has also been extended to fermionic CFTs, whose fermionic degrees of freedom obey Fermi--Dirac statistics \cite{Petkova:1988halfInteger,Furlan:1990fusion, Aasen:2017ubm,
Zhou:2021ulc, Chang:2022hud,Runkel:2022fzi}. Fermionic CFTs on spin surfaces with boundaries and defects, and the boundary states and anomalous symmetries of fermionic minimal models, were studied in refs.~\cite{Runkel:2020zgg,BoyleSmith:2021fermionic}. A distinguished property of TDLs in the fermionic CFTs is that fermionic defect operators can ``live" on junctions of TDLs, or ``move along" some of them. With this novel feature, TDLs in fermionic systems have much richer and more involved structures from various physics and mathematical aspects. For example, a well-known example is the 2d massless Majorana theory obtained after fermionization from the Ising CFT, in which the famous non-invertible Kramers-Wannier duality $\mathcal N$-line in the Ising model is mapped to an \emph{invertible but anomalous} TDL, denoted by $(-1)^{F_L}$, which is counting the left-moving fermionic number in the Majorana theory. Because of the existence of a fermionic defect operator living on $(-1)^{F_L}$, it thus admits a $\mathbb Z_8$-classification corresponding to ${\rm Hom}(\Omega^{\rm spin}_{3}(B\mathbb Z_2),\,{\rm U}(1))$ \cite{Gu:2013azn,Kapustin:2014dxa, Wang:2018pdc, Grigoletto:2021zyv}. The mathematical structure behind TDLs in fermionic CFTs has been identified as a super-fusion category encoding extra data of fermionic defect operators~\cite{Gu:2013gma,GuWangWen, Aasen:2017ubm, usher2018fermionic,Zhou:2021ulc}. In a parallel manner, fermionic TDLs can be lifted to 3d TFTs in terms of the so-called fermionic anyon condensation, which has received extensive attention because of its intimate relation to the classification of fermionic topological orders.

Motivated by this progress, we study topological defect lines in two-dimensional $\mathbb Z_N$-parafermionic CFTs, with fermionic TDLs arising as the special case $N=2$. The $\mathbb Z_N$ parafermions were first introduced as continuum limits of lattice models by Fateev and Zamolodchikov \cite{Fateev:1985mm}; see also \cite{Yao:2020dqx} for a modern perspective. The spectra of local operators and their bosonization have also been studied extensively \cite{Thorngren:2019iar}. TDLs of the corresponding bosonized theories were recently investigated in \cite{Haghighat:2023sax}. However, except for $N=2$, the properties of TDLs in parafermionic CFTs have not yet been systematically established. As in the fermionic case, parafermionic defect operators can live on TDLs and obey fractional statistics. We therefore introduce a mathematical structure, called a \emph{para-fusion category}, that describes these TDLs and their parafermionic defect operators. In addition to the two-dimensional CFT construction, we study parafermionic anyon condensation from a three-dimensional perspective. This construction leads to generalized pentagon identities, which we call \emph{para-pentagons}. We illustrate the resulting para-fusion categories with explicit CFT examples and obtain their $F$-symbols by solving the para-pentagon equations. In particular, we find a class of para-fusion categories, denoted by \pfTY, obtained either by parafermionic anyon condensation of the $\mathbb Z_N$ Tambara--Yamagami category or by parafermionizing a bosonic CFT with $\mathbb Z_N$ self-duality. By solving the para-pentagon equations, we give a full classification of \pfTY for arbitrary $N$.

The paper is organized as follows. In Section~\ref{2}, we review orbifolding, fermionization, and parafermionization, and then introduce the defining properties of TDLs in parafermionic CFTs. In Section~\ref{3}, we relate the two-dimensional TDL data to the three-dimensional Turaev--Viro--Levin--Wen description and introduce parafermionic anyon condensation, or para-condensation. We use para-condensation to verify the properties proposed in Section~\ref{2}. In Sections~\ref{4} and \ref{5}, we present examples of TDLs in parafermionic CFTs. Finally, the appendices collect the mathematical background used in our construction of $\mathbb Z_N$ para-fusion categories.\\

\noindent
\paragraph{Note:} 
While completing this paper, we learned that Zhihao Duan, Qiang Jia, and Sungjay Lee were working on a related subject \cite{DJL}. We thank them for coordinating the submission.

\section{Topological Defect Lines in Parafermionic CFTs}
\label{2}
\subsection{Orbifolding, Fermionization and Parafermionization}
\label{sec:orbifolding-fermionization-parafermionization}
In this subsection, we briefly review orbifolding, fermionization, and parafermionization of a two-dimensional bosonic CFT with a non-anomalous zero-form symmetry $G=\mathbb Z_N$.

\paragraph{Orbifolding}
Let $\mathcal T$ be a two-dimensional CFT with a non-anomalous symmetry $G=\mathbb Z_N$. Put the theory on a torus with spatial coordinate $x\sim x+L_x$ and Euclidean-time coordinate $t\sim t+L_t$. For $g,h\in G$, impose the boundary conditions
\begin{align}
\mathcal O(x+L_x,t)=g\cdot \mathcal O(x,t)\,,\qquad
\mathcal O(x,t+L_t)=h\cdot\mathcal O(x,t)\,,
\label{eq:g_action}
\end{align} 
The spatial boundary condition defines the $g$-twisted Hilbert space $\mathcal H_g$. The temporal boundary condition is implemented by inserting the symmetry operator $\mathcal L_h$ in the trace. Equivalently, the symmetry TDLs $\mathcal L_g$ and $\mathcal L_h$ run along the Euclidean-time and spatial cycles, respectively \cite{Gaiotto:2014kfa,Chang:2018iay,Bhardwaj:2017xup}. The corresponding twisted partition function is
\begin{align}
Z_{\mathcal T}[g,h]=\Tr_{\mathcal H_g}\!\left(\mathcal L_h q^{L_0-\frac{c}{24}}\bar q^{\bar L_0-\frac{c}{24}}\right)\,.
\end{align}
Here $\tau\in\mathbb H$ is the complex modulus of the torus, $q=e^{2\pi i\tau}$, and $\bar q=e^{-2\pi i\bar\tau}$. The operators $L_0$ and $\bar L_0$ are the left- and right-moving Virasoro zero modes, and $c$ is the central charge.

For fixed $g$, a single insertion $\mathcal L_h$ does not by itself project onto a charge sector. The charge projectors are obtained by taking character-weighted sums over $h$. Thus the $N$ spatially twisted Hilbert spaces, each decomposed into $N$ charge sectors, give $N^2$ twisted-sector blocks. Together, $(g,h)$ specifies the holonomies of a flat background gauge field $S\in H^1(\mathbb T^2,G)\simeq G\times G$. On a general Riemann surface $M$, such a background is represented by a network of symmetry TDLs; in this paper we use only $M=\mathbb T^2$.

The orbifold theory $\mathcal T^\prime\equiv\mathcal T/G$ is obtained by promoting the flat background field $S$ to a dynamical field $s$ and summing over its holonomies. It has an emergent Pontryagin-dual symmetry $\widehat G$ \cite{Vafa:1989ih,Bhardwaj:2017xup}. The corresponding symmetry lines are labeled by the irreducible representations of $G$. For $G=\mathbb Z_N$,
\begin{align}
\widehat G={\rm Hom}(\mathbb Z_N,{\rm U}(1))\simeq\mathbb Z_N\,.
\end{align}
Coupling $\mathcal T^\prime$ to a flat background $\widehat S\in H^1(M,\widehat G)$ gives the discrete Fourier transform
\begin{align}
Z_{\mathcal T^\prime}[\widehat S]=\frac{1}{\sqrt{\left|H^1(M,G)\right|}}\sum_{s\in H^1(M,G)}e^{i\langle\widehat S,s\rangle}Z_{\mathcal T}[s]\,,
\end{align}
where $\langle\widehat S,s\rangle\in\mathbb R/2\pi\mathbb Z$ is the canonical pairing between $H^1(M,\widehat G)$ and $H^1(M,G)$. Gauging the dual symmetry returns the original theory. For $M=\mathbb T^2$ and $G=\mathbb Z_N$, the transform becomes
\begin{align} 
Z_{\mathcal T^\prime}[b_1, b_2]=\frac{1}{N}\sum_{a_1,a_2\in \mathbb Z_N}\omega^{a_1 b_2-a_2 b_1} Z_{\mathcal T}[a_1, a_2]\,,
\end{align}
where $\omega\equiv e^{2\pi i/N}$. The pair $a=(a_1,a_2)\in\mathbb Z_N^2$ gives the holonomies of the dynamical gauge field $s$ around the spatial and Euclidean-time cycles of $\mathbb T^2$, while $b=(b_1,b_2)\in\mathbb Z_N^2$ gives the corresponding holonomies of the dual background field $\widehat S$.

\paragraph{Main example.}
A primary example of such a theory with $\mathbb Z_N$-symmetry is the $\frac{{\rm SU}(2)_N}{U(1)}$ coset CFT with central charge $c = \frac{2(N-1)}{N+2}$. The theory consists of $\frac{N(N+1)}{2}$ primaries labeled by two integers $(l,m)$ which vary in the range
\begin{equation}
    0 \leq l \leq N, \quad -l + 2 \leq m \leq l, \quad l -m \in 2 \mathbb{Z},
\end{equation}
with the following conformal weights
\begin{equation}
    \Delta = \bar{\Delta} = \frac{l(l+2)}{4(N+2)} - \frac{m^2}{4N}.
\end{equation}
The partition function in the main sector is given by
\begin{equation}
    Z[0,0] = \sum_{l,m} \chi_{l,m}(\tau) \overline{\chi_{l,m}(\tau)},
\end{equation}
where the characters $\chi_{l,m}$ satisfy the relations
\begin{equation}
    \chi_{l,m} = \chi_{l,m+2N}, \quad \chi_{l,m} = \chi_{l,-m}, \quad \chi_{l,m} = \chi_{N-l,N+m}.
\end{equation}
The twisted sectors are given by
\begin{equation}
    Z[a_1,a_2] = e^{-2\pi i a_1 a_2/N} \sum_{l,m}e^{2\pi i a_2 m /N} \chi_{l,m} \overline{\chi_{l,m-2a_1}}.
\end{equation}
They satisfy the following $S$- and $T$-transformation properties
\begin{equation}
    Z[a_1,a_2](\tau+1)=Z[a_1,a_2-a_1](\tau),\qquad
    Z[a_1,a_2](-1/\tau)=Z[a_2,-a_1](\tau).
\end{equation}

\paragraph{Fermionization}
In addition to orbifolding, a bosonic CFT with a non-anomalous $\mathbb Z_2\subset\mathbb Z_N$ symmetry can be fermionized. The best-known example is the duality between the Ising model and a Majorana fermion realized by the Jordan--Wigner transformation. This construction has been generalized to an arbitrary CFT with a non-anomalous $\mathbb Z_2$ symmetry \cite{Hsieh:2020uwb}. The standard procedure stacks a Kitaev chain with fermion-parity symmetry $\mathbb Z_{2f}$ with the CFT and gauges the diagonal subgroup of $\mathbb Z_2\times\mathbb Z_{2f}$:
\begin{align}
\mathcal T_f=\mathcal T\times {\rm Kitaev}/\mathbb Z_2\,.
\end{align} 
After fermionization, $\mathcal T_f$ has an emergent $\mathbb Z_2$ fermion-parity symmetry generated by $(-1)^F$.

% From the perspective of TDLs, the appearance of $(-1)^F$ can be understood in the following\cite{BSW}: The low energy physics of the Kitaev chain is a 1d TQFT phase. After gauging the diagonal of $\mathbb Z_2\times \mathbb Z_{2f}$ symmetry, the two formerly decoupled CFT and the Kitaev chain turn out to be coupled, and the 1d TQFT becomes a TDL in the resulting fermionic theory $\mathcal T_f$, corresponding to the $(-1)^F$ symmetry.

There is one subtlety in coupling $\mathcal T_f$ to a background gauge field $A$ for $(-1)^F$. A fermionic theory depends on a spin structure $\rho$ on $M$, and the background field shifts it as $\rho\rightarrow\rho+A$. The background field can therefore be absorbed into a redefinition of $\rho$. The partition function of $\mathcal T_f$ depends on $\rho$ and is related to the bosonic partition functions by
\begin{align}
Z_{\mathcal T_f}[\rho]=\frac{1}{\sqrt{\left|H^1(M, G)\right|}}\sum_{a\in H^1(M,\mathbb Z_2)}(-1)^{{\rm Arf}(\rho+a)}Z_{\mathcal T}[a]\,,
\label{eq:fermionization_Arf}
\end{align}
where the spin structure $\rho$ along a cycle of $M$ can take values of $\{0, 1\}$ corresponding to the NS or Ramond sectors, respectively. In the case of $M=\mathbb T^2$, one can make this more explicit,
\begin{align}
Z_{\mathcal T_f}[\rho_1, \rho_2]=\frac{1}{2}\sum_{a_1,a_2\in \mathbb Z_2}(-1)^{(\rho_1+a_1)(\rho_2+a_2)}Z_{\mathcal T}[a_1,a_2]\,,
\end{align}
or
\begin{align}
Z_{\mathcal T}[a_1, a_2]=\frac{1}{2}\sum_{\rho_1,\rho_2\in \mathbb Z_2}(-1)^{(\rho_1+a_1)(\rho_2+a_2)}Z_{\mathcal T_f}[\rho_1,\rho_2]\,.
\end{align}
Notice that, from eq.\,(\ref{eq:fermionization_Arf}), the Arf invariant is just the non-trivial topological order of the Kitaev Majorana chain. It precisely means that our fermionization is defined by first stacking the Kitaev chain onto the bosonic theory and then gauging the diagonal $\mathbb Z_2$ of the two systems.

One can combine the orbifolding and fermionization operations together, so that we can first orbifold the theory $\mathcal T$ to $\mathcal T^\prime$, and successively fermionize it to $\mathcal T^\prime_f$, and thus have
\begin{align}
Z_{\mathcal T^\prime_f}[\rho]=\frac{1}{\sqrt{\left|H^1(M, G)\right|}}\sum_{a\in H^1(M,\mathbb Z_2)}(-1)^{{\rm Arf}(\rho+a)}Z_{\mathcal T'}[a]\,,
\label{eq:fermionization_Arf2}
\end{align}
or on $\mathbb T^2$,
\begin{align}
Z_{\mathcal T^\prime_f}[\rho_1,\rho_2]=\frac{1}{2}\sum_{a_1,a_2\in \mathbb Z_2}(-1)^{(\rho_1+a_1)(\rho_2+a_2)}Z_{\mathcal T'}[a_1,a_2]\,.
\end{align}
Furthermore, from eq.\,(\ref{eq:fermionization_Arf}) and (\ref{eq:fermionization_Arf2}), we have the relations between $\mathcal T_f^\prime$ and $\mathcal T_f$ as
\begin{align}
Z_{\mathcal T_f^\prime}[\rho]=(-1)^{{\rm Arf}(\rho)}Z_{\mathcal T_f}[\rho]\,,
\label{eq:Tfp_Tf}
\end{align}
which implies that $\mathcal T_f^\prime$ differs from $\mathcal T_f$ by the additional SPT phase $(-1)^{{\rm Arf}(\rho)}$. These relations can be summarized by the following diagram:
\begin{align}
\begin{tikzpicture}
\node at (-0.5,0.5) {$\mathcal T$};
\node at (4.5,0.5) {$\mathcal T_f$};
\node at (-0.5,-2.5) {$\mathcal T^\prime$};
\node at (4.5,-2.5) {$\mathcal T_f^\prime$};
\draw[<->] (0,.5)--(4,.5);
\draw[<->] (0,-2.5)--(4,-2.5);
\draw[<->] (-.5,0)--(-.5,-2);
\draw[<->] (4.5,0)--(4.5,-2);
\node at (2,.75) {fermionize};
\node at (2,0.25) {bosonize};
\node at (2,-2.25) {fermionize};
\node at (2,-2.75) {bosonize};
\node at (-1,-.8) {gauge};
\node at (-1,-1.3) {$\mathbb Z_2$};
\node at (5,-.8) {stack};
\node at (5,-1.3) {Arf};
\end{tikzpicture}
\notag
\end{align}

\paragraph{Parafermionization}
The idea of fermionization can be further generalized to parafermionization, so long as the bosonic CFT admits a non-anomalous $\mathbb Z_N$ symmetry. Parallel to the discussion of fermionization where we stack a Kitaev Majorana chain onto the bosonic system, here we also need an analogous $\mathbb Z_N$ chain to perform the parafermionization. The $\mathbb Z_N$ generalization of the Ising model has been long known \cite{Fateev:1985mm,Fateev:1991bv}, where Fadeev and Zamolodchikov studied the 1d lattice model with Hamiltonian
\begin{align}
H_N=-\sum_{j=1}^L\sum_{m=1}^{N-1}\alpha_m\left(\tau_j\right)^m-\sum_{j=1}^{L-1}\sum_{m=1}^{N-1}\beta_m\left(\sigma_j^\dagger\sigma_{j+1}\right)^m\,,
\label{eq:Z_n_chain}
\end{align}
where $\alpha_m$ and $\beta_m$ are certain parameters, and $L$ is the number of sites of the lattice. The $N\times N$ matrices $\tau_j$ and $\sigma_j$ are generalized Pauli matrices at site $j$, satisfying
\begin{align}
\sigma^N=\tau^N=1\,,\ \ \sigma^\dagger=\sigma^{N-1}\,,\ \ \tau^\dagger=\tau^{N-1}\,, \ \ {\rm and}\ \ \sigma\tau=\omega\,\tau\sigma\,,
\end{align}
with $\omega=e^{2\pi i/N}$. A representation of the above algebra can be given by diagonalizing $\tau$,
\begin{align}
\tau={\rm diag}\{1,\omega,\omega^2,\,\dots,\,\omega^{N-1}\}\,,\ \ \sigma_{rs}=\delta_{r+1,s}\,,
\end{align}
where $r,s\in\mathbb Z_N$ are the row and column indices. The Hamiltionian is invariant under a $\mathbb Z_N$ action: $\sigma_j\rightarrow\omega\,\sigma_j$, and similar to the Ising chain, the corresponding symmetry generator is defined as
\begin{align}
\omega^P\equiv\prod_{j=1}^L\tau_j^\dagger\,,
\label{eq:pf_parity}
\end{align}
satisfying $\left(\omega^P\right)^N=1$. The lattice model itself is interesting to study as it turns out to be integrable in a certain range of parameters $\alpha_m$ and $\beta_m$, as well as admits conformal phases described by the WZW coset model $\mathfrak {su}(2)_k/\mathfrak u(1)$. On the other hand, there is also a generalized Jordan-Wigner transformation, the Fradkin-Kadanoff transformation, that can recast the bosonic lattice model into a parafermionic chain, by defining
\begin{align}
\gamma_{2j-1}\equiv\sigma_j\prod_{i<j}\tau_i\,,
\ \ {\rm and}\ \ \gamma_{2j}\equiv\omega^{(N-1)/2}\sigma_j\prod_{i\leq j}\tau_i\,.
\end{align}
With these new variables, one can show that the operators $\gamma_j$ satisfy the following statistics:
\begin{align}
\gamma^N_j=1\,,\ \ 
\gamma_j^\dagger=\gamma_j^{-1}\,,\ \ {\rm and}\ \ \gamma_j\gamma_k=\omega^{{\rm sgn}(k-j)}\,\gamma_k\gamma_j\,,
\ \ {\rm for}\ \ j\neq k\,.
\label{eq:parafermion_statistics}
\end{align}
The last commutation relation will reappear below as part of the braiding data. We now recall the parafermion-chain attachment construction of refs.~\cite{Yao:2020dqx,Thorngren:2021yso} at the level of torus partition functions. For a bosonic theory $\mathcal T$ with a non-anomalous $\mathbb Z_N$ symmetry, this construction gives an invertible linear relation between the $\mathbb Z_N$-twisted bosonic partition functions and the parafermionic partition functions on the torus.

Let $a=(a_1,a_2)\in\mathbb Z_N^2$ denote the holonomies of the background $\mathbb Z_N$ gauge field around the spatial and Euclidean-time cycles of the torus, respectively. Following refs.~\cite{Yao:2020dqx,Thorngren:2021yso}, we write $\rho=(\rho_1,\rho_2)\in\mathbb Z_N^2$ for the corresponding pair of paraspin labels on the two cycles. The complex modulus of the torus is denoted by $\tau\in\mathbb H$, where $\mathbb H$ is the upper half-plane. We collect the $N^2$ partition functions into the column vectors
\begin{align}
\boldsymbol Z_{B}(\tau)&\equiv
\bigl(Z_{\mathcal T}[a_1,a_2](\tau)\bigr)_{a\in\mathbb Z_N^2}\,,\notag\\
\boldsymbol Z_{PF}^{(m)}(\tau)&\equiv
\bigl(Z_{\mathcal T_{PF}}[\rho_1,\rho_2;m](\tau)\bigr)_{\rho\in\mathbb Z_N^2}\,.
\end{align}
The subscripts $B$ and $PF$ denote the bosonic and parafermionic theories. Here $m$ belongs to the group of units $\mathbb Z_N^\times=\{m\in\mathbb Z_N\mid\gcd(m,N)=1\}$ and specifies the non-degenerate bicharacter $\chi_m(x,y)=\omega^{mxy}$ for $x,y\in\mathbb Z_N$. Define the $N^2\times N^2$ matrix $F_m$, with row index $\rho=(\rho_1,\rho_2)$ and column index $a=(a_1,a_2)$, by
\begin{align}
(F_m)_{\rho a}&\equiv\frac{1}{N}
\omega^{m(\rho_1+a_1)(\rho_2+a_2)}\,.
\end{align}
The parafermionization formula and its inverse are then
\begin{align}
Z_{\mathcal T_{PF}}[\rho_1,\rho_2;m](\tau)
&=\sum_{a\in\mathbb Z_N^2}(F_m)_{\rho a}
Z_{\mathcal T}[a_1,a_2](\tau)\,,\notag\\
Z_{\mathcal T}[a_1,a_2](\tau)
&=\sum_{\rho\in\mathbb Z_N^2}(F_m^\dagger)_{a\rho}
Z_{\mathcal T_{PF}}[\rho_1,\rho_2;m](\tau)\,,\notag\\
(F_m^\dagger)_{a\rho}&=\frac{1}{N}
\omega^{-m(\rho_1+a_1)(\rho_2+a_2)}\,.
\label{eq:parafermionization_Arfn}
\end{align}
In vector notation, these two relations are $\boldsymbol Z_{PF}^{(m)}=F_m\boldsymbol Z_B$ and $\boldsymbol Z_B=F_m^\dagger\boldsymbol Z_{PF}^{(m)}$, where $\dagger$ denotes Hermitian conjugation. Character orthogonality gives $F_m^\dagger F_m=F_mF_m^\dagger=I_{N^2}$ for $m\in\mathbb Z_N^\times$, where $I_{N^2}$ is the $N^2\times N^2$ identity matrix. In particular, for $N=5$, the four choices $m=1,2,3,4$ all give invertible transforms. Invertibility alone does not distinguish four physical theories, since automorphisms of $\mathbb Z_N$ and relabelings of the background holonomies may relate the corresponding matrices $F_m$. We follow the convention $m=1$ used in refs.~\cite{Yao:2020dqx,Thorngren:2021yso}. For $N=2$, $\rho$ is the usual spin-structure label; for $N>2$, it is the torus paraspin label used in these references.

Let $\mathcal T^\prime=\mathcal T/\mathbb Z_N$, where the orbifold sum uses the same bicharacter $\chi_m$. Parafermionizing $\mathcal T^\prime$ gives~\cite{Thorngren:2021yso}
\begin{align}
Z_{\mathcal T^\prime_{PF}}[\rho_1,\rho_2;m]
=\omega^{m\rho_1\rho_2}
Z_{\mathcal T_{PF}}[\rho_1,-\rho_2;m]\,.
\label{eq:parafermionization_Arfn2}
\end{align}
This relation combines the phase $\omega^{m\rho_1\rho_2}$ with the reversal $\rho_2\mapsto-\rho_2$. We return to the origin of the latter after discussing modular covariance.

As a simple example of parafermionization, consider the three-state Potts model described by the $D$-type modular invariant of the minimal model $\mathcal M(5,6)$. With respect to the Virasoro algebra, the Potts model has eight modules with conformal weights $h=\{0,\, \frac{2}{5},\, \frac{7}{5},\, 3,\, \frac{1}{15},\, \frac{1}{15}^*,\, \frac{2}{3},\, \frac{2}{3}^*\}$. The superscript $*$ distinguishes pairs of modules with the same conformal weight that are exchanged by charge conjugation $\mathcal C$. Equivalently, the model is diagonal with respect to the $W_3$ algebra. The Virasoro modules labeled by $\left(0,\,3\right)$ and $\left(\frac{2}{5},\, \frac{7}{5}\right)$ combine into irreducible $W_3$ modules, leaving six $W_3$ modules in total. At the level of the partition function, this gives
\begin{align}
Z_{\mathcal T}=|\chi_0|^2+|\chi_{2/5}|^2+|\chi_{1/15}|^2+|\chi_{1/15^*}|^2+|\chi_{2/3}|^2+|\chi_{2/3^*}|^2\,,
\end{align}
where $\chi_h$ labels the character of a $W_3$-module of conformal weight $h$. Associated to these primaries, we have six Verlinde lines $\mathcal L_h$, and among them the lines $\left\{\mathcal L_0,\, \mathcal L_{2/3},\, \mathcal L_{2/3^*}\right\}$ are corresponding to the $\mathbb Z_3$ symmetry of the 3-Potts. Now using eq.\,(\ref{eq:parafermionization_Arfn}), we can obtain nine partition functions of the parafermionized theory $\mathcal T_{PF}$, one for each paraspin label $\rho=(\rho_1,\rho_2)\in\mathbb Z_3^2$. For simplicity, we only display the partition function for $\rho=(0,0)$,
\begin{align}
Z_{\mathcal T_{PF}}[0,0]=\left(\chi_0+\chi_{2/3}+\chi_{2/3^*}\right)\bar\chi_0+\left(\chi_{2/5}+\chi_{1/15}+\chi_{1/15^*}\right)\bar\chi_{2/5}\,,
\end{align}
from which, one can easily read off operator spectra in the $(0,0)$-sector of $\mathcal T_{PF}$: They are 
\begin{align}
\phi_{0,0},\quad \phi_{\frac{2}{3},0},\quad \phi^*_{\frac{2}{3},0},\quad \phi_{\frac{2}{5},\frac{2}{5}},\quad \phi_{\frac{1}{15},\frac{2}{5}},\quad \phi^*_{\frac{1}{15},\frac{2}{5}}\,.
\end{align}
One can see that some of them have spin-$\frac{2}{3}$ (modulo 1), satisfying the unusual parafermionic statistics. The other partition functions with different $\rho$'s will also display similar features, and they are all related by modular transformations as summarized below following reference \cite{Yao:2020dqx}.

\paragraph{Torus modular covariance.}

Throughout this paper, we use the parafermionic-chain construction of refs.~\cite{Yao:2020dqx,Thorngren:2021yso} at the level of the finite transform between twisted partition functions on the torus. The discussion below therefore concerns torus sectors and their modular transformations; no extension to para-spin structures on general Riemann surfaces will be needed.

We next define the matrices that implement modular transformations on the two vectors of partition functions. For $\gamma=\left(\begin{smallmatrix}u&v\\ w&x\end{smallmatrix}\right)\in {\rm SL}_2(\mathbb Z)$, let $\gamma\tau=(u\tau+v)/(w\tau+x)$ and define the $N^2\times N^2$ matrix $R_B(\gamma)$ by
\begin{align}
Z_{\mathcal T}[a_1,a_2](\gamma\tau)
=\sum_{b\in\mathbb Z_N^2}[R_B(\gamma)]_{ab}
Z_{\mathcal T}[b_1,b_2](\tau)\,,
\qquad a,b\in\mathbb Z_N^2\,.
\end{align}
For the generators $T:\tau\mapsto\tau+1$ and $S:\tau\mapsto-1/\tau$, the bosonic twisted partition functions obey
\begin{align}
[R_B(T)]_{ab}&=\delta_{b_1,a_1}\delta_{b_2,a_2-a_1}\,,\notag\\
[R_B(S)]_{ab}&=\delta_{b_1,a_2}\delta_{b_2,-a_1}\,,
\end{align}
where the Kronecker deltas are understood modulo $N$. Similarly, $R_{PF}^{(m)}(\gamma)$ is the $N^2\times N^2$ matrix defined by
\begin{align}
Z_{\mathcal T_{PF}}[\rho_1,\rho_2;m](\gamma\tau)
=\sum_{\rho'\in\mathbb Z_N^2}
[R_{PF}^{(m)}(\gamma)]_{\rho\rho'}
Z_{\mathcal T_{PF}}[\rho'_1,\rho'_2;m](\tau)\,.
\end{align}
Using $\boldsymbol Z_{PF}^{(m)}=F_m\boldsymbol Z_B$, its matrix elements are
\begin{align}
R_{PF}^{(m)}(\gamma)&=F_mR_B(\gamma)F_m^\dagger\,,\notag\\
[R_{PF}^{(m)}(\gamma)]_{\rho\rho'}
&=\sum_{a,b\in\mathbb Z_N^2}(F_m)_{\rho a}
[R_B(\gamma)]_{ab}(F_m^\dagger)_{b\rho'}\,.
\label{eq:pf_modularity}
\end{align}
Thus $R_{PF}^{(m)}$ is the modular representation obtained from $R_B$ by the change of basis $F_m$. It satisfies the same modular-group relations, with the same projective phases if the representation is projective. For $N>2$, the matrices generally mix partition functions with different paraspin labels rather than permuting them, in agreement with the modular covariance derived from the lattice construction in ref.~\cite{Yao:2020dqx}.

\paragraph{Multiplication by $\omega^{\ell\rho_1\rho_2}$ and inverse parafermionization.}

Let $D_\ell$ be the $N^2\times N^2$ diagonal matrix
\begin{align}
(D_\ell)_{\rho\rho'}&\equiv
\delta_{\rho_1,\rho'_1}\delta_{\rho_2,\rho'_2}
\omega^{\ell\rho_1\rho_2}\,,\qquad \ell\in\mathbb Z_N\,,\notag\\
(D_\ell\boldsymbol Z_{PF}^{(m)})_\rho
&=\omega^{\ell\rho_1\rho_2}
Z_{\mathcal T_{PF}}[\rho_1,\rho_2;m]\,.
\end{align}
In particular, $D_0=I_{N^2}$. Applying the inverse parafermionization transform to $D_\ell\boldsymbol Z_{PF}^{(m)}$ gives $F_m^\dagger D_\ell F_m\boldsymbol Z_B$, whose matrix elements are
\begin{align}
(F_m^\dagger D_\ell F_m)_{ab}
=\frac{\omega^{m(b_1b_2-a_1a_2)}}{N^2}
\sum_{\rho_1,\rho_2\in\mathbb Z_N}
\omega^{\ell\rho_1\rho_2
+m\rho_1(b_2-a_2)+m\rho_2(b_1-a_1)}\,.
\label{eq:stacking_sum}
\end{align}
For prime $N$ and $\ell\neq0$, the sum evaluates to
\begin{align}
(F_m^\dagger D_\ell F_m)_{ab}
=\frac{1}{N}\omega^{m(b_1b_2-a_1a_2)
-m^2\ell^{-1}(b_1-a_1)(b_2-a_2)}\,,
\label{eq:stacking_kernel}
\end{align}
where $\ell^{-1}$ is the multiplicative inverse of $\ell$ in $\mathbb Z_N$. In the component $a=(0,0)$, which corresponds to trivial background holonomies, the coefficient multiplying $b_1b_2$ in the exponent is $m-m^2\ell^{-1}$. Under the Dehn twist $b_2\mapsto b_2+b_1$, this term produces the additional phase $\omega^{(m-m^2\ell^{-1})b_1^2}$. The component with trivial background holonomies can therefore be modular covariant for general bosonic twisted partition functions only if $m-m^2\ell^{-1}=0$. Besides $\ell=0$, for which $D_0$ is the identity matrix, the nonzero solution is
\begin{align}
\ell=m\,.
\end{align}
For the convention $m=1$, multiplication by $\omega^{\rho_1\rho_2}$ followed by the inverse parafermionization transform therefore gives the $\mathbb Z_N$-orbifold partition function in the component with trivial background holonomies. Other values of $\ell$ generically retain the additional phase found above.

\paragraph{Charge conjugation and orbifolding.}

The reversal of the second paraspin label in eq.~\eqref{eq:parafermionization_Arfn2} can be written as the permutation matrix $C$ with matrix elements
\begin{align}
C_{\rho\rho'}&\equiv
\delta_{\rho'_1,\rho_1}\delta_{\rho'_2,-\rho_2}\,,\notag\\
(C\boldsymbol Z_{PF}^{(m)})_{\rho}
&=Z_{\mathcal T_{PF}}[\rho_1,-\rho_2;m]\,.
\end{align}
This is the charge-conjugation action on the second $\mathbb Z_N$ label. We also write the discrete Fourier transform appearing in the $\mathbb Z_N$ orbifold formula as the $N^2\times N^2$ matrix $O_m$,
\begin{align}
(O_m)_{ab}&\equiv\frac{1}{N}
\omega^{m(b_1a_2-b_2a_1)}\,,\notag\\
(O_m\boldsymbol Z_B)_a&=\frac{1}{N}
\sum_{b\in\mathbb Z_N^2}
\omega^{m(b_1a_2-b_2a_1)}Z_{\mathcal T}[b_1,b_2]\,.
\end{align}
A direct finite sum gives
\begin{align}
\boxed{F_m^\dagger D_mCF_m=O_m}\,.
\label{eq:conjugation_orbifold}
\end{align}
The reversal implemented by $C$ changes the finite sum so that its kernel contains the alternating pairing $b_1a_2-b_2a_1$ of the two background holonomies, as required by the orbifold formula. Without $C$, the same finite sum instead gives
\begin{align}
(F_m^\dagger D_mF_m\boldsymbol Z_B)_{a}
=\omega^{-2ma_1a_2}(O_m\boldsymbol Z_B)_{(-a_1,a_2)}\,.
\end{align}
The two expressions agree for $a=(0,0)$ but differ as transformations of the full vector of twisted partition functions. The distinction is invisible for $N=2$ and is essential for $N>2$.

Only the torus relations \eqref{eq:pf_modularity}--\eqref{eq:conjugation_orbifold} will be used in the remainder of the paper. In particular, the matrix $D_\ell$ has been defined only on the $N^2$-component vector of torus partition functions. An extension to paraspin structures on higher-genus surfaces would require additional state spaces and sewing conditions. The categorical discussion below instead starts from specified $\mathbb Z_N$-graded braiding data.

\subsection{TDLs in parafermionic CFTs}
We now characterize topological defect lines in a parafermionic CFT. In addition to the general properties of TDLs \cite{Chang:2018iay}, we focus on features specific to parafermionic theories. For $\mathbb Z_2$ parafermions, namely ordinary fermions, these features were studied in detail in \cite{Chang:2022hud}.

\subsubsection{Parafermion defect operators}
A distinguishing feature of parafermionic theories, compared with bosonic theories, is the presence of parafermionic zero modes. The simplest example appears in the fermionized Ising spin chain. After the Jordan--Wigner transformation, the Ising chain is rewritten in terms of Majorana degrees of freedom, which are nonlocal string operators in the original model. With open boundary conditions, one finds two normalized zero-energy modes localized at the ends of the chain. Similarly, the parafermionized spin chain in Eq.~(\ref{eq:Z_n_chain}) admits edge zero modes \cite{Fendley:2012vv}. We call these modes parafermionic defect operators; they obey the statistics in Eq.~(\ref{eq:parafermion_statistics}). These defect operators originate from the parafermionic-chain degrees of freedom, while the corresponding torus partition functions are related to the bosonic twisted partition functions by the parafermionization formula (\ref{eq:parafermionization_Arfn}).

The properties of $\mathbb Z_2$-parafermionic defect operators in fermionic theories were discussed in detail in \cite{Chang:2022hud}. We generalize this analysis to $\mathbb Z_N$ parafermions. In a $\mathbb Z_N$-parafermionic theory, parafermionic defect operators may occur at junctions of TDLs. According to Eq.~(\ref{eq:parafermion_statistics}), the defect Hilbert space $\mathcal H_{\mathcal L_1, \mathcal L_2, \mathcal L_3}$ associated with a trivalent junction containing a $\mathbb Z_N$-parafermionic defect operator,
\begin{align}
\begin{gathered}
\begin{tikzpicture}[scale=1.5]
    % \draw [lightgray] (0,0)  -- (0,2) -- (2,2) -- (2,0) -- (0,0) ;
    \draw[-<-=.5] (1,1) -- (0,2);
    \draw[-<-=.5] (1,1) -- (2,2);
    \draw[->-=.5] (1,1) -- (1,0);
    \draw node at (1.25, 0.9) {$\psi_k$};
    \draw node at (0.5, 1.75) {$\mathcal L_1$};
\draw node at (1.5, 1.75) {$\mathcal L_2$};
\draw node at (0.8, 0.25) {$\mathcal L_3$};
\node at (1,1)[circle,fill,inner sep=1.5pt,red,opacity=1]{};
\end{tikzpicture}
\end{gathered}
\end{align}
is, in general, $\mathbb Z_N$-graded. We label a junction state by the type $\psi_a$, with $a=0,\,\dots,\,N-1$, of its parafermionic defect operator; we call $a$ the color of the junction. These junctions inherit the fractional statistics in Eq.~(\ref{eq:parafermion_statistics}): exchanging two defect operators at the junctions multiplies the graph by a phase,
\begin{align}
\begin{gathered}
    \begin{tikzpicture}[scale=1]
% \draw [lightgray] (0,0)  -- (0,3) -- (3,3) -- (3,0) -- (0,0) ;
% \fill (1.5,1) circle (2pt);
% \fill (1.5,2) circle (2pt);
    \draw (1.5,1)--(1.5,2);
    \draw (1.5,1)--(0,0);
    \draw (1.5,1)--(3,0);
    \draw (1.5,2)--(3,3);
    \draw (1.5,2)--(0,3);    
    \draw node at (1.5,2.35) {$\psi_a$};
    \draw node at (1.5,0.65) {$\psi_b$};
    \draw[->] (1.65,1.05) arc (-45:45:0.65);
    \draw[<-] (1.35,1.05) arc (-135:-225:0.65);
    \node at (1.5,1)[circle,fill,inner sep=1.5pt,red,opacity=1]{};
    \node at (1.5,2)[circle,fill,inner sep=1.5pt,red,opacity=1]{};
    \end{tikzpicture}
\end{gathered}
    \quad=\quad e^{2\pi i\theta(a,b)}\quad
\begin{gathered}
    \begin{tikzpicture}[scale=1]
% \draw [lightgray] (0,0)  -- (0,3) -- (3,3) -- (3,0) -- (0,0) ;
% \fill (1.5,1) circle (2pt);
% \fill (1.5,2) circle (2pt);
    \draw (1.5,1)--(1.5,2);
    \draw (1.5,1)--(0,0);
    \draw (1.5,1)--(3,0);
    \draw (1.5,2)--(3,3);
    \draw (1.5,2)--(0,3);    
    \draw node at (1.5,2.35) {$\psi_b$};
    \draw node at (1.5,0.65) {$\psi_a$};
    \node at (1.5,1)[circle,fill,inner sep=1.5pt,red,opacity=1]{};
    \node at (1.5,2)[circle,fill,inner sep=1.5pt,red,opacity=1]{};
    %\draw[->] (1.65,1.05) arc (-45:45:0.65);
    %\draw[<-] (1.35,1.05) arc (-135:-225:0.65);
    \end{tikzpicture}
\end{gathered}\,,
\label{eq:defect_operator_statistics}
\end{align}
where the phase $\theta(a,b)$ modulo unity is uniquely determined by the colors $a$ and $b$. Similarly, for the tensor product of two TDLs dressed with parafermionic defect operators, we also have
\begin{align}
    \begin{gathered}
    \begin{tikzpicture}[scale=0.75]
% \draw [lightgray] (0,0)  -- (0,3) -- (3,3) -- (3,0) -- (0,0) ;
    \draw (-.75,2)--(-.75,-2);
    \draw (.75,2)--(.75,-2);
    \draw node at (0,0) {\small $\otimes$ };
    \draw node at (-1.1,1) {\tiny$\psi_a$};
    \draw node at (1.1,-1) {\tiny$\psi_b$};
    %\draw[->] (1.65,1.05) arc (-45:45:0.65);
    %\draw[<-] (1.35,1.05) arc (-135:-225:0.65);
    \node at (-.75,1)[circle,fill,inner sep=1pt,red,opacity=1]{};
    \node at (.75,-1)[circle,fill,inner sep=1pt,red,opacity=1]{};
    \end{tikzpicture}
\end{gathered}
\quad=\quad e^{2\pi i\theta(a,b)}\quad
\begin{gathered}
    \begin{tikzpicture}[scale=0.75]
% \draw [lightgray] (0,0)  -- (0,3) -- (3,3) -- (3,0) -- (0,0) ;
    \draw (-.75,2)--(-.75,-2);
    \draw (.75,2)--(.75,-2);
    \draw node at (0,0) {\small $\otimes$ };
    \draw node at (-1.1,-1) {\tiny$\psi_a$};
    \draw node at (1.1,1) {\tiny$\psi_b$};
    %\draw[->] (1.65,1.05) arc (-45:45:0.65);
    %\draw[<-] (1.35,1.05) arc (-135:-225:0.65);
    \node at (-.75,-1)[circle,fill,inner sep=1pt,red,opacity=1]{};
    \node at (.75,1)[circle,fill,inner sep=1pt,red,opacity=1]{};
    \end{tikzpicture}
\end{gathered}
\label{eq:defect_operator_tensor_exchange}
\end{align}
In the case of $N=2$, it is known that
\begin{align}
e^{2\pi i\theta(a,b)}=(-1)^{ab}\,,\quad{\rm for}\quad a,b=0,1\,,
\end{align}
corresponding to the fermionic statistics of the 1d Majorana zero modes when $a,\, b=1$. For generic $N$, the parafermionic statistics are labeled by an element $m\in\mathbb Z_N^\times$. We write
\begin{align}
    e^{2\pi i\theta(a,b)}=\omega_{N,m}^{ab}\,,\quad{\rm for}\quad a,b=0,1,\dots,N-1\,,
    \label{eq:phase}
\end{align}
where $\omega_{N,m}=e^{2\pi i m/N}$ and $\mathbb Z_N^\times=\{m\in\mathbb Z_N\mid\gcd(m,N)=1\}$. Thus $\omega_{N,m}$ is a primitive $N$-th root of unity. The label $m$ can be read from the conformal spin of the generator $\psi_1$. In the notation of Appendix~\ref{app:premetric-group}, a general cyclic pointed braided category is labeled by $k\in\mathbb Z_{\gcd(2,N)N}$, and its charge-$a$ object has conformal spin
\begin{align}
    h_a=\frac{k}{\gcd(2,N)N}a^2\mod 1\,.
\end{align}
The parafermionic statistics in Eq.~\eqref{eq:phase} correspond to $k=\gcd(2,N)m$. Consequently, the defect operator $\psi_a$ has fractional spin
\begin{align}
s_a=h_a=\frac{m}{N}a^2\mod 1\,.
\end{align} 
Exchanging two defect operators therefore gives
\begin{align}
\psi_a\,\psi_b=e^{2\pi i \frac{m}{N}ab}\psi_b\,\psi_a\,.
\end{align}
After parafermionization, the $\mathbb Z_N$ symmetry lines are gauged, while the operators $\psi_a$ remain as parafermionic defect operators living on TDLs and obeying Eq.~\eqref{eq:phase}.

Clearly, due to \eqref{eq:phase}, these parafermionic defect operators $\psi_a$ cannot be treated as usual local operators except for the case of $\mathbb Z_2$-parafermions. It can be easily seen that, by swapping two defect operators twice, there will be a phase
\begin{align}
e^{4\pi i \theta(a,b)}=\omega_{N,m}^{2ab}\neq 1\,,
\label{eq:double_phase}
\end{align} 
that is generically non-unity, except for $N=2$. In this sense, the parafermionic defects cannot be regarded as local operators. In contrast, it would be necessary to always imagine that they are ``stringy-like objects" that the strings attached to them are braided after swapping, and double braided for swapping twice to give the non-trivial phase in \eqref{eq:double_phase}. In this picture, the ways, clockwise or counter-clockwise, to switch two defect operators also need to been taken into account, i.e.
\begin{align}
\begin{gathered}
    \begin{tikzpicture}[scale=0.75]
% \draw [lightgray] (0,0)  -- (0,3) -- (3,3) -- (3,0) -- (0,0) ;
    \draw (1.5,1)--(1.5,2);
    \draw (1.5,1)--(0,0);
    \draw (1.5,1)--(3,0);
    \draw (1.5,2)--(3,3);
    \draw (1.5,2)--(0,3);    
    \draw node at (1.5,2.35) {\tiny$\psi_a$};
    \draw node at (1.5,0.65) {\tiny$\psi_b$};
    \draw[->] (1.65,1.05) arc (-45:45:0.65);
    \draw[<-] (1.35,1.05) arc (-135:-225:0.65);
    \node at (1.5,1)[circle,fill,inner sep=1pt,red,opacity=1]{};
    \node at (1.5,2)[circle,fill,inner sep=1pt,red,opacity=1]{};
    \end{tikzpicture}
\end{gathered}
   \ =\ \omega_{N,m}^{a b}\ 
\begin{gathered}
    \begin{tikzpicture}[scale=0.75]
% \draw [lightgray] (0,0)  -- (0,3) -- (3,3) -- (3,0) -- (0,0) ;
    \draw (1.5,1)--(1.5,2);
    \draw (1.5,1)--(0,0);
    \draw (1.5,1)--(3,0);
    \draw (1.5,2)--(3,3);
    \draw (1.5,2)--(0,3);    
    \draw node at (1.5,2.35) {\tiny$\psi_b$};
    \draw node at (1.5,0.65) {\tiny$\psi_a$};
    \node at (1.5,1)[circle,fill,inner sep=1pt,red,opacity=1]{};
    \node at (1.5,2)[circle,fill,inner sep=1pt,red,opacity=1]{};
    %\draw[->] (1.65,1.05) arc (-45:45:0.65);
    %\draw[<-] (1.35,1.05) arc (-135:-225:0.65);
    \end{tikzpicture}
\end{gathered}\,,
\ \ {\rm or}\ \ 
\begin{gathered}
    \begin{tikzpicture}[scale=0.75]
% \draw [lightgray] (0,0)  -- (0,3) -- (3,3) -- (3,0) -- (0,0) ;
    \draw (1.5,1)--(1.5,2);
    \draw (1.5,1)--(0,0);
    \draw (1.5,1)--(3,0);
    \draw (1.5,2)--(3,3);
    \draw (1.5,2)--(0,3);    
    \draw node at (1.5,2.35) {\tiny$\psi_a$};
    \draw node at (1.5,0.65) {\tiny$\psi_b$};
    \draw[<-] (1.65,1.05) arc (-45:45:0.65);
    \draw[->] (1.35,1.05) arc (-135:-225:0.65);
    \node at (1.5,1)[circle,fill,inner sep=1pt,red,opacity=1]{};
    \node at (1.5,2)[circle,fill,inner sep=1pt,red,opacity=1]{};
    \end{tikzpicture}
\end{gathered}
   \ =\ \omega_{N,m}^{-a b}\ 
\begin{gathered}
    \begin{tikzpicture}[scale=0.75]
% \draw [lightgray] (0,0)  -- (0,3) -- (3,3) -- (3,0) -- (0,0) ;
    \draw (1.5,1)--(1.5,2);
    \draw (1.5,1)--(0,0);
    \draw (1.5,1)--(3,0);
    \draw (1.5,2)--(3,3);
    \draw (1.5,2)--(0,3);    
    \draw node at (1.5,2.35) {\tiny$\psi_b$};
    \draw node at (1.5,0.65) {\tiny$\psi_a$};
    \node at (1.5,1)[circle,fill,inner sep=1pt,red,opacity=1]{};
    \node at (1.5,2)[circle,fill,inner sep=1pt,red,opacity=1]{};
    %\draw[->] (1.65,1.05) arc (-45:45:0.65);
    %\draw[<-] (1.35,1.05) arc (-135:-225:0.65);
    \end{tikzpicture}
\end{gathered}
\label{eq:defect_operator_oriented_exchange}
\end{align}
In the note, we will stick to the convention of swapping two parafermionic defect operators in the counter-clockwise fashion, and thus produce a phase as in \eqref{eq:phase}. Since a category of 2d TDLs does not have a canonical braiding, in Section~\ref{3} we characterize these fractional exchange phases using the half-braiding of the corresponding object in the Drinfeld center of the boundary fusion category.

\subsubsection{\boldmath $\mathbb Z_N$ Para-fusion category}
\label{parafusion}
Now we are ready to propose an axiomatic description of a $\mathbb Z_N$ para-fusion category $\mathcal C_{N,p}$ for a chosen finite, semisimple collection of TDLs in a parafermionic CFT that is closed under fusion and duals. We do not assume that $\mathcal C_{N,p}$ exhausts all topological lines of the theory. Most features of parafermionic TDLs are analogous to their bosonic counterparts. On the other hand, in contrast to ordinary fusion categories, there are two kinds of TDLs, called $m$-type and $q$-type, distinguished by their parafermionic defect operators.
\begin{enumerate}
\item {\bf TDLs (Objects):} An object $\mathcal L\in\mathcal C_{N,p}$ in a $\mathbb Z_N$ para-fusion category, is an oriented topological line operator defined on a path $C$, $\mathcal L(C)$, whose expectation value can be computed by inserting it in the path integral, and the dependence on $C$ is topological. 
\item {\bf Defect Operators (Morphisms):} The morphisms in $\mathcal C_{N,p}$ correspond to topological defect operators between two oriented lines $\mathcal L$ and $\mathcal K$. In contrast to ordinary fusion categories where the defect operators are required to be local, we here extend the notion to include the set of non-local parafermionic defect operators $\{\psi_a\}$. The defect operators form a (graded) vector space denoted by ${\rm Hom}(\mathcal L,\, \mathcal K)$. In addition, given defect operators $\mu\in {\rm Hom} (\mathcal L,\,\mathcal K)$ and $\nu\in {\rm Hom} (\mathcal K,\,\mathcal J)$, there is a composition of defect operators $\mu$ and $\nu$, denoted as $\nu\circ \mu\in {\rm Hom}(\mathcal L,\,\mathcal J)$, that can be thought to shrink the segment of topological line $\mathcal K$ to bring the two defect operator $\mu$ and $\nu$ together. Especially, for the composition of two parafermionic defect operators, we have
\begin{align}
\psi_a\circ\psi_b=\psi_{a+b\!\!\!\!\mod N}\,.
\label{eq:psi_OPE}
\end{align}
\item {\bf Additive Structure:} Given two TDLs $\mathcal L$ and $\mathcal K$, we define a new TDL as the sum of $\mathcal L$ and $\mathcal K$, denoted by $\mathcal L\oplus\mathcal K$, such that
\begin{align}
\left\langle\cdots\left(\mathcal L\oplus\mathcal K\right)(C)\cdots\right\rangle
=\left\langle\cdots\mathcal L(C)\cdots\right\rangle+\left\langle\cdots\mathcal K(C)\cdots\right\rangle\,.
\end{align}
\item {\bf Simplicity, Semisimplicity, and Finiteness:} A TDL in a para-fusion category $\mathcal C_{N,p}$ is simple if it cannot be decomposed as a direct sum of other TDLs. The category $\mathcal C_{N,p}$ is semisimple if every TDL is isomorphic to a direct sum of simple lines. Finally, we require $\mathcal C_{N,p}$ to contain only finitely many isomorphism classes of simple lines.

\item {\bf $\boldmath m$-type and $\boldmath q$-type of TDLs:} A simple $m$-type TDL $\mathcal L_m$ is defined as a simple line whose defect space ${\rm Hom}(\mathcal L_m,\,\mathcal L_m)$ only contains the trivial parafermionic defect $\psi_0$, and thus we have
\begin{align}
{\rm Hom}(\mathcal L_m,\,\mathcal L_m)={\rm Span}\{\psi_0\}\simeq \mathbb C^{\overbrace{1|0|\cdots|0}^{N}}\,,
\end{align}
where the Hom space is $N$-graded and the subscript of $\mathbb C$ labels the type, or color, of the parafermionic defect operator. A simple $q$-type TDL $\mathcal L_q$ is a simple line whose defect space ${\rm Hom}(\mathcal L_q,\,\mathcal L_q)$ contains at least one nontrivial parafermionic defect $\psi_a$ with $a\neq 0$. This definition has the following consequence. Given $\psi_{a\neq 0}\in {\rm Hom}(\mathcal L_q,\,\mathcal L_q)$, consider $\mathcal L_q$ dressed with two such defect operators. Shrinking the segment between them and applying \eqref{eq:psi_OPE}, one finds
\begin{align}
\psi_a\circ\psi_a=\psi_{2a\!\!\!\!\mod N}\in {\rm Hom}(\mathcal L_q,\,\mathcal L_q)\,.
\end{align} 
Repeating this procedure, we can find a collection of $\left\{\psi_a\right\}$, with $a\in \mathbb Z_M$ and $M$ dividing $N$, in ${\rm Hom}(\mathcal L_q,\,\mathcal L_q)$. In general, for ${\rm Hom}(\mathcal L_q,\,\mathcal L_q)$ spanned by a collection of $S=\{\psi_a\}$, one can always find a smallest $\psi_M$ of order $M$ generating the whole set $S=\left\langle\psi_M\right\rangle$. We thus denote such a TDL as a ``$\,\mathbb Z_M$ $q$-type TDL". For example, for a $\mathbb Z_N$ $q$-type TDL $\mathcal L_q$, its Hom space is given by
\begin{align}
{\rm Hom}(\mathcal L_q,\,\mathcal L_q)={\rm Span}\{\psi_1\}\simeq \mathbb C^{\overbrace{1|\cdots|1}^{N}}\,.
\end{align}
\item {\bf Fusion:} Given two TDLs $\mathcal J$ and $\mathcal K$, one can bring them close enough to fuse them as a single line, denoted by $\mathcal J\otimes\mathcal K$. By semisimplicity, $\mathcal J\otimes\mathcal K$ can be decomposed as a sum of simple TDLs $\{\mathcal L_i\}$, which is determined by the Hom space, ${\rm Hom}(\mathcal J\otimes\mathcal K,\,\mathcal L)$. Obviously parafermionic defect operators $\{\psi_a\}$ in ${\rm Hom}(\mathcal J\otimes\mathcal K,\,\mathcal L)$ define the colors of the 3-way junction.
\item{\bf Associativity Structure:} The fusion operation is associative in the sense that $\left(\mathcal I\otimes\mathcal J\right)\otimes\mathcal K$ and $\mathcal I\otimes\left(\mathcal J\otimes\mathcal K\right)$ are isomorphic by an associator
\begin{align}
\mathcal F^{\mathcal I,\,\mathcal J,\,\mathcal K}\in {\rm Hom}\left(\left(\mathcal I\otimes\mathcal J\right)\otimes\mathcal K,\,\mathcal I\otimes\left(\mathcal J\otimes\mathcal K\right)\right)\,,
\end{align}
where the associator $\mathcal F$ is also called F-symbol. In the para-fusion category $\mathcal C_{N,p}$, it is colored because of the four 3-way junctions in ${\rm Hom}\left(\left(\mathcal I\otimes\mathcal J\right)\otimes\mathcal K,\,\mathcal I\otimes\left(\mathcal J\otimes\mathcal K\right)\right)$. Expanded in a base $S=\{\mathcal L_i\}$ of simple lines, we have
\begin{align}
\begin{gathered}
\begin{tikzpicture}[scale=1]
\draw [line] (-1,1) -- (-2,2);
\draw [line] (-1,1) --(0,2);
\draw [line] (0,0) --(2,2);
\draw [line] (0,0)  -- (0,-1.5);
\draw [line] (0,0) -- (-1,1) ;
\node at (-1,1)[circle,fill,inner sep=1.5pt,red,opacity=1]{};
\node at (0,0)[circle,fill,inner sep=1.5pt,red,opacity=1]{};
\draw node at (-1.2, .8) {\tiny $\psi_a$};
\draw node at (-.2, -.2) {\tiny $\psi_b$};
\draw node at (-2.2, 2) {\tiny $\mathcal I$};
\draw node at (-.2, 2) {\tiny $\mathcal J$};
\draw node at (1.8, 2) {\tiny $\mathcal K$};
\draw node at (-.2, -1.5) {\tiny $\mathcal L_k$};
\draw node at (-.3, .7) {\tiny $\mathcal L_i$};
\end{tikzpicture}
\end{gathered}
=\quad \sum_{j;\,c,d}\,\mathcal F^{\mathcal I,\,\mathcal J,\,\mathcal K}_{\mathcal L_k}\left(\mathcal L_i,\,\mathcal L_j;\,a,\,b,\,c,\,d\right)\!\!\!\!\!\!\!\!\!\!\!\!\!\!\!\!\!\!\!\!
\begin{gathered}
\begin{tikzpicture}[scale=1]
\draw [line] (-2,2) -- (0,0);
\draw [line] (0,2) --(1,1);
\draw [line] (1,1) --(2,2);
\draw [line] (0,0)  -- (1,1);
\draw [line] (0,0) -- (0,-1.5) ;
\node at (1,1)[circle,fill,inner sep=1.5pt,red,opacity=1]{};
\node at (0,0)[circle,fill,inner sep=1.5pt,red,opacity=1]{};
\draw node at (1.3, .8) {\tiny $\psi_c$};
\draw node at (.3, -.2) {\tiny $\psi_d$};
\draw node at (-2.2, 2) {\tiny $\mathcal I$};
\draw node at (-.2, 2) {\tiny $\mathcal J$};
\draw node at (1.8, 2) {\tiny $\mathcal K$};
\draw node at (-.2, -1.5) {\tiny $\mathcal L_k$};
\draw node at (.3, .6) {\tiny $\mathcal L_j$};
\end{tikzpicture}
\end{gathered}\,,
\end{align}
where we have used $\{\mathcal L_i\}$ to decompose the fusions of $\mathcal I\otimes\mathcal J$, $\mathcal L_i\otimes\mathcal K$, $\mathcal J\otimes\mathcal K$, and $\mathcal I\otimes\mathcal L_j$\,. In addition, there is a selection rule imposed on the colors of a given F-symbol element because of the emergent $\mathbb Z_N$-symmetry \eqref{eq:pf_parity}, that
\begin{align}
\mathcal F^{\mathcal I,\,\mathcal J,\,\mathcal K}_{\mathcal L_k}\left(\mathcal L_i,\,\mathcal L_j;\,a,\,b,\,c,\,d\right)=0\,,\quad\quad {\rm if}\quad\quad a+b\neq c+d\mod{N}\,.
\end{align}
\item {\bf Para-fusion Pentagons:} As in an ordinary fusion category, the F-symbols of a para-fusion category satisfy pentagon identities. Here each trivalent junction carries a color, and exchanging colored junctions produces the phases in \eqref{eq:phase}. For $\mathbb Z_2$ parafermions, this phase is $-1$ and the pentagons become super-pentagons \cite{Aasen:2017ubm,Wang:2018pdc}. We generalize these identities to arbitrary $\mathbb Z_N$ parafermions. Consider the following diagrams with five external TDLs:
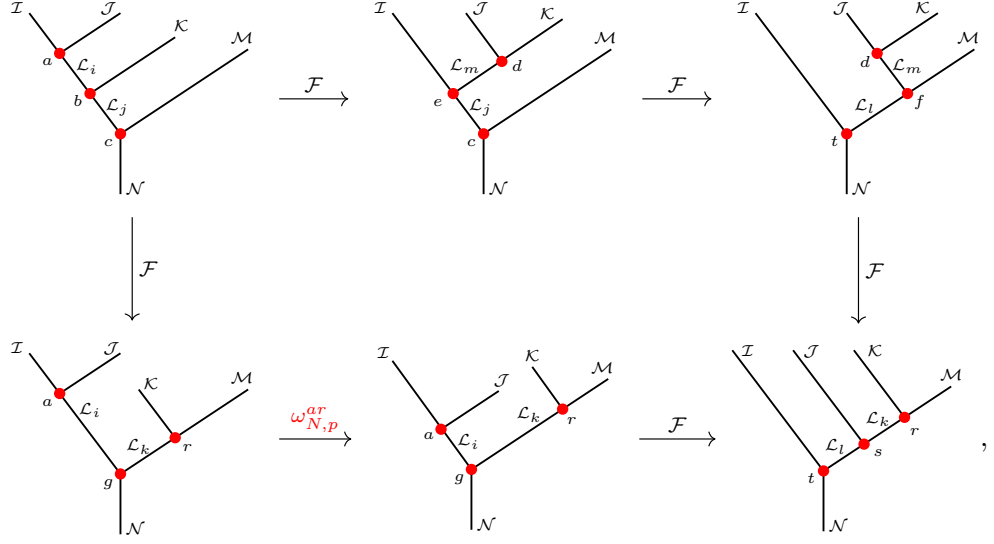
\begin{equation}
\label{eqn:FTchain}
\begin{tikzcd}
&
\begin{gathered}
\begin{tikzpicture}[scale=.8]
%\draw [line,lightgray] (0,0)  -- (0,3) -- (3,3) -- (3,0) -- (0,0) ;
\draw [line] (0,3) --(1.5,1);
\draw [line] (3.6,2.4) --(1.5,1);
\draw [line] (1.5,0) --(1.5,1);
\draw [line] (2.4,2.6) --(1,1.667);
\draw [line] (1.5,3) --(0.5,2.333);
\node at (1,1.667)[circle,fill,inner sep=1.5pt,red,opacity=1]{};
\node at (0.5,2.333)[circle,fill,inner sep=1.5pt,red,opacity=1]{};
\node at (1.5,1)[circle,fill,inner sep=1.5pt,red,opacity=1]{};
\draw node at (.8,1.467) {\tiny $b$};
\draw node at (.3,2.133) {\tiny $a$};
\draw node at (1.3,0.8) {\tiny $c$};
\draw node at (-.2, 3) {\tiny $\mathcal I$};
\draw node at (1.3, 3) {\tiny $\mathcal J$};
\draw node at (2.5, 2.7) {\tiny $\mathcal K$};
\draw node at (3.5, 2.5) {\tiny $\mathcal M$};
\draw node at (1.75, 0) {\tiny $\mathcal N$};
\draw node at (.95, 2.05) {\tiny $\mathcal L_i$};
\draw node at (1.45, 1.4) {\tiny $\mathcal L_j$};
\end{tikzpicture}
\end{gathered}
\arrow[r, "\mathcal F"]
\arrow[dd, "\mathcal F"]
&
\begin{gathered}
\begin{tikzpicture}[scale=.8]
%\draw [line,lightgray] (0,0)  -- (0,3) -- (3,3) -- (3,0) -- (0,0) ;
\draw [line] (0,3) --(1.5,1);
\draw [line] (3.6,2.4) --(1.5,1);
\draw [line] (1.5,0) --(1.5,1);
\draw [line] (2.8,2.9) --(1,1.667);
%\draw [line] (1.5,3) --(2,2.333);
\draw [line] (1.8,2.2) --(1.21,3);
\node at (1,1.667)[circle,fill,inner sep=1.5pt,red,opacity=1]{};
\node at (1.8,2.2) [circle,fill,inner sep=1.5pt,red,opacity=1]{};
\node at (1.5,1)[circle,fill,inner sep=1.5pt,red,opacity=1]{};
\draw node at (.75,1.467) {\tiny $e$};
\draw node at (2.05,2.033) {\tiny $d$};
\draw node at (1.3,0.8) {\tiny $c$};
\draw node at (-.2, 3) {\tiny $\mathcal I$};
\draw node at (1.45, 3) {\tiny $\mathcal J$};
\draw node at (2.55, 2.9) {\tiny $\mathcal K$};
\draw node at (3.5, 2.5) {\tiny $\mathcal M$};
\draw node at (1.75, 0) {\tiny $\mathcal N$};
\draw node at (1.2, 2.05) {\tiny $\mathcal L_m$};
\draw node at (1.45, 1.4) {\tiny $\mathcal L_j$};
\end{tikzpicture}
\end{gathered}
\arrow[r, "\mathcal F"]
&
\begin{gathered}
\begin{tikzpicture}[scale=.8]
%\draw [line,lightgray] (0,0)  -- (0,3) -- (3,3) -- (3,0) -- (0,0) ;
\draw [line] (0,3) --(1.5,1);
\draw [line] (3.6,2.4) --(1.5,1);
\draw [line] (1.5,0) --(1.5,1);
\draw [line] (3,3) --(2,2.333);
\draw [line] (1.5,3) --(2,2.333);
\draw [line] (2,2.333)--(2.5,1.667);
\node at (2,2.333)[circle,fill,inner sep=1.5pt,red,opacity=1]{};
\node at (2.5,1.667)[circle,fill,inner sep=1.5pt,red,opacity=1]{};
\node at (1.5,1)[circle,fill,inner sep=1.5pt,red,opacity=1]{};
\draw node at (2.7,1.467) {\tiny $f$};
\draw node at (1.8,2.1) {\tiny $d$};
\draw node at (1.3,0.8) {\tiny $t$};
\draw node at (-.2, 3) {\tiny $\mathcal I$};
\draw node at (1.4, 3) {\tiny $\mathcal J$};
\draw node at (2.75, 3) {\tiny $\mathcal K$};
\draw node at (3.5, 2.5) {\tiny $\mathcal M$};
\draw node at (1.75, 0) {\tiny $\mathcal N$};
\draw node at (2.5, 2.05) {\tiny $\mathcal L_m$};
\draw node at (1.8, 1.4) {\tiny $\mathcal L_l$};
\end{tikzpicture}
\end{gathered}
\arrow[dd, "\mathcal F"]
\\
\\
&
\begin{gathered}
\begin{tikzpicture}[scale=.8]
%\draw [line,lightgray] (0,0)  -- (0,3) -- (3,3) -- (3,0) -- (0,0) ;
\draw [line] (0,3) --(1.5,1); %I
\draw [line] (3.6,2.4) --(1.5,1); %M
\draw [line] (1.5,0) --(1.5,1); %N
\draw [line] (1.5,3) --(0.5,2.333); %J
\draw [line] (1.8, 2.4) -- (2.4, 1.6); %K
\node at (0.5,2.333)[circle,fill,inner sep=1.5pt,red,opacity=1]{};
\node at (2.4, 1.6)[circle,fill,inner sep=1.5pt,red,opacity=1]{};
\node at (1.5,1)[circle,fill,inner sep=1.5pt,red,opacity=1]{};
\draw node at (.3,2.133) {\tiny $a$};
\draw node at (2.6, 1.388) {\tiny $r$};
\draw node at (1.3,0.8) {\tiny $g$};
\draw node at (-.2, 3) {\tiny $\mathcal I$};
\draw node at (1.35, 3) {\tiny $\mathcal J$};
\draw node at (2, 2.4) {\tiny $\mathcal K$};
\draw node at (3.5, 2.5) {\tiny $\mathcal M$};
\draw node at (1.75, 0) {\tiny $\mathcal N$};
\draw node at (1.0, 2.0) {\tiny $\mathcal L_i$};
\draw node at (1.8, 1.4) {\tiny $\mathcal L_k$};
\end{tikzpicture}
\end{gathered}
\arrow[r, "{\color{red}\omega_{N,p}^{ar}}"]
&
\begin{gathered}
\begin{tikzpicture}[scale=.8]
%\draw [line,lightgray] (0,0)  -- (0,3) -- (3,3) -- (3,0) -- (0,0) ;
\draw [line] (0.2,2.8) --(1.5,1); %I
\draw [line] (3.75,2.5) --(1.5,1); %M
\draw [line] (1.5,0) --(1.5,1); %N
\draw [line] (1.95,2.3) --(1,1.666); %J
\draw [line] (2.5,2.7) -- (3, 2); %K
\node at (1,1.666)[circle,fill,inner sep=1.5pt,red,opacity=1]{};
\node at (3, 2)[circle,fill,inner sep=1.5pt,red,opacity=1]{};
\node at (1.5,1)[circle,fill,inner sep=1.5pt,red,opacity=1]{};
\draw node at (.8,1.5) {\tiny $a$};
\draw node at (3.15, 1.75) {\tiny $r$};
\draw node at (1.3,0.8) {\tiny $g$};
\draw node at (0.05, 2.85) {\tiny $\mathcal I$};
\draw node at (2, 2.4) {\tiny $\mathcal J$};
\draw node at (2.5, 2.8) {\tiny $\mathcal K$};
\draw node at (3.65, 2.6) {\tiny $\mathcal M$};
\draw node at (1.75, 0) {\tiny $\mathcal N$};
\draw node at (1.45, 1.4) {\tiny $\mathcal L_i$};
\draw node at (2.45, 1.85) {\tiny $\mathcal L_k$};
\end{tikzpicture}
\end{gathered}
\arrow[r, "\mathcal F"]
&
\begin{gathered}
\begin{tikzpicture}[scale=.8]
%\draw [line,lightgray] (0,0)  -- (0,3) -- (3,3) -- (3,0) -- (0,0) ;
\draw [line] (0,3) --(1.5,1);
\draw [line] (3.6,2.4) --(1.5,1);
\draw [line] (1.5,0) --(1.5,1);
\draw [line] (1,3) -- (2.167, 1.444);
\draw [line] (2,3) -- (2.833, 1.888);
\node at (2.167, 1.444)[circle,fill,inner sep=1.5pt,red,opacity=1]{};
\node at (2.833, 1.888)[circle,fill,inner sep=1.5pt,red,opacity=1]{};
\node at (1.5,1)[circle,fill,inner sep=1.5pt,red,opacity=1]{};
\draw node at (2.4, 1.244) {\tiny $s$};
\draw node at (3, 1.588) {\tiny $r$};
\draw node at (1.3,0.8) {\tiny $t$};
\draw node at (.25, 2.9) {\tiny $\mathcal I$};
\draw node at (1.3, 2.9) {\tiny $\mathcal J$};
\draw node at (2.35, 2.9) {\tiny $\mathcal K$};
\draw node at (3.6, 2.5) {\tiny $\mathcal M$};
\draw node at (1.75, 0) {\tiny $\mathcal N$};
\draw node at (1.7, 1.35) {\tiny $\mathcal L_l$};
\draw node at (2.4, 1.8) {\tiny $\mathcal L_k$};
\end{tikzpicture}
\end{gathered}
\,,\notag
\end{tikzcd}
\end{equation}
where, when switching the position of colors ``$a$" and ``$r$" from the first to the second diagram of the second line, we have employed \eqref{eq:defect_operator_statistics} and \eqref{eq:phase} to pick up a phase $\omega_{N,p}^{ar}$. The two routes, to bring the above diagrams from left top corner to the right bottom one, give us constraints on the F-symbols, denoted as the para-pentagon equations in $\mathcal C_{N,p}$. In the base $S=\{\mathcal L_i\}$, the para-pentagons of above diagrams spell as
\begin{align}
\!\!\!\!
&\sum_{m;\,d,e,f}\mathcal F^{\mathcal I,\,\mathcal J,\,\mathcal K}_{\mathcal L_j}\left(\mathcal L_i,\mathcal L_m;\,a,b,d,e\right)\,
\mathcal F^{\mathcal I,\,\mathcal L_m,\,\mathcal M}_{\mathcal N}\left(\mathcal L_j,\mathcal L_l;\,e,c,f,t\right)\,
\mathcal F^{\mathcal J,\,\mathcal K,\,\mathcal M}_{\mathcal L_l}\left(\mathcal L_m,\mathcal L_k;\,d,f,r,s\right)\notag\\\notag\\
&=\sum_{g}\omega_{N,p}^{ar}\,
\mathcal F^{\mathcal L_i,\,\mathcal K,\,\mathcal M}_{\mathcal N}\left(\mathcal L_j,\mathcal L_k;\,b,c,r,g\right)\,
\mathcal F^{\mathcal I,\,\mathcal J,\,\mathcal L_k}_{\mathcal N}\left(\mathcal L_i,\mathcal L_l;\,a,g,s,t\right)\,,
\label{eq:parapentagons}
\end{align}
where we have projected the para-pentagon equations to the diagram with internal TDLs $\left(\mathcal L_k,\mathcal L_l\right)$, and colors $\left(r,\,s,\,t\right)$.
%%%
%%%
%%%
%%%
%%%
%%%
\end{enumerate}

\section{Categorical Parafermionic Anyon Condensation}
\label{3}

In previous sections, we discussed parafermionization in 2d CFTs. In this section, we lift this 2d discussion to a 3d Turaev-Viro-Levin-Wen string-net framework~\cite{TV1992,LW2005}. The point of the lift is not to define a separate operation unrelated to the 2d story, but to make the origin of the parafermionic defect operators and their fractional statistics more transparent. The 3d bulk keeps track of the anyonic line that is condensed, while the 2d boundary records the resulting TDLs and the defect operators living on their junctions.
We will explore how parafermionic anyon condensation\footnote{In the fusion-category construction below, ``anyon condensation in $\C$'' means condensation of an algebra object in $\C$, as in Appendix~\ref{app:bosonic-condensation}. The lift to $Z(\C)$ provides the half-braiding, while the condensation is performed in $\C$. We distinguish this use from ordinary anyon condensation in a braided category.} in the 3d bulk is related to the parafermionization of the 2d boundary theory. Parafermionic anyon condensation provides us with a categorical understanding of parafermionization, allowing us to better understand the relation between the TDLs before and after parafermionization.

The bosonic condensation reviewed in Appendix~\ref{app:bosonic-condensation} provides the template for the following construction. Given an algebra $A$ in a fusion category $\C$, the right module category $\C_A$ is generated by objects of the form $X\otimes A$, and the Hom spaces are controlled by the adjunction relation
\begin{align}
\Hom_{\C_A}(X\otimes A,M)\simeq \Hom_\C(X,\mathrm{Forget}(M)).
\end{align}
For $M=Y\otimes A$, this says that the morphisms in the condensed category are obtained by expanding $Y\otimes A$ in the parent category. Fermionic and parafermionic condensation use the same mechanism, but the summands of $A$ become nontrivial graded morphism sectors. Thus the ordinary Hom spaces in the bosonic condensed category are replaced by $\Z_2$-graded Hom spaces in the fermionic case and by $\Z_N$-graded Hom spaces in the parafermionic case.

The main result is that, if the TDLs in the original 2d theory are described by a fusion category $\mathcal{C}$, then after condensing a suitable algebra $A$, the TDLs of the condensed theory are described by the corresponding module category with extra structure. Ordinary bosonic condensation gives a fusion category such as $_A\mathcal C_A$ or, in the central commutative case, $\mathcal C_A$. Fermionic condensation replaces the Hom spaces by $\Z_2$-graded ones and produces a super-fusion category $\mathcal C_{A^f}$. Parafermionic condensation further replaces them by $\Z_N$-graded Hom spaces and produces a para-fusion category $\mathcal C_{A^{pf}}$. From the point of view of 3d topological orders, the generalized Turaev-Viro-Levin-Wen string-net model before and after condensation are constructed with input data $\mathcal C$ and the corresponding fusion/super-fusion/para-fusion category, respectively. There is a gapped domain wall between these two topological orders.

The remainder of this section is divided into four parts. Subsection~\ref{sec:fcond} reviews fermionic anyon condensation in both a unitary modular tensor category (UMTC) and a unitary fusion category (UFC). Subsection~\ref{sec:pfcond} extends this construction to parafermionic anyon condensation within a UMTC. Subsection~\ref{sec:pfcond2} focuses on the case of a UFC, where a parafermion is defined as an object that can be lifted to the Drinfeld center as a parafermion in the UMTC sense, followed by a discussion of its condensation in this UFC. Finally, Subsection~\ref{sec:fusion2parafusion} provides a detailed procedure for deriving the para-fusion category from a parent UFC through the parafermion condensation. This includes an in-depth explanation of how the $F$-symbols in the condensed theory are obtained from those in the original theory. For a foundational discussion on bosonic anyon condensation, please refer to Appendix~\ref{app:bosonic-condensation}.

\subsection{Fermionic anyon condensation}
\label{sec:fcond}

In this subsection, we will review fermionic anyon condensation in both a UMTC and a UFC separately. This will establish a conceptual framework and set of conventions, which will serve as a basis for comparison with the generalized parafermionic anyon condensation discussed in the following sections.

\subsubsection{Fermionic anyon condensation in UMTC $\B$}

A topological order can encompass a fermion $f$ as an object characterized by nontrivial self-braiding. In a braided tensor category $\B$, it's possible to carry out fermionic anyon condensation following the methodology detailed in Ref.~\cite{wan_fermion_2017}.

To perform fermion condensation in a UMTC $\B$, we introduce another minimal fermionic system $\F_0$ with two objects ${1,\psi}$, where $\psi$ is a simple fermion with fusion rule $\psi\times\psi=1$. We then consider the stacked model $\B\boxtimes\F_0$ of $\B$ and $\F_0$, where a new boson $(f,\psi)$ arises from the fusion of the fermion $f$ in $\B$ and $\psi$ in $\F_0$. The fermionic anyon condensation is defined as the bosonic anyon condensation introduced previously by modding out the boson $(f,\psi)$ in $\B\boxtimes\F_0$. In other words, we choose the condensable algebra to be $A=(1,1)\oplus(f,\psi)$ in $\B\boxtimes\F_0$, and the condensed phase is given by the category $(\B\boxtimes\F_0)_A^0$. This condensed phase is not modular because the fermion $\psi$ is deconfined in the condensation process and has trivial full braidings with all anyons. The total quantum dimension of the condensed phase is related to that of the original phase by 
\begin{align}\label{dim_f}
\dim((\B\boxtimes\F_0)_A^0)=\frac{\dim\B\times \dim\F_0}{(\dim A)^2} = \frac{1}{2}\dim\B,
\end{align}
where $\dim\F_0=2$ and $\dim A = 2$.

\subsubsection{Fermionic anyon condensation in UFC $\C$}

%\cite{ostrik2003module}

Let's now explore the process of condensing a fermion within a fusion category $\C$. In this context, a fermion $f$ in $\C$ means an object whose lift to the Drinfeld center $Z(\C)$ exhibits fermionic behavior. Equivalently, $f$ is an invertible object of $\C$ equipped with half-braiding data inherited from $Z(\C)$, and this half-braiding is what allows us to speak about its fermionic self-statistics even though $\C$ itself is not braided in general.

The challenge arises due to the nontrivial self-braiding of fermions, preventing their arbitrary creation or annihilation within a vacuum. However, we can circumvent this issue by employing fermion creation and annihilation operators to terminate fermionic worldlines. This process elevates a fermionic object to the status of a fermionic morphism. Consequently, we can perform fermion condensation~\cite{Aasen:2017ubm,Zhou:2021ulc} within a fusion category, resulting in a super-fusion category — a category featuring $\mathbb{Z}_2$-graded Hom spaces.

If we consider the fusion category as the input for a Levin-Wen string-net model, the condensed super-fusion category becomes the input data in constructing a 3D fermionic string-net model. Hence, fermion condensation serves as a method to generate fermionic topological order from a bosonic one. The study of fermionic string-net models originated with the fermionic toric code model~\cite{Gu:2013gma} and its extensions~\cite{GuWangWen}. This concept evolved further to encompass Majorana-type fermionic string-net models~\cite{Aasen:2017ubm,Zhou:2021ulc}. It's worth noting that fermionic string-net models can also be interpreted as gauging  of the bosonic subgroup of a fermionic symmetry-protected topological phase (FSPT). The classification of FSPT is addressed using group super-cohomology theory~\cite{Gu:2012ib,Wang:2017moj,PhysRevX.10.031055,wang2021exactly} or spin cobordism~\cite{Kapustin:2014dxa}, revealing both complex fermion and Majorana fermion layers within the FSPT classification.

In a super-fusion category, the Hom spaces exhibit a $\Z_2$-graded structure. The even and odd components correspond to bosonic and fermionic operators, respectively. Specifically, for an object $X$ within the super-fusion category, the $\End(X)$ space becomes a super division algebra, with only two distinct irreducible types: the complex numbers themselves $\mathbb C$ and the Clifford algebra $\mathbb{C}l_1=\mathbb C^{1|1}$. Objects possessing $\End(X) = \mathbb{C}$ and $\mathbb{C}l_1$ are referred to as $m$-type and $q$-type objects, respectively~\cite{Aasen:2017ubm}.

Now, let's assume that there exists a fermion $f$ within a fusion category $\C$. This fermion has the fusion rule $f \times f = 1$. The algebra we intend to condense is $A^f = 1 \oplus f$. This algebra should be associative, meaning that the $\Z_2$ symmetry generated by $f$ is non-anomalous. Upon completion of the condensation process, a super-fusion category denoted as $\C_{A^f}$ emerges.

At the object level, the fermion $f$ is effectively identified with the vacuum object $1$. In general, an object $X$ should be identified with $X\otimes f$ in the super-fusion category $\C_{A^f}$. So an object $[X]$ within $\C_{A^f}$ corresponds to $X \otimes A^f$ in terms of the original fusion category $\C$.

At the level of morphisms, analogous to the bosonic anyon condensation given by Eq.~(\ref{adjoint}), we should have the following relation for fermionic anyon condensation:
\begin{align}\nonumber\label{ajoint_f}
\Hom_{\C_{A^f}}(X \otimes A^f, Y \otimes A^f)
&= \Hom_{\C}(X, Y \otimes A^f)\\
&= \Hom_{\C}(X, Y) \oplus \Hom_{\C}(X, Y \otimes f).
\end{align}
This equation connects the Hom spaces of the condensed super-fusion category to those of the original fusion category. From this relation, we can discern two types of objects that arise after the condensation. The first type encompasses objects $a \in \C$ such that $a \not\simeq a \times f$. In this scenario, the endomorphism $\End_{\C_{A^f}}([X]) = \Hom_{\C_{A^f}}(X \otimes A^f, X \otimes A^f)$ is simply $\mathbb C$. This corresponds to the $m$-type object within $\C_{A^f}$. Conversely, if $X$ remains unchanged under the fusion with $f$, then both terms on the right-hand side of Eq.~(\ref{ajoint_f}) contribute. In this case, the endomorphism $\End_{\C_{A^f}}([X])$ corresponds to $\mathbb{C}l_1 = \mathbb C^{1|1}$. Thus, the fixed point object $X$ with $X\simeq X\otimes f$ will lead to the condensation of a $q$-type object within $\C_{A^f}$.

At the next level of natural transformations involving the associator, we can proceed to construct the $F$ symbols of the super-fusion category using the $F$ symbols of the original fusion category $\C$. A comprehensive explanation of this construction is provided in section~\ref{sec:fusion2parafusion} in the general $\Z_N$ case.

As an example, we can consider $\C$ to be the Ising category with objects $1,\psi$ and $\sigma$. The endomorphism of $[\sigma]=\sigma\otimes A^f$ after the condensation of $A^f=1\oplus \psi$ is
\begin{align}\nonumber
\End_{\C_{A^f}}([\sigma])
&=\Hom_{\C_{A^f}} (\sigma\otimes A^f,\sigma\otimes A^f)
=\Hom_{\C} (\sigma,\sigma\otimes A^f)
=\Hom_{\C} (\sigma,\sigma\otimes (1\oplus\psi))\\
&=\Hom_{\C} (\sigma,\sigma \oplus (\sigma\otimes\psi))
=\Hom_{\C} (\sigma,\sigma) \oplus \Hom_{\C} (\sigma,\sigma\otimes\psi)
=\mathbb C^{1|1}.
\end{align}
So $[\sigma]=\sigma\otimes A\in\C_{A^f}$ is a $q$-type object. The condensed super-fusion category has only two simple objects $[1]$ and $[\sigma]$.

The total quantum dimension of the super-fusion category can be defined as
\begin{align}
\dim (\C_{A^f}) = \sum_{a\in \C_{A^f}} \frac{d_a^2}{\dim\End(a)} 
= \sum_{a\in m\text{ type}} d_a^2 + \frac{1}{2} \sum_{a\in q\text{ type}} d_a^2.
\end{align}
By separating the objects in $\C$ into those that are either non-fixed or fixed points under fusion with $f$, it becomes evident that the dimensions before and after the fermion condensation are related by
\begin{align}
\dim(\C_{A^f}) = \frac{1}{2}\dim(\C).
\end{align}
It exhibits similarities to the dimension of fermion condensation in a UMTC $\B$, as seen in Eq.~(\ref{dim_f}).

% \subsubsection{Excitations or Drinfeld center of $\C_{A^f}$}

% \subsubsection{Relation between condensations in $\B$ and $\C$}

We can also explore fermion condensation in a fusion category from the perspective of 3d TQFT. Similar to previous discussions, the original fusion category $\C$ serves as the input for the Hamiltonian of the Levin-Wen string-net model. After fermion condensation, the resulting super-fusion category $\C_{A^f}$ can be employed to construct a fermionic string-net model~\cite{Zhou:2021ulc}. The excitations in this model should be described by a fermionic analog of the Drinfeld center, denoted as $Z_f(\C_{A^f})$. In practice, tube algebras can be used as tools to analyze the properties of these fermionic excitations~\cite{Gu:2013gma,Aasen:2017ubm}.

The fermionic condensation from the fusion category $\C$ to the super-fusion category $\C_{A^f}$ implies the existence of a gapped domain wall between the bosonic Levin-Wen string-net model $Z(\C)$ and the fermionic string-net model $Z_f(\C_{A^f})$. This raises the question of whether fermion condensation in the fusion category $\C$ can be connected to fermion condensation in the bulk UMTC $Z(\C)$. In the special case where $\C=\B$ is a UMTC with a fermion, we have a relation~\cite{Aasen:2017ubm}:
\begin{align}
Z_f(\B_{A^f})=\B\boxtimes (\B_{A^f}),
\end{align}
However, in the general scenario of a fusion category, this relation remains to be established.

\subsection{Parafermionic anyon condensation in UMTC $\B$}
\label{sec:pfcond}

In a manner similar to fermion condensation in a braided category~\cite{wan_fermion_2017}, we extend the construction from $\Z_2$ fermions to $\Z_N$ parafermions.

Let $\B$ be a UMTC. In this paper, a $\Z_N$ parafermionic line of type $m$ is a simple invertible object $f\in\B$ of exact fusion order $N$ whose topological spin is $h_f=m/N$ modulo one, where $m\in\mathbb Z_N^\times$. Equivalently,
\begin{align}\label{eq:parafermionic-line-definition}
f^{\otimes N}\simeq 1,\qquad
f^{\otimes j}\not\simeq 1\quad (1\leq j<N),\qquad
\theta_f=e^{2\pi i h_f}=e^{2\pi i m/N}.
\end{align}
Here $\mathbb Z_N^\times=\{m\in\mathbb Z_N\mid\gcd(m,N)=1\}$, so $\theta_f$ is a primitive $N$-th root of unity. The powers $f^a$, with $a\in\mathbb Z_N$, generate a pointed braided fusion subcategory $\langle f\rangle\subset\B$ with fusion $f^a\otimes f^b\simeq f^{a+b}$. Its twists are
\begin{align}\label{eq:parafermionic-line-twists}
\theta_{f^a}=e^{2\pi i m a^2/N},\qquad a\in\mathbb Z_N.
\end{align}
The same coprime label $m$ specifies the exchange phase $\omega_{N,m}$ in Eq.~\eqref{eq:phase}. In particular, for $N=3$, the two choices $m=1,2$ give topological spins $h_f=1/3$ and $2/3$ modulo one, respectively; both are included in the definition.

Appendix~\ref{app:premetric-group} describes a general pointed braided category with fusion ring $\Z_N$ by the pre-metric group $\C(\Z_N,q_k)$, where $k$ is defined modulo $\gcd(2,N)N$. If $d:=\gcd(2,N)$, the parafermionic line in Eq.~\eqref{eq:parafermionic-line-definition} corresponds to
\begin{align}\label{eq:k-m-dictionary}
k=dm\pmod{dN},\qquad q_k(a)=\theta_{f^a}=e^{2\pi i m a^2/N}.
\end{align}
The associator of this pointed subcategory determines whether the $\Z_N$ orbit can be condensed. In the notation of Appendix~\ref{app:premetric-group},
\begin{align}\label{non-anomalous}
\text{non-anomalous subcategory $\C(\Z_N,q_k)$} &\Longleftrightarrow \text{$N$ is odd or $k$ is even}.
\end{align}
For $k=dm$, this condition is automatic: $N$ is either odd, or $N$ is even and $k=2m$ is even. The associator condition is nevertheless conceptually distinct from the spin definition in Eq.~\eqref{eq:parafermionic-line-definition}; it is the condition that makes the orbit algebra associative. We can then introduce the inverse pointed braided category $\F_0=\C(\Z_N,q_{-k})$, whose simple objects are $\{1,\psi,\ldots,\psi^{N-1}\}$. The generator $\psi$ has exact fusion order $N$ and twists opposite to those of $f$. The product category $\B\boxtimes\F_0$ therefore contains a diagonal bosonic $\Z_N$ sector generated by $(f,\psi)$.

The role of the auxiliary system $\F_0$ is to cancel the fractional statistics of the original parafermion. For $i=0,1,\dots,N-1$, the object $f^i$ in $\B$ has the opposite topological spin to $\psi^i$ in $\F_0$, so
\begin{align}
\theta_{(f^i,\psi^i)}=\theta_{f^i}\theta_{\psi^i}=1.
\end{align}
Moreover, for $i,j=0,1,\dots,N-1$, their monodromy phases also cancel,
\begin{align}
M_{(f^i,\psi^i),(f^j,\psi^j)}
 =
M_{f^i,f^j}M_{\psi^i,\psi^j}
 =1.
\end{align}
Thus the diagonal sector has trivial self- and mutual statistics. In this sense, parafermionic condensation in a UMTC is reduced to ordinary bosonic condensation after stacking with the inverse parafermionic theory $\F_0$.

The concept of parafermionic anyon condensation is established by employing the same principles as bosonic anyon condensation, but applied to the diagonal bosonic algebra. Although the diagonal sector is generated by the single object $(f,\psi)$, the algebra to be condensed must contain all of its powers in order to be closed under fusion:
\begin{align}
A=A_\Delta=\bigoplus_{i=0}^{N-1}(f^i,\psi^i)
\end{align}
within $\B\boxtimes\F_0$. Indeed, $(f,\psi)^i=(f^i,\psi^i)$ and $(f,\psi)^N=(1,1)$, so $A_\Delta$ is the full group algebra of this diagonal $\Z_N$ sector. The previous cancellation of topological spins and monodromies is precisely what makes $A_\Delta$ a commutative condensable algebra. The non-anomalous condition ensures that its multiplication can be chosen associatively. Condensing $A_\Delta$ then gives the condensed phase characterized by the category $(\B\boxtimes\F_0)_A^0$. By a straightforward calculation, the total quantum dimension of the condensed phase is given by the formula:
\begin{align}\label{dim_pf}
\dim((\B\boxtimes\F_0)_A^0)=\frac{\dim\B\times \dim\F_0}{(\dim A)^2} = \frac{1}{N}\dim\B,
\end{align}
where $\dim\F_0=N$ and $\dim A = N$. It generalizes the fermion case in a straightforward manner.

A special case arises when the pointed subcategory $\C(\Z_N,q_k)$ is modular. As shown in Appendix~\ref{app:premetric-group},
\begin{align}
\text{$\C(\Z_N,q_k)$ is modular $\Longleftrightarrow$ $q_k$ is non-degenerate $\Longleftrightarrow$ $\gcd(k,N)=1$.}
\end{align}
For the parafermionic types $k=dm$ considered here, this condition holds for odd $N$. For even $N$, $k=2m$ and the pointed subcategory is not modular. In the modular case, the UMTC $\B$ can be decomposed as follows:
\begin{align}
\B = \C(\Z_N,q_k) \boxtimes \B',
\end{align}
where $\B'$ is another subcategory of $\B$ that is also modular. Since the anyons in $\C(\Z_N,q_k)$ and $\B'$ are unrelated and have trivial mutual braidings, the parafermion condensation is independent of $\B'$. Therefore, the condensed phase is solely determined by the parafermion condensation within the category $\C(\Z_N,q_k)$:
\begin{align}
(\B\boxtimes\F_0)_A^0 
=(\C(\Z_N,q_k)\boxtimes\B'\boxtimes\F_0)_A^0 
=(\C(\Z_N,q_k)\boxtimes\F_0)_A^0 \boxtimes\B'
=\B'.
\end{align}
In the last step, the diagonal algebra identifies $f^i$ with the dual object $\psi^{-i}=\psi^{N-i}$ for $1 \leq i \leq N-1$. At the level of right $A$-modules, this allows one to move the $\psi$-charge into the $\C(\Z_N,q_k)$ factor, so $(\C(\Z_N,q_k)\boxtimes\F_0)_A$ is equivalent, as a module category, to $\C(\Z_N,q_k)$. The deconfined sector, however, is the subcategory of local $A$-modules. Since $\C(\Z_N,q_k)$ is modular, every nontrivial simple object has nontrivial monodromy with some object in the condensed diagonal sector. Hence only the vacuum is local, and $(\C(\Z_N,q_k)\boxtimes\F_0)_A^0$ is trivial. In summary, when the $\Z_N$ part of $\B$ is modular, the condensed phase simplifies to $\B'$ which can be considered orthogonal to $\Z_N$ due to the decomposition $\B=\C(\Z_N,q_k)\boxtimes \B'$.

\subsection{Parafermionic anyon condensation in UFC $\C$}
\label{sec:pfcond2}

In this section, we will explore parafermionic anyon condensation in a fusion category $\C$. Similar to fermion condensation, a $\Z_N$ parafermion of type $m$ in $\C$ consists of a simple invertible object $f$ of exact fusion order $N$ together with a chosen central lift $\widetilde f\in Z(\C)$ that satisfies the UMTC definition in Eq.~\eqref{eq:parafermionic-line-definition}. This lift consists of $f$ and half-braiding data with every object of $\C$. Since $\C$ is not braided in general, the fractional statistics of $f$ are not intrinsic to the object $f\in\C$ alone; they are part of the extra central data carried by $\widetilde f$. The pointed braided subcategory generated by $\widetilde f$ is $\C(\Z_N,q_k)$ with $k=\gcd(2,N)m$.

This is the parafermionic counterpart of the central commutative algebra discussed in Appendix~\ref{app:bosonic-condensation}. In the bosonic case, an algebra $A$ in $\C$ becomes condensable because its lift to $Z(\C)$ is commutative. In the parafermionic case, the object $f$ is not bosonic after the lift. Instead, the lift supplies the braided data of $\C(\Z_N,q_k)$, while the object-level algebra condensed in $\C$ is
\begin{align}
A^{pf}=1\oplus f\oplus\cdots\oplus f^{N-1},
\end{align}
and the central lift controls the $\Z_N$-graded morphism sectors and the phases that appear after condensation. In particular, the half-braiding between $f^i$ and the other objects of $\C$ is the categorical origin of the fractional statistics of the parafermionic defect operators.

The parafermion condensation can be understood as elevating the parafermionic object $f$ to a parafermionic or $\Z_N$-graded Hom space. Consequently, after the condensation of $A^{pf}$, a fusion category $\C$ transforms into a para-fusion category denoted as $\C_{A^{pf}}$. The main change from bosonic condensation is not the object-level module construction, which is still based on objects of the form $X\otimes A^{pf}$, but the morphism spaces: the different summands $f^i$ of $A^{pf}$ become different $\Z_N$-graded components of Hom spaces.

Before delving into the concept of condensation, it's important to establish an understanding of the simple objects within a para-fusion category. In a super-fusion category, we encountered both $m$-type and $q$-type simple objects, each associated with a distinct super division algebra for their endomorphisms. For para-fusion categories, the simple objects should possess $\Z_N$-graded division algebras as their endomorphism algebras. As outlined in Appendix~\ref{app:graded-division-algebra}, these objects can be classified using $\mathbb C[\Z_M]$, where $M$ is a divisor of $N$. This classification introduces various layers of $q$-type objects within the para-fusion category.

For such a parafermion, the algebra we intend to condense is $A^{pf}=1\oplus f\oplus\cdots\oplus f^{N-1}$, whose lift to $Z(\C)$ is controlled by the pre-metric group $\C(\Z_N,q_k)$ with $k=\gcd(2,N)m$. Its associativity follows from the non-anomalous condition in Eq.~(\ref{non-anomalous}).

At the object level, the parafermion $f$ should be effectively treated as the vacuum object $1$. In this context, any object $X$ gets identified with all objects in its $\Z_N$ orbit under fusion with $f$,
\begin{align}
X\sim X\otimes f\sim X\otimes f^2\sim\cdots\sim X\otimes f^{N-1}.
\end{align}
This means that an object $[X]$ within $\C_{A^{pf}}$ essentially corresponds to the right $A^{pf}$-module $X \otimes A^{pf}$ in the original fusion category $\C$. In this notation, $[X]$ records the whole orbit of $X$ under the condensed parafermion line.

At the level of morphisms, similar to the bosonic anyon condensation discussed in Eq.~(\ref{adjoint}), the relation for parafermionic anyon condensation can be expressed as follows:
\begin{align}\nonumber
&\quad\Hom_{\C_{A^{pf}}}(X \otimes A^{pf}, Y \otimes A^{pf})\\\nonumber
&= \Hom_{\C}(X, Y \otimes A^{pf})\\
&= \Hom_{\C}(X, Y) \oplus \Hom_{\C}(X, Y \otimes f) \oplus ...\oplus \Hom_{\C}(X,Y\otimes f^{N-1}).
\label{eq:pf_Hom_cond}
\end{align}
This equation establishes a connection between the Hom spaces of the condensed para-fusion category and those of the original fusion category. More precisely, the summand
\begin{align}
\Hom_{\C_{A^{pf}}}([X],[Y])_i= \Hom_\C(X,Y\otimes f^i)
\end{align}
is the degree-$i$ homogeneous component of the $\Z_N$-graded Hom space. The parafermionic defect operator $\psi_i$ at a junction in the condensed category is therefore the remnant of the condensed line $f^i$ in the parent category. In the diagrammatic language used above, these degrees are the junction labels often called colors.

The grading is also compatible with composition. If $i,j\in\Z_N$ and
\begin{align}
u\in \Hom_{\C_{A^{pf}}}([X],[Y])_i,\qquad
v\in \Hom_{\C_{A^{pf}}}([Y],[Z])_j,
\end{align}
then their composition has degree $i+j$ modulo $N$,
\begin{align}
v\circ u\in \Hom_{\C_{A^{pf}}}([X],[Z])_{i+j}.
\end{align}
Indeed, after lifting the two morphisms back to $\C$, choose representatives
\begin{align}
\widetilde u:X\rightarrow Y\otimes f^i,\qquad
\widetilde v:Y\rightarrow Z\otimes f^j .
\end{align}
The representative of the composite is obtained by inserting $\widetilde v$ before the condensed $f^i$ line is multiplied into the new degree:
\begin{align}
X &\xrightarrow{\widetilde u} Y\otimes f^i
\xrightarrow{\widetilde v\otimes \mathrm{id}_{f^i}} (Z\otimes f^j)\otimes f^i
\simeq Z\otimes (f^j\otimes f^i)
\xrightarrow{\mathrm{id}_Z\otimes m_{j,i}} Z\otimes f^{i+j}.
\end{align}
Since the grading group is $\Z_N$, the final degree is $j+i=i+j$ modulo $N$. The half-braiding of the lifted parafermion is needed whenever these condensed lines pass through other parent-category lines or when different parenthesizations of homogeneous junctions are compared. This is the categorical origin of the fractional phases in the para-pentagon equations.

This formula also explains the emergence of the different $q_M$-type objects. For a simple object $X\in\C$, define its stabilizer under fusion with $f$ by
\begin{align}
H_X=\{i\in \Z_N\mid X\otimes f^i\simeq X\}.
\end{align}
If $H_X$ is trivial, then only the $i=0$ summand contributes to $\End_{\C_{A^{pf}}}([X])$, and $[X]$ is an $m$-type object. If $H_X\simeq \Z_M$ is nontrivial, then
\begin{align}
\End_{\C_{A^{pf}}}([X])
=\bigoplus_{i\in H_X}\Hom_\C(X,X\otimes f^i)
\simeq \mathbb C[\Z_M],
\end{align}
after choosing isomorphisms $X\simeq X\otimes f^i$ for the stabilizer elements. Thus $[X]$ becomes a $q_M$-type object in the para-fusion category. The usual fermionic $q$-type object is recovered when $N=2$ and $M=2$.

Equivalently, the simple objects of the condensed para-fusion category are organized by the orbits of the $\Z_N$ action generated by $f$, together with the stabilizer of each orbit. A free orbit gives an ordinary $m$-type object because no nontrivial degree contributes to its endomorphism algebra. A partially fixed orbit gives a $q_M$-type object because the stabilizer degrees label additional homogeneous endomorphisms. This orbit-stabilizer viewpoint is the category-theoretic reason why the examples below can contain, for instance, both $\Z_2$ and $\Z_4$ q-type objects in a single $\Z_N$ para-fusion category.

Moving to the next step, we delve into the realm of natural transformations that involve the associator. In this phase, we construct the $F$ symbols of the para-fusion category using the pre-existing $F$ symbols derived from the original fusion category $\C$, together with the half-braiding data of the condensed parafermion. The half-braiding is necessary because the condensed lines $f^i$ have to be moved through other parent-category lines when one lifts a homogeneous junction back to $\C$. To fully comprehend this process, we consolidate the entire construction in section~\ref{sec:fusion2parafusion}.

Similar to the super-fusion category, we can define the total quantum dimension of a para-fusion category as follows:
\begin{align}
\dim (\C_{A^{pf}}) = \sum_{a\in \C_{A^{pf}}} \frac{d_a^2}{\dim\End(a)}.
\end{align}
The connection between the dimensions before and after parafermion condensation can be expressed as:
\begin{align}
\dim(\C_{A^{pf}}) = \frac{\dim(\C)}{N}.
\end{align}

% \subsubsection{Excitations or Drinfeld center of $\C_{A^f}$}

% \subsubsection{Relation between condensations in $\B$ and $\C$}

From a 3d TQFT perspective, we can represent the Levin-Wen string-net model on the left using the fusion category $\C$ as input data. On the right-hand side, the para-fusion category $\C_{A^{pf}}$ corresponds to a parafermionic string-net model. Between these two models, there exists a gapped domain wall. This domain wall serves as a boundary separating the two phases, so it is a $(\C,\C_{A^{pf}})$-bimodule category. However, the specific relationship between anyon condensation in the fusion category $\C$ and the analogous process in $Z(\C)$ still requires further investigation and study.

% \subsection{Relation between parafermionic anyon condensations in $\B$ and $\C$}

% \subsection{Parafermionic Levin-Wen model in 3d bulk}

\subsection{A condensation functor from fusion category to para-fusion category}
\label{sec:fusion2parafusion}

%[wqr]add distion to object and morphism levels

Now using parafermionic condensation, we can establish a relationship between the $F$-symbols of the para-fusion category and those of its parent fusion category. The construction follows the same philosophy as Eq.~(\ref{eq:pf_Hom_cond}). A homogeneous junction in the condensed category is first lifted to an ordinary junction in the parent category, together with extra $f^i$ lines that record its $\Z_N$ degree. We then apply the ordinary $F$-moves of the parent fusion category, use the half-braiding data in $Z(\C)$ whenever a condensed $f^i$ line has to pass through another line, and finally condense the result back to the para-fusion category. More specifically, consider the following move:
\begin{align}
\begin{gathered}
\begin{tikzpicture}[scale=.75]
% \draw [lightgray] (-2,-2)  -- (-2,2) -- (2,2) -- (2,-2) -- (-2,-2) ;
\draw [-<-=.5, line] (-1,1) -- (-2,2);
\draw [-<-=.5,line] (-1,1) --(0,2);
\draw [-<-=.5,line] (0,0) --(2,2);
\draw [->-=.5,line] (0,0)  -- (0,-2);
\draw [-<-=.5,line] (0,0) -- (-1,1) ;
\node at (-1,1)[circle,fill,inner sep=1.5pt,red,opacity=1]{};
\node at (0,0)[circle,fill,inner sep=1.5pt,red,opacity=1]{};
\draw node at (-1.3, .7) {\tiny $a$};
\draw node at (-.3, -.3) {\tiny $b$};
\draw node at (-2, 2.2) {\tiny $[\mathcal I]$};
\draw node at (0, 2.2) {\tiny $[\mathcal J]$};
\draw node at (2, 2.2) {\tiny $[\mathcal K]$};
\draw node at (-.4, -1.8) {\tiny $[\mathcal L]$};
\draw node at (-.9, .3) {\tiny $[\mathcal M]$};
\end{tikzpicture}
\end{gathered}
\quad=\quad\sum_{[\mathcal N];\,c,d}\mathcal F^{[\mathcal I],\,[\mathcal J],\,[\mathcal K]}_{[\mathcal L]}\left( [\mathcal M], [\mathcal N];\, a,b,c,d\right) 
\begin{gathered}
\begin{tikzpicture}[scale=.75]
% \draw [lightgray] (-2,-2)  -- (-2,2) -- (2,2) -- (2,-2) -- (-2,-2) ;
\draw [->-=.5,line] (-2,2) -- (0,0);
\draw [->-=.5,line] (0,2) --(1,1);
\draw [-<-=.5,line] (1,1) --(2,2);
\draw [-<-=.5,line] (0,0)  -- (1,1);
\draw [->-=.5,line] (0,0) -- (0,-2) ;
\node at (1,1)[circle,fill,inner sep=1.5pt,red,opacity=1]{};
\node at (0,0)[circle,fill,inner sep=1.5pt,red,opacity=1]{};
\draw node at (1.4, .7) {\tiny $c$};
\draw node at (.4, -.3) {\tiny $d$};
\draw node at (-2, 2.2) {\tiny $[\mathcal I]$};
\draw node at (0, 2.2) {\tiny $[\mathcal J]$};
\draw node at (2, 2.2) {\tiny $[\mathcal K]$};
\draw node at (-.4, -1.8) {\tiny $[\mathcal L]$};
\draw node at (.9, .3) {\tiny $[\mathcal N]$};
\end{tikzpicture}
\end{gathered}\,,
\label{eq:para-F}
\end{align}
where lines of $[\mathcal I]$ to $[\mathcal N]$ are TDLs in the para-fusion category condensed from lines of $\mathcal I$ to $\mathcal N$ in the parent fusion category, and the parafermionic defects $a,b,c$ and $d$ are remnants of the condensed $\mathbb Z_N$ anyons in the parent theory. We aim to find how $\mathcal F^{[\mathcal I],\,[\mathcal J],\,[\mathcal K]}_{[\mathcal L]}$ can be obtained from the data of its parent fusion category. The strategy is to lift the LHS of \eqref{eq:para-F} back to the original category, then perform a series of $F$-moves and half-braiding moves, and finally condense back to the para-fusion category in the RHS of \eqref{eq:para-F}:
\begin{align}
\begin{gathered}
\begin{tikzpicture}[scale=.5]
% \draw [lightgray] (-2,-2)  -- (-2,2) -- (2,2) -- (2,-2) -- (-2,-2) ;
\draw [-<-=.5, line] (-1,1) -- (-2,2);
\draw [-<-=.5,line] (-1,1) --(0,2);
\draw [-<-=.5,line] (0,0) --(2,2);
\draw [->-=.5,line] (0,0)  -- (0,-2);
\draw [-<-=.5,line] (0,0) -- (-1,1) ;
\node at (-1,1)[circle,fill,inner sep=1pt,red,opacity=1]{};
\node at (0,0)[circle,fill,inner sep=1pt,red,opacity=1]{};
\draw node at (-1.3, .7) {\tiny $a$};
\draw node at (-.3, -.3) {\tiny $b$};
\draw node at (-2, 2.4) {\tiny $[\mathcal I]$};
\draw node at (0, 2.4) {\tiny $[\mathcal J]$};
\draw node at (2, 2.4) {\tiny $[\mathcal K]$};
\draw node at (-.6, -1.8) {\tiny $[\mathcal L]$};
%\draw node at (-.9, .3) {\tiny $[\mathcal M]$};
\end{tikzpicture}
\end{gathered}
&\quad\overset{\rm lift}{\Longrightarrow}\quad
\begin{gathered}
\begin{tikzpicture}[scale=.5]
% \draw [lightgray] (-2,-2)  -- (-2,2) -- (2,2) -- (2,-2) -- (-2,-2) ;
\draw [-<-=.5, line] (-1,1) -- (-2,2);
\draw [-<-=.5,line] (-1,1) --(0,2);
\draw [-<-=.5,line] (0,0) --(2,2);
\draw [->-=.5,line] (0,0)  -- (0,-2);
\draw [-<-=.5,line] (0,0) -- (-1,1) ;
%\node at (-1,1)[circle,fill,inner sep=1.5pt,red,opacity=1]{};
%\node at (0,0)[circle,fill,inner sep=1.5pt,red,opacity=1]{};
%\draw node at (-1.3, .7) {\tiny $a$};
%\draw node at (-.3, -.3) {\tiny $b$};
\draw node at (-2, 2.4) {\tiny $\mathcal I$};
\draw node at (0, 2.4) {\tiny $\mathcal J$};
\draw node at (2, 2.4) {\tiny $\mathcal K$};
\draw node at (-.6, -1.8) {\tiny $\mathcal L$};
%\draw node at (-.9, .3) {\tiny $[\mathcal M]$};
\draw [-<-=.7, line, red] (-.5,.5) -- (-2, .65);
\draw [-<-=.7, line, red] (0,-1) -- (-2, -.85);
\draw node at (-1.6, .3) {\color{red}\tiny $-a$};
\draw node at (-1.6, -1.2) {\color{red}\tiny $-b$};
\end{tikzpicture}
\end{gathered}
\quad=\quad
\begin{gathered}
\begin{tikzpicture}[scale=.5]
% \draw [lightgray] (-2,-2)  -- (-2,2) -- (2,2) -- (2,-2) -- (-2,-2) ;
\draw [-<-=.5, line] (-1,1) -- (-2,2);
\draw [-<-=.5,line] (-1,1) --(0,2);
\draw [-<-=.5,line] (0,0) --(2,2);
\draw [->-=.5,line] (0,0)  -- (0,-2);
\draw [-<-=.5,line] (0,0) -- (-1,1) ;
%\node at (-1,1)[circle,fill,inner sep=1.5pt,red,opacity=1]{};
%\node at (0,0)[circle,fill,inner sep=1.5pt,red,opacity=1]{};
%\draw node at (-1.3, .7) {\tiny $a$};
%\draw node at (-.3, -.3) {\tiny $b$};
\draw node at (-2, 2.4) {\tiny $\mathcal I$};
\draw node at (0, 2.4) {\tiny $\mathcal J$};
\draw node at (2, 2.4) {\tiny $\mathcal K$};
\draw node at (-.6, -1.8) {\tiny $\mathcal L$};
%\draw node at (-.9, .3) {\tiny $[\mathcal M]$};
\draw [-<-=.7, line, red] (-.5,.5) -- (-2, .65);
\draw [-<-=.7, line, red] (0,-.8) -- (-2, -.65);
\draw [-<-=.7, line, red] (0,-1.2) -- (-2, -1.05);
\draw node at (-1.6, .3) {\color{red}\tiny $-a$};
\draw node at (-1.6, -.4) {\color{red}\tiny $-b+d$};
\draw node at (-1.6, -1.45) {\color{red}\tiny $-d$};
\end{tikzpicture}
\end{gathered}
\quad\overset{\mathcal F}{\longrightarrow}\quad
\begin{gathered}
\begin{tikzpicture}[scale=.5]
% \draw [lightgray] (-2,-2)  -- (-2,2) -- (2,2) -- (2,-2) -- (-2,-2) ;
\draw [-<-=.5, line] (-1,1) -- (-2,2);
\draw [-<-=.5,line] (-1,1) --(0,2);
\draw [-<-=.5,line] (0,0) --(2,2);
\draw [->-=.5,line] (0,0)  -- (0,-2);
\draw [-<-=.5,line] (0,0) -- (-1,1) ;
%\node at (-1,1)[circle,fill,inner sep=1.5pt,red,opacity=1]{};
%\node at (0,0)[circle,fill,inner sep=1.5pt,red,opacity=1]{};
%\draw node at (-1.3, .7) {\tiny $a$};
%\draw node at (-.3, -.3) {\tiny $b$};
\draw node at (-2, 2.4) {\tiny $\mathcal I$};
\draw node at (0, 2.4) {\tiny $\mathcal J$};
\draw node at (2, 2.4) {\tiny $\mathcal K$};
\draw node at (-.6, -1.8) {\tiny $\mathcal L$};
%\draw node at (-.9, .3) {\tiny $[\mathcal M]$};
\draw [-<-=.7, line, red] (-.5,.5) -- (-2, .65);
%\draw [-<-=.7, line, red] (-.65,.665) -- (-2, .8);
\draw [-<-=.7, line, red] (-.115,.12) -- (-2, .3);
%\draw [-<-=.7, line, red] (-.5,.3) -- (-2, .45);
\draw [-<-=.7, line, red] (0,-1) -- (-2, -.85);
\draw node at (-1.6, .9) {\color{red}\tiny $-a$};
\draw node at (-1.6, -0.15) {\color{red}\tiny $-b+d$};
\draw node at (-1.6, -1.2) {\color{red}\tiny $-d$};
\end{tikzpicture}
\end{gathered}
\notag\\
&\quad=\quad
\begin{gathered}
\begin{tikzpicture}[scale=.5]
% \draw [lightgray] (-2,-2)  -- (-2,2) -- (2,2) -- (2,-2) -- (-2,-2) ;
\draw [-<-=.5, line] (-1,1) -- (-2,2);
\draw [-<-=.5,line] (-1,1) --(0,2);
\draw [-<-=.5,line] (0,0) --(2,2);
\draw [->-=.5,line] (0,0)  -- (0,-2);
\draw [-<-=.5,line] (0,0) -- (-1,1) ;
%\node at (-1,1)[circle,fill,inner sep=1.5pt,red,opacity=1]{};
%\node at (0,0)[circle,fill,inner sep=1.5pt,red,opacity=1]{};
%\draw node at (-1.3, .7) {\tiny $a$};
%\draw node at (-.3, -.3) {\tiny $b$};
\draw node at (-2, 2.4) {\tiny $\mathcal I$};
\draw node at (0, 2.4) {\tiny $\mathcal J$};
\draw node at (2, 2.4) {\tiny $\mathcal K$};
\draw node at (-.6, -1.8) {\tiny $\mathcal L$};
%\draw node at (-.9, .3) {\tiny $[\mathcal M]$};
\draw [-<-=.7, line, red] (-.5,.5) -- (-2, .65);
\draw [-<-=.7, line, red] (0,-1) -- (-2, -.85);
\draw node at (-1.6, .3) {\color{red}\tiny $-c$};
\draw node at (-1.6, -1.2) {\color{red}\tiny $-d$};
\end{tikzpicture}
\end{gathered}
\quad\overset{\mathcal F}{\longrightarrow}\quad
\begin{gathered}
\begin{tikzpicture}[scale=.5]
% \draw [lightgray] (-2,-2)  -- (-2,2) -- (2,2) -- (2,-2) -- (-2,-2) ;
\draw [-<-=.5, line] (-1,1) -- (-2,2);
\draw [-<-=.5,line] (-.6,.6) --(.8,2);
\draw [-<-=.5,line] (0,0) --(2,2);
\draw [->-=.5,line] (0,0)  -- (0,-2);
\draw [-<-=.5,line] (0,0) -- (-1,1) ;
%\node at (-1,1)[circle,fill,inner sep=1.5pt,red,opacity=1]{};
%\node at (0,0)[circle,fill,inner sep=1.5pt,red,opacity=1]{};
%\draw node at (-1.3, .7) {\tiny $a$};
%\draw node at (-.3, -.3) {\tiny $b$};
\draw node at (-2, 2.4) {\tiny $\mathcal I$};
\draw node at (.8, 2.4) {\tiny $\mathcal J$};
\draw node at (2, 2.4) {\tiny $\mathcal K$};
\draw node at (-.6, -1.8) {\tiny $\mathcal L$};
%\draw node at (-.9, .3) {\tiny $[\mathcal M]$};
\draw [-<-=.7, line, red] (-.9, .89) -- (-2, 1);
\draw [-<-=.7, line, red] (0,-1) -- (-2, -.85);
\draw node at (-1.6, .5) {\color{red}\tiny $-c$};
\draw node at (-1.6, -1.2) {\color{red}\tiny $-d$};
\end{tikzpicture}
\end{gathered}
\quad\overset{\mathcal F}{\longrightarrow}\quad
\begin{gathered}
\begin{tikzpicture}[scale=.5]
% \draw [lightgray] (-2,-2)  -- (-2,2) -- (2,2) -- (2,-2) -- (-2,-2) ;
\draw [-<-=.5, line] (-1,1) -- (-2,2);
\draw [-<-=.5,line] (1,1) --(0,2);
\draw [-<-=.5,line] (0,0) --(1,1);
\draw [->-=.5,line] (2,2) --(1,1);
\draw [->-=.5,line] (0,0)  -- (0,-2);
\draw [line] (0,0) -- (-1,1) ;
%\node at (-1,1)[circle,fill,inner sep=1.5pt,red,opacity=1]{};
%\node at (0,0)[circle,fill,inner sep=1.5pt,red,opacity=1]{};
%\draw node at (-1.3, .7) {\tiny $a$};
%\draw node at (-.3, -.3) {\tiny $b$};
\draw node at (-2, 2.4) {\tiny $\mathcal I$};
\draw node at (0, 2.4) {\tiny $\mathcal J$};
\draw node at (2, 2.4) {\tiny $\mathcal K$};
\draw node at (-.6, -1.8) {\tiny $\mathcal L$};
%\draw node at (-.9, .3) {\tiny $[\mathcal M]$};
\draw [-<-=.7, line, red] (-.5,.5) -- (-2, .65);
\draw [-<-=.7, line, red] (0,-1) -- (-2, -.85);
\draw node at (-1.6, .3) {\color{red}\tiny $-c$};
\draw node at (-1.6, -1.2) {\color{red}\tiny $-d$};
\end{tikzpicture}
\end{gathered}
\notag\\
&\quad\overset{{\color{red} \beta}}{\longrightarrow}\quad
\begin{gathered}
\begin{tikzpicture}[scale=0.5]
% \draw [lightgray] (-2,-2)  -- (-2,2) -- (2,2) -- (2,-2) -- (-2,-2) ;
\draw [-<-=.5, line] (-1,1) -- (-2,2);
\draw [-<-=.5,line] (1,1) --(0,2);
\draw [-<-=.5,line] (0,0) --(1,1);
\draw [->-=.5,line] (2,2) --(1,1);
\draw [->-=.5,line] (0,0)  -- (0,-2);
\draw [line] (0,0) -- (-1,1) ;
\draw [-<-=.7, line, red] (-.8,.6) -- (-2, .75);
%\draw [-<-=.7, line, red] (-.5,.5) -- (-2, .65);
\draw [red, line width=1.05] (-.3, .5) arc (90:-50:.15);
\draw [-<-=.7, line, red] (0,-1) -- (-2, -.85);
%\node at (-1,1)[circle,fill,inner sep=1.5pt,red,opacity=1]{};
%\node at (0,0)[circle,fill,inner sep=1.5pt,red,opacity=1]{};
%\draw node at (-1.3, .7) {\tiny $a$};
%\draw node at (-.3, -.3) {\tiny $b$};
%\draw node at (1.4, 1.7) {\tiny $\frac{1}{15}$};
%\draw node at (-1.4, 1.7) {\tiny $\frac{1}{15}$};
\draw node at (-2, 2.4) {\tiny $\mathcal I$};
\draw node at (0, 2.4) {\tiny $\mathcal J$};
\draw node at (2, 2.4) {\tiny $\mathcal K$};
\draw node at (-.6, -1.8) {\tiny $\mathcal L$};
%\draw node at (-.9, .3) {\tiny $[\mathcal M]$};
\draw node at (-1.6, .3) {\color{red}\tiny $-c$};
\draw node at (-1.6, -1.2) {\color{red}\tiny $-d$};
\end{tikzpicture}
\end{gathered}
\quad\overset{\mathcal F}{\longrightarrow}\quad
\begin{gathered}
\begin{tikzpicture}[scale=0.5]
% \draw [lightgray] (-2,-2)  -- (-2,2) -- (2,2) -- (2,-2) -- (-2,-2) ;
\draw [-<-=.5, line] (-1,1) -- (-2,2);
\draw [-<-=.5,line] (1,1) --(0,2);
\draw [-<-=.5,line] (0,0) --(1,1);
\draw [->-=.5,line] (2,2) --(1,1);
\draw [->-=.5,line] (0,0)  -- (0,-2);
\draw [line] (0,0) -- (-1,1) ;
\draw [-<-=.7, line, red] (-.8,.63) -- (-2, .75);
\draw [line, red] (.6,.6) -- (-.25, .65);
%\draw [-<-=.7, line, red] (-.5,.5) -- (-2, .65);
%\draw [red, line width=1.05] (-.3, .5) arc (90:-50:.15);
\draw [-<-=.7, line, red] (0,-1) -- (-2, -.85);
%\node at (-1,1)[circle,fill,inner sep=1.5pt,red,opacity=1]{};
%\node at (0,0)[circle,fill,inner sep=1.5pt,red,opacity=1]{};
%\draw node at (-1.3, .7) {\tiny $a$};
%\draw node at (-.3, -.3) {\tiny $b$};
%\draw node at (1.4, 1.7) {\tiny $\frac{1}{15}$};
%\draw node at (-1.4, 1.7) {\tiny $\frac{1}{15}$};
\draw node at (-2, 2.4) {\tiny $\mathcal I$};
\draw node at (0, 2.4) {\tiny $\mathcal J$};
\draw node at (2, 2.4) {\tiny $\mathcal K$};
\draw node at (-.6, -1.8) {\tiny $\mathcal L$};
%\draw node at (-.9, .3) {\tiny $[\mathcal M]$};
\draw node at (-1.6, .3) {\color{red}\tiny $-c$};
\draw node at (-1.6, -1.2) {\color{red}\tiny $-d$};
\end{tikzpicture}
\end{gathered}
\quad\overset{\rm cond.}{\Longrightarrow}\quad
\begin{gathered}
\begin{tikzpicture}[scale=.5]
% \draw [lightgray] (-2,-2)  -- (-2,2) -- (2,2) -- (2,-2) -- (-2,-2) ;
\draw [->-=.5,line] (-2,2) -- (0,0);
\draw [->-=.5,line] (0,2) --(1,1);
\draw [-<-=.5,line] (1,1) --(2,2);
\draw [-<-=.5,line] (0,0)  -- (1,1);
\draw [->-=.5,line] (0,0) -- (0,-2) ;
\node at (1,1)[circle,fill,inner sep=1pt,red,opacity=1]{};
\node at (0,0)[circle,fill,inner sep=1pt,red,opacity=1]{};
\draw node at (1.4, .7) {\tiny $c$};
\draw node at (.4, -.3) {\tiny $d$};
\draw node at (-2, 2.4) {\tiny $[\mathcal I]$};
\draw node at (0, 2.4) {\tiny $[\mathcal J]$};
\draw node at (2, 2.4) {\tiny $[\mathcal K]$};
\draw node at (-.6, -1.8) {\tiny $[\mathcal L]$};
%\draw node at (.9, .3) {\tiny $[\mathcal N]$};
\end{tikzpicture}
\end{gathered}\,,
\label{eq:morphism_F}
\end{align}
where the braiding operation ``$\beta$" in the third line requires the knowledge of half braidings between the to-be-condensed $\mathbb Z_N$-anyons and the others defined through a morphism
\begin{align}
    \beta\in{\rm Hom}\left(a\otimes \mathcal L,\,\mathcal L\otimes a\right)\,,
\end{align}
i.e.
\begin{align}
\begin{gathered}
\begin{tikzpicture}[scale=0.5]
\draw [line,red] (0,2) -- (2,0);
\draw [line,red] (0,0) -- (.8,.8);
\draw [line,red] (1.2,1.2) -- (2,2);
\draw node at (0, -.3) {\tiny $a$};
\draw node at (2, -.3) {\tiny $b$};
\end{tikzpicture}
\end{gathered}
\quad\equiv\beta_a(b)\quad
\begin{gathered}
\begin{tikzpicture}[scale=0.5]
\draw [line,red] (0,2) -- (2,0);
\draw [line,red] (.4,0) -- (1.2,.8);
\draw [line,red] (1.6,2) -- (.8,1.2);
\draw node at (0, -.3) {\tiny $a$};
\draw node at (2, -.3) {\tiny $b$};
\draw node at (1.9,2) {\tiny $a$};
\end{tikzpicture}
\end{gathered}
\qquad{\rm and}\qquad
%%%%
\begin{gathered}
\begin{tikzpicture}[scale=0.5]
\draw [line] (0,2) -- (2,0);
\draw [line,red] (0,0) -- (.8,.8);
\draw [line,red] (1.2,1.2) -- (2,2);
\draw node at (0, -.3) {\tiny $a$};
\draw node at (2, -.3) {\tiny $\mathcal L$};
\end{tikzpicture}
\end{gathered}
\quad\equiv\beta_a(\mathcal L)\quad
\begin{gathered}
\begin{tikzpicture}[scale=0.5]
\draw [line] (0,2) -- (2,0);
\draw [line,red] (.4,0) -- (1.2,.8);
\draw [line,red] (1.6,2) -- (.8,1.2);
\draw node at (0, -.3) {\tiny $a$};
\draw node at (2, -.3) {\tiny $\mathcal L$};
\draw node at (1.9,2) {\tiny $a$};
\end{tikzpicture}
\end{gathered}\,.
\end{align}
The half braiding data are encoded in the Drinfeld center $Z(\mathcal C)$ of the parent fusion category $\mathcal C$. Therefore, to sum up, using \eqref{eq:morphism_F} we establish a map from the $F$-symbols and half-braiding data of the fusion category $\mathcal C$ to the $F$-symbols in the para-condensed fusion category $\mathcal C/\mathbb Z_N$. In practice, we can either apply \eqref{eq:morphism_F} or directly solve para-pentagon equations \eqref{eq:parapentagons} to find the $F$-moves in the para-fusion category. It would be interesting to show the equivalence between \eqref{eq:parapentagons} and \eqref{eq:morphism_F} that we will leave in the future work. On the other hand, in the following sections, we use both of them to study various examples and demonstrate the consistency of the two approaches case by case.

%[wqr]show it satisfies the para-pentagon eq

\section{Examples of Parafermionic TDLs}
\label{4}
\subsection{Parafermionic $\mathbb Z_3$}
The first example for us to demonstrate the para-fusion category is the Verlinde lines in the parafermionization of 3-Potts model. As discussed at the end of Section~\ref{sec:orbifolding-fermionization-parafermionization}, the 3-Potts model has partition function
\begin{align}
Z_{\mathcal T}=|\chi_0|^2+|\chi_{2/5}|^2+|\chi_{1/15}|^2+|\chi_{1/15^*}|^2+|\chi_{2/3}|^2+|\chi_{2/3^*}|^2\,.
\end{align}
Correspondingly, one has six Verlinde lines, denoted by 
\begin{align}
\mathcal C=\left\{\mathcal L_{0},\,\mathcal L_{2/3},\,\mathcal L_{2/3^*},\,\mathcal L_{2/5},\,\mathcal L_{1/15},\,\mathcal L_{1/15^*}\right\}\,,
\end{align}
where the first three lines form the $\mathbb Z_3$ symmetry, while the rest are non-invertible lines. The fusion rules in $\mathcal C$ can be easily obtained by Verlinde formula, from which one can study the orbits of $\left\{\mathcal L_{2/5},\,\mathcal L_{1/15},\,\mathcal L_{1/15^*}\right\}$ under the $\mathbb Z_3$-action. It turns out that
\begin{align}
\mathcal L_{2/3}\cdot \mathcal L_{2/5}=\mathcal L_{1/15}\,,\quad
\mathcal L_{2/3}\cdot \mathcal L_{1/15}=\mathcal L_{1/15^*}\,,\quad
\mathcal L_{2/3}\cdot \mathcal L_{1/15^*}=\mathcal L_{2/5}\,,
\label{eq:condensation}
\end{align}
implying that the set $\left\{\mathcal L_{2/5},\,\mathcal L_{1/15},\,\mathcal L_{1/15^*}\right\}$ form a 3-orbit, i.e. an orbit with 3 steps, under the $\mathbb Z_3$-action. Therefore, after parafermionization, the three Verlinde lines will condense to a single line in the $\mathbb Z_3$-parafermionic theory, whose fusion rule can be once again determined by its parent 3-Potts theory. Notice that
\begin{align}
\mathcal L_{2/5}\cdot\mathcal L_{2/5}=\mathcal L_0+\mathcal L_{2/5}\,.
\end{align}
The sub-category
\begin{align}
\mathcal C^\prime=\left\{\mathcal L_0,\,\mathcal L_{2/5}\right\}
\end{align}
is of Fibonacci type. If we choose $\mathcal L_0$ and $\mathcal L_{2/5}$ as the representatives after condensing the $\mathbb Z_3$ lines, the para-fusion category
\begin{align}
\widetilde{\mathcal C}_1\equiv\left\{\left[\mathcal L_0\right],\,\left[\mathcal L_{2/5}\right]\right\}
\end{align}
is isomorphic to $\mathcal C^\prime$ with the fusion rule
\begin{align}
\left[\mathcal L_{2/5}\right]\cdot\left[\mathcal L_{2/5}\right]=\mathbb C^{1|0|0}\left[\mathcal L_0\right]+\mathbb C^{1|0|0}\left[\mathcal L_{2/5}\right]\,,
\label{eq:F1}
\end{align}
where $\mathbb C^{1|0|0}$ implies that the TDL $\left[\mathcal L_{2/5}\right]$ is of $m$-type and that only the trivial parafermionic defect $\psi_0$ appears at its trivalent fusion junctions. For this choice of representatives, the F-symbols therefore satisfy the ordinary pentagon identities, and the only nontrivial F-symbol is $\mathcal F^{\left[\mathcal L_{2/5}\right],\left[\mathcal L_{2/5}\right],\left[\mathcal L_{2/5}\right]}_{\left[\mathcal L_{2/5}\right]}$, given by
\begin{align}
\begin{pmatrix}
\begin{gathered}
\begin{tikzpicture}[scale=.5]
\draw [line] (-1,1) -- (-2,2);
\draw [line] (-1,1) --(0,2);
\draw [line] (0,0) --(2,2);
\draw [line] (0,0)  -- (0,-1);
\draw [line,dotted] (0,0) -- (-1,1) ;
\node at (-1,1)[circle,fill,inner sep=1.5pt,red,opacity=1]{};
\node at (0,0)[circle,fill,inner sep=1.5pt,red,opacity=1]{};
\draw node at (-1.3, .7) {\tiny $\psi_0$};
\draw node at (-.3, -.3) {\tiny $\psi_0$};
\end{tikzpicture}
\end{gathered}
\\
\begin{gathered}
\begin{tikzpicture}[scale=.5]
\draw [line] (-1,1) -- (-2,2);
\draw [line] (-1,1) --(0,2);
\draw [line] (0,0) --(2,2);
\draw [line] (0,0)  -- (0,-1);
\draw [line] (0,0) -- (-1,1) ;
\node at (-1,1)[circle,fill,inner sep=1.5pt,red,opacity=1]{};
\node at (0,0)[circle,fill,inner sep=1.5pt,red,opacity=1]{};
\draw node at (-1.3, .7) {\tiny $\psi_0$};
\draw node at (-.3, -.3) {\tiny $\psi_0$};
\end{tikzpicture}
\end{gathered}
\end{pmatrix}
\quad=\quad
\begin{pmatrix}
\zeta^{-1} & 1 \\
\zeta^{-1} & -\zeta^{-1}
\end{pmatrix}
\cdot
\begin{pmatrix}
\begin{gathered}
\begin{tikzpicture}[scale=.5]
\draw [line] (0,0) -- (-2,2);
\draw [line] (1,1) --(0,2);
\draw [line] (1,1) --(2,2);
\draw [line] (0,0)  -- (0,-1);
\draw [line,dotted] (0,0) -- (1,1) ;
\node at (1,1)[circle,fill,inner sep=1.5pt,red,opacity=1]{};
\node at (0,0)[circle,fill,inner sep=1.5pt,red,opacity=1]{};
\draw node at (1.4, .7) {\tiny $\psi_0$};
\draw node at (.4, -.3) {\tiny $\psi_0$};
\end{tikzpicture}
\end{gathered}
\\
\begin{gathered}
\begin{tikzpicture}[scale=.5]
\draw [line] (0,0) -- (-2,2);
\draw [line] (1,1) --(0,2);
\draw [line] (1,1) --(2,2);
\draw [line] (0,0)  -- (0,-1);
\draw [line] (0,0) -- (1,1) ;
\node at (1,1)[circle,fill,inner sep=1.5pt,red,opacity=1]{};
\node at (0,0)[circle,fill,inner sep=1.5pt,red,opacity=1]{};
\draw node at (1.4, .7) {\tiny $\psi_0$};
\draw node at (.4, -.3) {\tiny $\psi_0$};
\end{tikzpicture}
\end{gathered}
\end{pmatrix}\,,
\end{align}
where the dashed and solid lines denote the TDLs $\left[\mathcal L_{0}\right]$ and $\left[\mathcal L_{2/5}\right]$ respectively, $\zeta=\frac{\sqrt{5}+1}{2}$, and we have uses the gauge freedom to fix the entry of the F-symbol,
\begin{align} \label{eq:FibF}
\mathcal F^{\left[\mathcal L_{2/5}\right],\left[\mathcal L_{2/5}\right],\left[\mathcal L_{2/5}\right]}_{\left[\mathcal L_{2/5}\right]}\left(\left[\mathcal L_{0}\right],\,\left[\mathcal L_{2/5}\right]\right)=1\,.
\end{align}
The F-symbol in \eqref{eq:FibF} is related to the standard F-symbol of Fibonacci Anyons by a similarity transformation (see for example \cite{Gu:2021utd}).

Instead, one can choose different representatives, for example $\left[\mathcal L_0\right]$ and $\left[\mathcal L_{1/15}\right]$, denoted by
\begin{align}
\widetilde{\mathcal C}_2\equiv\left\{\left[\mathcal L_0\right],\,\left[\mathcal L_{1/15}\right]\right\}
\end{align}
Notice the fusion rule
\begin{align}
\mathcal L_{1/15}\cdot\mathcal L_{1/15}=\mathcal L_{1/15^*}+\mathcal L_{2/3^*}=\mathcal L_{2/3}\cdot\mathcal L_{1/15}+\mathcal L_{2/3^*}\cdot\mathcal L_{0}\,,
\end{align}
implying the following fusion rule after condensation
\begin{align}
\begin{gathered}
\begin{tikzpicture}[scale=1.5]
    % \draw [lightgray] (0,0)  -- (0,2) -- (2,2) -- (2,0) -- (0,0) ;
    \fill (1,1) circle (1pt);
    \draw[-<-=.5] (1,1) -- (0,2);
    \draw[-<-=.5] (1,1) -- (2,2);
    \draw[->-=.5] (1,1) -- (1,0);
    %\draw node at (1.25, 0.9) {};
    \draw node at (0.55, 1.75) {\tiny $\mathcal L_{1/15}$};
\draw node at (1.45, 1.75) {\tiny $\mathcal L_{1/15}$};
\draw node at (0.8, 0.25) {\tiny $\mathcal L_{2/3^*}$};
\end{tikzpicture}
\end{gathered}
\quad=\quad
\begin{gathered}
\begin{tikzpicture}[scale=1.5]
    % \draw [lightgray] (0,0)  -- (0,2) -- (2,2) -- (2,0) -- (0,0) ;
    \fill (1,1) circle (1pt);
    \fill (1,.7) circle (1pt);
    \draw[red,->-=.6] (1,.7) -- (0,1);
    \draw[-<-=.5] (1,1) -- (0,2);
    \draw[-<-=.5]  (1,1) -- (2,2);
    \draw[->-=.6] (1,1) -- (1,0.7);
    \draw[-<-=.6] (1,0) -- (1,0.7);
    %\draw node at (1.25, 0.9) {};
    \draw node at (0.35, 1.05) {\tiny $\mathcal L_{2/3^*}$};
    \draw node at (0.55, 1.75) {\tiny $\mathcal L_{1/15}$};
\draw node at (1.45, 1.75) {\tiny $\mathcal L_{1/15}$};
\draw node at (0.8, 0.25) {\tiny $\mathcal L_{0}$};
\draw node at (1.3, .85) {\tiny $\mathcal L_{2/3^*}$};
\end{tikzpicture}
\end{gathered}
\quad\overset{\rm cond.}{\Longrightarrow}\quad
\begin{gathered}
\begin{tikzpicture}[scale=1.5]
    % \draw [lightgray] (0,0)  -- (0,2) -- (2,2) -- (2,0) -- (0,0) ;
    %\fill [red,opacity=2] (1,1) circle (1.5pt);
    %\fill (1,.7) circle (1pt);
    %\draw[red] (1,.7) -- (0,1);
    \draw[-<-=.5] (1,1) -- (0,2);
    \draw[-<-=.5] (1,1) -- (2,2);
    \draw[->-=.5] (1,1) -- (1,0);
    %\draw node at (1.25, 0.9) {};
    %\draw node at (0.35, 1.05) {\tiny $\mathcal L_{2/3^*}$};
    \draw node at (0.7, 1.75) {\tiny $\left[\mathcal L_{1/15}\right]$};
\draw node at (1.7, 1.25) {\tiny $\left[\mathcal L_{1/15}\right]$};
\draw node at (0.8, 0.25) {\tiny $\left[\mathcal L_{0}\right]$};
\draw node at (.8, 1) {\tiny $\psi_1$};
    \node at (1,1)[circle,fill,inner sep=1.5pt,red,opacity=1]{};
\end{tikzpicture}
\end{gathered}
\end{align}
and
\begin{align}
\begin{gathered}
\begin{tikzpicture}[scale=1.5]
    % \draw [lightgray] (0,0)  -- (0,2) -- (2,2) -- (2,0) -- (0,0) ;
    \fill (1,1) circle (1pt);
    \draw[-<-=.5] (1,1) -- (0,2);
    \draw[-<-=.5] (1,1) -- (2,2);
    \draw[->-=.5] (1,1) -- (1,0);
    %\draw node at (1.25, 0.9) {};
    \draw node at (0.55, 1.75) {\tiny $\mathcal L_{1/15}$};
\draw node at (1.45, 1.75) {\tiny $\mathcal L_{1/15}$};
\draw node at (0.7, 0.25) {\tiny $\mathcal L_{1/15^*}$};
\end{tikzpicture}
\end{gathered}
\quad=\quad
\begin{gathered}
\begin{tikzpicture}[scale=1.5]
    % \draw [lightgray] (0,0)  -- (0,2) -- (2,2) -- (2,0) -- (0,0) ;
    \fill (1,1) circle (1pt);
    \fill (1,.7) circle (1pt);
    \draw[red,->-=.6] (1,.7) -- (0,1);
    \draw[-<-=.5] (1,1) -- (0,2);
    \draw[-<-=.5]  (1,1) -- (2,2);
    \draw[->-=.6] (1,1) -- (1,0.7);
    \draw[-<-=.6] (1,0) -- (1,0.7);
    %\draw node at (1.25, 0.9) {};
    \draw node at (0.35, 1.05) {\tiny $\mathcal L_{2/3}$};
    \draw node at (0.55, 1.75) {\tiny $\mathcal L_{1/15}$};
\draw node at (1.45, 1.75) {\tiny $\mathcal L_{1/15}$};
\draw node at (0.7, 0.25) {\tiny $\mathcal L_{1/15}$};
\draw node at (1.35, .85) {\tiny $\mathcal L_{1/15^*}$};
\end{tikzpicture}
\end{gathered}
\quad\overset{\rm cond.}{\Longrightarrow}\quad
\begin{gathered}
\begin{tikzpicture}[scale=1.5]
    % \draw [lightgray] (0,0)  -- (0,2) -- (2,2) -- (2,0) -- (0,0) ;
    %\fill [red,opacity=2] (1,1) circle (1.5pt);
    %\fill (1,.7) circle (1pt);
    %\draw[red] (1,.7) -- (0,1);
    \draw[-<-=.5] (1,1) -- (0,2);
    \draw[-<-=.5] (1,1) -- (2,2);
    \draw[->-=.5] (1,1) -- (1,0);
    %\draw node at (1.25, 0.9) {};
    %\draw node at (0.35, 1.05) {\tiny $\mathcal L_{2/3^*}$};
    \draw node at (0.7, 1.75) {\tiny $\left[\mathcal L_{1/15}\right]$};
\draw node at (1.7, 1.25) {\tiny $\left[\mathcal L_{1/15}\right]$};
\draw node at (0.65, 0.25) {\tiny $\left[\mathcal L_{1/15}\right]$};
\draw node at (.8, 1) {\tiny $\psi_2$};
\node at (1,1)[circle,fill,inner sep=1.5pt,red,opacity=1]{};
\end{tikzpicture}
\end{gathered}\,,
\end{align}
where the orientations of the red lines must be reversed before applying the condensation prescription in \eqref{eq:condensation}. The red dots $\psi_i$ denote the one-dimensional parafermionic defects that remain after condensing the $\mathbb Z_3$ symmetry. They specify the morphisms ${\rm Hom}(\mathcal L_i\otimes\mathcal L_j,\,\mathcal L_k)$ associated with the fusion of two TDLs in the parafermionic theory. In the case at hand, we have
\begin{align}
\left[\mathcal L_{1/15}\right]\cdot\left[\mathcal L_{1/15}\right]=\mathbb C^{0|1|0}\left[\mathcal L_0\right]+\mathbb C^{0|0|1}\left[\mathcal L_{1/15}\right]\,.
\end{align}
With the above fusion rule, we can solve the para-pentagon equation with the choice of 
\begin{align}
\omega=e^{\frac{2\pi i}{3}}\,,
\end{align} 
and find the solutions to the non-trivial F-symbols $\mathcal F$,
\begin{align}
\begin{gathered}
\begin{tikzpicture}[scale=0.5]
\draw [line] (-1,1) -- (-2,2);
\draw [line] (-1,1) --(0,2);
\draw [line] (0,0) --(2,2);
\draw [line,dotted] (0,0)  -- (0,-1);
\draw [line] (0,0) -- (-1,1) ;
\node at (-1,1)[circle,fill,inner sep=1.5pt,red,opacity=1]{};
\node at (0,0)[circle,fill,inner sep=1.5pt,red,opacity=1]{};
\draw node at (-1.3, .7) {\tiny $\psi_2$};
\draw node at (-.3, -.3) {\tiny $\psi_1$};
\end{tikzpicture}
\end{gathered}
\quad=\quad \omega
\begin{gathered}
\begin{tikzpicture}[scale=0.5]
\draw [line] (-2,2) -- (0,0);
\draw [line] (0,2) --(1,1);
\draw [line] (1,1) --(2,2);
\draw [line] (0,0)  -- (1,1);
\draw [line,dotted] (0,0) -- (0,-1) ;
\node at (1,1)[circle,fill,inner sep=1.5pt,red,opacity=1]{};
\node at (0,0)[circle,fill,inner sep=1.5pt,red,opacity=1]{};
\draw node at (1.4, .7) {\tiny $\psi_2$};
\draw node at (.4, -.3) {\tiny $\psi_1$};
\end{tikzpicture}
\end{gathered}\,,
\label{eq:LLL1}
\end{align}
and
\begin{align}
\begin{pmatrix}
\begin{gathered}
\begin{tikzpicture}[scale=.5]
\draw [line] (-1,1) -- (-2,2);
\draw [line] (-1,1) --(0,2);
\draw [line] (0,0) --(2,2);
\draw [line] (0,0)  -- (0,-1);
\draw [line,dotted] (0,0) -- (-1,1) ;
\node at (-1,1)[circle,fill,inner sep=1.5pt,red,opacity=1]{};
\node at (0,0)[circle,fill,inner sep=1.5pt,red,opacity=1]{};
\draw node at (-1.3, .7) {\tiny $\psi_1$};
\draw node at (-.35, -.3) {\tiny $\psi_0$};
\end{tikzpicture}
\end{gathered}
\\
\begin{gathered}
\begin{tikzpicture}[scale=.5]
\draw [line] (-1,1) -- (-2,2);
\draw [line] (-1,1) --(0,2);
\draw [line] (0,0) --(2,2);
\draw [line] (0,0)  -- (0,-1);
\draw [line] (0,0) -- (-1,1) ;
\node at (-1,1)[circle,fill,inner sep=1.5pt,red,opacity=1]{};
\node at (0,0)[circle,fill,inner sep=1.5pt,red,opacity=1]{};
\draw node at (-1.3, .7) {\tiny $\psi_2$};
\draw node at (-.35, -.3) {\tiny $\psi_2$};
\end{tikzpicture}
\end{gathered}
\end{pmatrix}
\quad=\quad
\begin{pmatrix}
\omega^{-1}\zeta^{-1} & 1 \\
\zeta^{-1} & -\omega\,\zeta^{-1}
\end{pmatrix}
\cdot
\begin{pmatrix}
\begin{gathered}
\begin{tikzpicture}[scale=.5]
\draw [line] (0,0) -- (-2,2);
\draw [line] (1,1) --(0,2);
\draw [line] (1,1) --(2,2);
\draw [line] (0,0)  -- (0,-1);
\draw [line,dotted] (0,0) -- (1,1) ;
\node at (1,1)[circle,fill,inner sep=1.5pt,red,opacity=1]{};
\node at (0,0)[circle,fill,inner sep=1.5pt,red,opacity=1]{};
\draw node at (1.4, .7) {\tiny $\psi_1$};
\draw node at (.4, -.3) {\tiny $\psi_0$};
\end{tikzpicture}
\end{gathered}
\\
\begin{gathered}
\begin{tikzpicture}[scale=.5]
\draw [line] (0,0) -- (-2,2);
\draw [line] (1,1) --(0,2);
\draw [line] (1,1) --(2,2);
\draw [line] (0,0)  -- (0,-1);
\draw [line] (0,0) -- (1,1) ;
\node at (1,1)[circle,fill,inner sep=1.5pt,red,opacity=1]{};
\node at (0,0)[circle,fill,inner sep=1.5pt,red,opacity=1]{};
\draw node at (1.4, .7) {\tiny $\psi_2$};
\draw node at (.4, -.3) {\tiny $\psi_2$};
\end{tikzpicture}
\end{gathered}
\end{pmatrix}\,,
\label{eq:F2}
\end{align}
where the dashed and solid lines stand for the TDLs $\left[\mathcal L_{0}\right]$ and $\left[\mathcal L_{1/15}\right]$ respectively. Solving the para-pentagons, one can find two solutions with $\zeta=\frac{1\pm\sqrt{5}}{2}$. Spin selection rules \cite{Chang:2018iay} imply that $\zeta=\frac{1+\sqrt{5}}{2}$ and $\zeta=\frac{1-\sqrt{5}}{2}$ correspond to the existence of defect operators with spin $s=\pm \frac{2}{5}$ and $s=\pm \frac{1}{5}$ respectively. Therefore the solution with $\zeta=\frac{1+\sqrt{5}}{2}$ need to be picked up.
\paragraph{Quantum dimensions.}
The quantum dimensions of $[\mathcal{L}_{2/5}]$ and $[\mathcal{L}_{1/15}]$ can be determined through the largest eigenvalue of their fusion matrices, respectively. In the case of $\mathcal{L}_{2/5}$ this is straightforward and gives $d_{[2/5]} = \zeta$. In the case of $\mathcal{L}_{1/15}$ the fusion matrix is
\begin{equation}
    {N_{[\mathcal{L}_{1/15}]}}^a_{~b} = \left(\begin{array}{cc}
        0 & 1\\
        \omega & \omega^2
    \end{array}\right), \quad a,b \in \{\mathcal{L}_0, \mathcal{L}_{1/15}\},
\end{equation}
and hence the eigenvalues are $\omega^{-1} \frac{1+\sqrt{5}}{2}$ and $\omega^{-1} \frac{1-\sqrt{5}}{2}$. The absolute value of the largest eigenvalue then gives
\begin{equation}
    d_{\mathcal{L}_{1/15}} = \frac{1+\sqrt{5}}{2} =\zeta. 
\end{equation}

\paragraph{Isomorphism of $\mathcal C_1$ and $\mathcal C_2$}

As $\mathcal C_1$ and $\mathcal C_2$ are different only due to different choices of representatives, they are expected to be isomorphic to each other. One can establish a natural transformation between them by lifting them to the parent fusion category and then condense back. For example, for \eqref{eq:LLL1}, we lift the LHS of the equation, perform a series of F-moves in the parent fusion category, and finally condense back to the RHS of \eqref{eq:LLL1}:
\begin{align}
\begin{gathered}
\begin{tikzpicture}[scale=0.5]
% \draw [lightgray] (-2,-2)  -- (-2,2) -- (2,2) -- (2,-2) -- (-2,-2) ;
\draw [line] (-1,1) -- (-2,2);
\draw [line] (-1,1) --(0,2);
\draw [line] (0,0) --(2,2);
\draw [line, dashed] (0,0)  -- (0,-2);
\draw [line] (0,0) -- (-1,1) ;
\node at (-1,1)[circle,fill,inner sep=1.5pt,red,opacity=1]{};
\node at (0,0)[circle,fill,inner sep=1.5pt,red,opacity=1]{};
\draw node at (-1.3, .7) {\tiny $2$};
\draw node at (-.3, -.3) {\tiny $1$};
\end{tikzpicture}
\end{gathered}
&\quad \overset{\rm lift}{\Longrightarrow} \quad
\begin{gathered}
\begin{tikzpicture}[scale=0.5]
% \draw [lightgray] (-2,-2)  -- (-2,2) -- (2,2) -- (2,-2) -- (-2,-2) ;
\draw [line] (-1,1) -- (-2,2);
\draw [line] (-1,1) --(0,2);
\draw [line] (0,0) --(2,2);
\draw [->-=.7, line, red] (0,0)  -- (0,-1);
\draw [line,dashed] (0,-1)  -- (0,-2);
\draw [line] (0,0) -- (-1,1) ;
\draw [-<-=.7, line, red] (-.5,.5) -- (-2, .65);
\draw [-<-=.7, line, red] (0,-1) -- (-2, -.85);
%\node at (-1,1)[circle,fill,inner sep=1.5pt,red,opacity=1]{};
%\node at (0,0)[circle,fill,inner sep=1.5pt,red,opacity=1]{};
%\draw node at (-1.3, .7) {\tiny $a$};
%\draw node at (-.3, -.3) {\tiny $b$};
%\draw node at (1.4, 1.7) {\tiny $\frac{1}{15}$};
%\draw node at (-1.4, 1.7) {\tiny $\frac{1}{15}$};
\draw node at (-.4, -.4) {\color{red}\tiny $2$};
\draw node at (-1.6, .3) {\color{red}\tiny $2$};
\draw node at (-1.6, -1.2) {\color{red}\tiny $1$};
\end{tikzpicture}
\end{gathered}
\quad=\quad
\begin{gathered}
\begin{tikzpicture}[scale=0.5]
% \draw [lightgray] (-2,-2)  -- (-2,2) -- (2,2) -- (2,-2) -- (-2,-2) ;
\draw [line] (-1,1) -- (-2,2);
\draw [line] (1,1) --(0,2);
\draw [line] (0,0) --(2,2);
\draw [->-=.7, line, red] (0,0)  -- (0,-1);
\draw [line,dashed] (0,-1)  -- (0,-2);
\draw [line] (0,0) -- (-1,1) ;
\draw [-<-=.7, line, red] (-.5,.5) -- (-2, .65);
\draw [-<-=.7, line, red] (0,-1) -- (-2, -.85);
%\node at (-1,1)[circle,fill,inner sep=1.5pt,red,opacity=1]{};
%\node at (0,0)[circle,fill,inner sep=1.5pt,red,opacity=1]{};
%\draw node at (-1.3, .7) {\tiny $a$};
%\draw node at (-.3, -.3) {\tiny $b$};
%\draw node at (1.4, 1.7) {\tiny $\frac{1}{15}$};
%\draw node at (-1.4, 1.7) {\tiny $\frac{1}{15}$};
\draw node at (-.4, -.4) {\color{red}\tiny $2$};
\draw node at (-1.6, .3) {\color{red}\tiny $2$};
\draw node at (-1.6, -1.2) {\color{red}\tiny $1$};
\end{tikzpicture}
\end{gathered}
\quad=\quad \omega\quad
\begin{gathered}
\begin{tikzpicture}[scale=0.5]
% \draw [lightgray] (-2,-2)  -- (-2,2) -- (2,2) -- (2,-2) -- (-2,-2) ;
\draw [line] (-1,1) -- (-2,2);
\draw [line] (1,1) --(0,2);
\draw [line] (0,0) --(2,2);
\draw [->-=.7, line, red] (0,0)  -- (0,-1);
\draw [line,dashed] (0,-1)  -- (0,-2);
\draw [line] (0,0) -- (-1,1) ;
\draw [-<-=.7, line, red] (-.8,.6) -- (-2, .75);
\draw [red, line width=1.05] (-.3, .5) arc (90:-50:.15);
\draw [-<-=.7, line, red] (0,-1) -- (-2, -.85);
%\node at (-1,1)[circle,fill,inner sep=1.5pt,red,opacity=1]{};
%\node at (0,0)[circle,fill,inner sep=1.5pt,red,opacity=1]{};
%\draw node at (-1.3, .7) {\tiny $a$};
%\draw node at (-.3, -.3) {\tiny $b$};
%\draw node at (1.4, 1.7) {\tiny $\frac{1}{15}$};
%\draw node at (-1.4, 1.7) {\tiny $\frac{1}{15}$};
\draw node at (-.4, -.4) {\color{red}\tiny $2$};
\draw node at (-1.6, .3) {\color{red}\tiny $2$};
\draw node at (-1.6, -1.2) {\color{red}\tiny $1$};
\end{tikzpicture}
\end{gathered}
\notag\\\notag\\
&\quad=\quad\omega\quad
\begin{gathered}
\begin{tikzpicture}[scale=0.5]
% \draw [lightgray] (-2,-2)  -- (-2,2) -- (2,2) -- (2,-2) -- (-2,-2) ;
\draw [line] (-1,1) -- (-2,2);
\draw [line] (1,1) --(0,2);
\draw [line] (0,0) --(2,2);
\draw [->-=.7, line, red] (0,0)  -- (0,-1);
\draw [line,dashed] (0,-1)  -- (0,-2);
\draw [line] (0,0) -- (-1,1) ;
\draw [-<-=.7, line, red] (-.8,.6) -- (-2, .75);
\draw [line, red] (-.45,.6) -- (.5,.5);
\draw [-<-=.7, line, red] (0,-1) -- (-2, -.85);
%\node at (-1,1)[circle,fill,inner sep=1.5pt,red,opacity=1]{};
%\node at (0,0)[circle,fill,inner sep=1.5pt,red,opacity=1]{};
%\draw node at (-1.3, .7) {\tiny $a$};
%\draw node at (-.3, -.3) {\tiny $b$};
%\draw node at (1.4, 1.7) {\tiny $\frac{1}{15}$};
%\draw node at (-1.4, 1.7) {\tiny $\frac{1}{15}$};
\draw node at (-.4, -.4) {\color{red}\tiny $2$};
\draw node at (-1.6, .3) {\color{red}\tiny $2$};
\draw node at (-1.6, -1.2) {\color{red}\tiny $1$};
\end{tikzpicture}
\end{gathered}
\quad \overset{\rm cond.}{\Longrightarrow} \quad
\omega
\begin{gathered}
\begin{tikzpicture}[scale=0.5]
\draw [line] (-2,2) -- (0,0);
\draw [line] (0,2) --(1,1);
\draw [line] (1,1) --(2,2);
\draw [line] (0,0)  -- (1,1);
\draw [line,dashed] (0,0) -- (0,-1) ;
\node at (1,1)[circle,fill,inner sep=1.5pt,red,opacity=1]{};
\node at (0,0)[circle,fill,inner sep=1.5pt,red,opacity=1]{};
\draw node at (1.4, .7) {\tiny $\psi_2$};
\draw node at (.4, -.3) {\tiny $\psi_1$};
\end{tikzpicture}
\end{gathered}\,,
\end{align}
where the prefactor ``$\omega$" is precisely captured by the half-braiding between lines $\mathcal L_{\frac{1}{15}}$ and $\mathcal L_{\frac{2}{3}^*}$. One can similarly work out the map from \eqref{eq:F1} to \eqref{eq:F2} that we will omit for brevity, and thus establish the isomorphism between $\mathcal C_1$ and $\mathcal C_2$. 

\subsection{Parafermionic $\mathbb Z_4$}
Our second example is the $\mathbb Z_4$ parafermionization of the $\mathbb Z_2$ orbifold theory of the $c=1$ compact boson at the special radius $R=\sqrt{6}$, which can also be realized as the ${\rm SU}(2)_4/{\rm U(1)}$ coset model. The original bosonic CFT is rational and has 10 primaries with conformal weights
\begin{align}
\left\{0,\,\frac{3}{4},\,1,\frac{3}{4}^*,\,\frac{1}{16},\,\frac{1}{16}^*,\,\frac{9}{16},\,\frac{9}{16}^*,\,\frac{1}{3},\,\frac{1}{12}\right\}\,,
\end{align}
where the ``$*$" denotes charge conjugation as before. The 10 primaries implies that there are 10 Verlinde lines labeled by the conformal weights as well,
\begin{align}
\mathcal C\equiv\left\{\mathcal L_0,\,\mathcal L_{\frac{3}{4}},\,\mathcal L_{1},\mathcal L_{\frac{3}{4}^*},\,\mathcal L_{\frac{1}{16}},\,\mathcal L_{\frac{1}{16}^*},\,\mathcal L_{\frac{9}{16}},\,\mathcal L_{\frac{9}{16}^*},\,\mathcal L_{\frac{1}{3}},\,\mathcal L_{\frac{1}{12}}\right\}\,.
\end{align}
By studying their fusion rules, one can find the first four lines in $\mathcal C$ form the to-be-condensed $\mathbb Z_4$ symmetry. On the other hand, the rest are non-invertible TDLs, and can be grouped by the orbits of the $\mathbb Z_4$-actions. One finds that the set
\begin{align}
\mathcal S_1=\left\{\mathcal L_{\frac{1}{16}},\,\mathcal L_{\frac{1}{16}^*},\,\mathcal L_{\frac{9}{16}},\,\mathcal L_{\frac{9}{16}^*}\right\}
\end{align}
form a 4-orbit, and the set
\begin{align}
\mathcal S_2=\left\{\mathcal L_{\frac{1}{3}},\,\mathcal L_{\frac{1}{12}}\right\}
\end{align}
form a 2-orbit, i.e. the two elements in $\mathcal S_2$ are $\mathbb Z_2$ fixed points out of $\mathbb Z_4$. 

Now, after we parafermionize the bosonic theory to the $\mathbb Z_4$ parafermionic one, we end up with a para-fusion category consisting of three TDLs,
\begin{align} 
\widetilde{\mathcal C}=\left\{\left[\mathcal L_0\right],\,\left[\mathcal L_{\frac{1}{3}}\right],\,\left[\mathcal L_{\frac{1}{16}}\right]\right\}\,,
\end{align}
where $\left[\mathcal L_0\right]$, $\left[\mathcal L_{\frac{1}{3}}\right]$ and $\left[\mathcal L_{\frac{1}{16}}\right]$ are representatives of the $Z_4$-symmetry category, lines in $\mathcal S_2$ and $\mathcal S_1$ respectively. Since $\mathcal S_2$ is a 2-orbit, the TDL $\left[\mathcal L_{\frac{1}{3}}\right]$ is a $\mathbb Z_2$ q-type object, whereas $\left[\mathcal L_{\frac{1}{16}}\right]$ is of m-type because elements in $\mathcal S_1$ are transitive with respect to $\mathbb Z_4$. The fusion rules of TDLs in $\widetilde{\mathcal C}$ can be read off from their parent category $\mathcal C$,
\begin{align}
\begin{cases}
\mathcal L_{\frac{1}{3}}\cdot\mathcal L_{\frac{1}{3}}=\mathcal L_{0}+\mathcal L_{1}+\mathcal L_{\frac{1}{3}}\,,\quad\mathcal L_{\frac{1}{3}}\cdot\mathcal L_{1}=\mathcal L_{\frac{1}{3}} \\
\mathcal L_{\frac{1}{3}}\cdot \mathcal L_{\frac{1}{16}}=\mathcal L_{\frac{1}{16}}+\mathcal L_{\frac{9}{16}}=\mathcal L_{\frac{1}{16}}+\mathcal L_{1}\cdot\mathcal L_{\frac{1}{16}}\,,\\
\mathcal L_{\frac{1}{16}}\cdot \mathcal L_{\frac{1}{16}}=\mathcal L_{\frac{1}{12}}+\mathcal L_{\frac{3}{4}^*}=\mathcal L_{\frac{3}{4}}\cdot\mathcal L_{\frac{1}{3}}+\mathcal L_{\frac{3}{4}^*}\cdot\mathcal L_0\,.
\end{cases}
\label{eq:Z4}
\end{align}
The first two equations in \,(\ref{eq:Z4}) simply imply that $\left[\mathcal L_{\frac{1}{3}}\right]$ is a $\mathbb Z_2$ q-type object after condensation. Put differently, the junctions between $\left[\mathcal L_{\frac{1}{3}}\right]$ and the other TDLs can have two types of parafermionic defect operators $\psi_0$ and $\psi_2$, i.e.
\begin{align}
\begin{gathered}
\begin{tikzpicture}[scale=1.5]
    % \draw [lightgray] (0,0)  -- (0,2) -- (2,2) -- (2,0) -- (0,0) ;
    %\fill [red,opacity=2] (1,1) circle (1.5pt);
    %\fill (1,.7) circle (1pt);
    %\draw[red] (1,.7) -- (0,1);
    \draw[-<-=.5] (1,1) -- (0,2);
    \draw[-<-=.5] (1,1) -- (2,2);
    \draw[->-=.5] (1,1) -- (1,0);
    %\draw node at (1.25, 0.9) {};
    %\draw node at (0.35, 1.05) {\tiny $\mathcal L_{2/3^*}$};
    \draw node at (0.7, 1.75) {\tiny $\left[\mathcal L_{1/3}\right]$};
\draw node at (1.7, 1.25) {\tiny $\left[\mathcal L_{i}\right]$};
\draw node at (0.8, 0.25) {\tiny $\left[\mathcal L_{j}\right]$};
\draw node at (.8, 1) {\tiny $\psi_0$};
    \node at (1,1)[circle,fill,inner sep=1.5pt,red,opacity=1]{};
\end{tikzpicture}
\end{gathered}
\qquad {\rm and} \qquad
\begin{gathered}
\begin{tikzpicture}[scale=1.5]
    % \draw [lightgray] (0,0)  -- (0,2) -- (2,2) -- (2,0) -- (0,0) ;
    %\fill [red,opacity=2] (1,1) circle (1.5pt);
    %\fill (1,.7) circle (1pt);
    %\draw[red] (1,.7) -- (0,1);
    \draw[-<-=.5] (1,1) -- (0,2);
    \draw[-<-=.5] (1,1) -- (2,2);
    \draw[->-=.5] (1,1) -- (1,0);
    %\draw node at (1.25, 0.9) {};
    %\draw node at (0.35, 1.05) {\tiny $\mathcal L_{2/3^*}$};
    \draw node at (0.7, 1.75) {\tiny $\left[\mathcal L_{1/3}\right]$};
\draw node at (1.7, 1.25) {\tiny $\left[\mathcal L_{i}\right]$};
\draw node at (0.8, 0.25) {\tiny $\left[\mathcal L_{j}\right]$};
\draw node at (.8, 1) {\tiny $\psi_2$};
    \node at (1,1)[circle,fill,inner sep=1.5pt,red,opacity=1]{};
\end{tikzpicture}
\end{gathered}\,,
\end{align}
where $\left[\mathcal L_{i}\right]$ and $\left[\mathcal L_{j}\right]$ are objects in $\widetilde{\mathcal C}$. The parafermionic defect operators $\psi_0$ and $\psi_2$ arise from the condensation of the lines $\mathcal L_0$ and $\mathcal L_1$. Similarly, the third equation in Eq.~(\ref{eq:Z4}) implies that
\begin{align}
\begin{gathered}
\begin{tikzpicture}[scale=1.5]
    % \draw [lightgray] (0,0)  -- (0,2) -- (2,2) -- (2,0) -- (0,0) ;
    \fill (1,1) circle (1pt);
    \draw[-<-=.5] (1,1) -- (0,2);
    \draw[-<-=.5] (1,1) -- (2,2);
    \draw[->-=.5] (1,1) -- (1,0);
    %\draw node at (1.25, 0.9) {};
    \draw node at (0.55, 1.75) {\tiny $\mathcal L_{1/16}$};
\draw node at (1.45, 1.75) {\tiny $\mathcal L_{1/16}$};
\draw node at (0.75, 0.25) {\tiny $\mathcal L_{1/12}$};
\end{tikzpicture}
\end{gathered}
\quad=\quad
\begin{gathered}
\begin{tikzpicture}[scale=1.5]
    % \draw [lightgray] (0,0)  -- (0,2) -- (2,2) -- (2,0) -- (0,0) ;
    \fill (1,1) circle (1pt);
    \fill (1,.7) circle (1pt);
    \draw[red,->-=.6] (1,.7) -- (0,1);
    \draw[-<-=.5] (1,1) -- (0,2);
    \draw[-<-=.5]  (1,1) -- (2,2);
    \draw[->-=.6] (1,1) -- (1,0.7);
    \draw[-<-=.6] (1,0) -- (1,0.7);
    %\draw node at (1.25, 0.9) {};
    \draw node at (0.35, 1.05) {\tiny $\mathcal L_{3/4}$};
    \draw node at (0.55, 1.75) {\tiny $\mathcal L_{1/16}$};
\draw node at (1.45, 1.75) {\tiny $\mathcal L_{1/16}$};
\draw node at (0.8, 0.25) {\tiny $\mathcal L_{1/3}$};
\draw node at (1.3, .85) {\tiny $\mathcal L_{1/12}$};
\end{tikzpicture}
\end{gathered}
\quad\overset{\rm cond.}{\Longrightarrow}\quad
\begin{gathered}
\begin{tikzpicture}[scale=1.5]
    % \draw [lightgray] (0,0)  -- (0,2) -- (2,2) -- (2,0) -- (0,0) ;
    %\fill [red,opacity=2] (1,1) circle (1.5pt);
    %\fill (1,.7) circle (1pt);
    %\draw[red] (1,.7) -- (0,1);
    \draw[-<-=.5] (1,1) -- (0,2);
    \draw[-<-=.5] (1,1) -- (2,2);
    \draw[->-=.5] (1,1) -- (1,0);
    %\draw node at (1.25, 0.9) {};
    %\draw node at (0.35, 1.05) {\tiny $\mathcal L_{2/3^*}$};
    \draw node at (0.7, 1.75) {\tiny $\left[\mathcal L_{1/16}\right]$};
\draw node at (1.7, 1.25) {\tiny $\left[\mathcal L_{1/16}\right]$};
\draw node at (0.7, 0.25) {\tiny $\left[\mathcal L_{1/3}\right]$};
\draw node at (.8, 1) {\tiny $\psi_3$};
    \node at (1,1)[circle,fill,inner sep=1.5pt,red,opacity=1]{};
\end{tikzpicture}
\end{gathered}
\label{eq:mmq}
\end{align}
and
\begin{align}
\begin{gathered}
\begin{tikzpicture}[scale=1.5]
    % \draw [lightgray] (0,0)  -- (0,2) -- (2,2) -- (2,0) -- (0,0) ;
    \fill (1,1) circle (1pt);
    \draw[-<-=.5] (1,1) -- (0,2);
    \draw[-<-=.5] (1,1) -- (2,2);
    \draw[->-=.5] (1,1) -- (1,0);
    %\draw node at (1.25, 0.9) {};
    \draw node at (0.55, 1.75) {\tiny $\mathcal L_{1/16}$};
\draw node at (1.45, 1.75) {\tiny $\mathcal L_{1/16}$};
\draw node at (0.7, 0.25) {\tiny $\mathcal L_{3/4^*}$};
\end{tikzpicture}
\end{gathered}
\quad=\quad
\begin{gathered}
\begin{tikzpicture}[scale=1.5]
    % \draw [lightgray] (0,0)  -- (0,2) -- (2,2) -- (2,0) -- (0,0) ;
    \fill (1,1) circle (1pt);
    \fill (1,.7) circle (1pt);
    \draw[red,->-=.6] (1,.7) -- (0,1);
    \draw[-<-=.5] (1,1) -- (0,2);
    \draw[-<-=.5]  (1,1) -- (2,2);
    \draw[->-=.6] (1,1) -- (1,0.7);
    \draw[-<-=.6] (1,0) -- (1,0.7);
    %\draw node at (1.25, 0.9) {};
    \draw node at (0.35, 1.05) {\tiny $\mathcal L_{3/4^*}$};
    \draw node at (0.55, 1.75) {\tiny $\mathcal L_{1/16}$};
\draw node at (1.45, 1.75) {\tiny $\mathcal L_{1/16}$};
\draw node at (0.8, 0.25) {\tiny $\mathcal L_{0}$};
\draw node at (1.35, .85) {\tiny $\mathcal L_{3/4^*}$};
\end{tikzpicture}
\end{gathered}
\quad\overset{\rm cond.}{\Longrightarrow}\quad
\begin{gathered}
\begin{tikzpicture}[scale=1.5]
    % \draw [lightgray] (0,0)  -- (0,2) -- (2,2) -- (2,0) -- (0,0) ;
    %\fill [red,opacity=2] (1,1) circle (1.5pt);
    %\fill (1,.7) circle (1pt);
    %\draw[red] (1,.7) -- (0,1);
    \draw[-<-=.5] (1,1) -- (0,2);
    \draw[-<-=.5] (1,1) -- (2,2);
    \draw[->-=.5] (1,1) -- (1,0);
    %\draw node at (1.25, 0.9) {};
    %\draw node at (0.35, 1.05) {\tiny $\mathcal L_{2/3^*}$};
    \draw node at (0.7, 1.75) {\tiny $\left[\mathcal L_{1/16}\right]$};
\draw node at (1.7, 1.25) {\tiny $\left[\mathcal L_{1/16}\right]$};
\draw node at (0.8, 0.25) {\tiny $\left[\mathcal L_{0}\right]$};
\draw node at (.8, 1) {\tiny $\psi_1$};
\node at (1,1)[circle,fill,inner sep=1.5pt,red,opacity=1]{};
\end{tikzpicture}
\end{gathered}\,.
\end{align}
Notice also in eq.\,(\ref{eq:mmq}), the TDL $\mathcal L_{1/3}$ is in addition invariant by fusing $\mathcal L_{1}$,
\begin{align}
\mathcal L_{\frac{3}{4}}\cdot\mathcal L_{\frac{1}{3}}=\mathcal L_{\frac{3}{4}^*}\cdot\mathcal L_{\frac{1}{3}}
\end{align}
Therefore, after condensation, the parafermionic defect $\psi_1$ can also be located at the junction, namely
\begin{align}
\begin{gathered}
\begin{tikzpicture}[scale=1.5]
    % \draw [lightgray] (0,0)  -- (0,2) -- (2,2) -- (2,0) -- (0,0) ;
    %\fill [red,opacity=2] (1,1) circle (1.5pt);
    %\fill (1,.7) circle (1pt);
    %\draw[red] (1,.7) -- (0,1);
    \draw[-<-=.5] (1,1) -- (0,2);
    \draw[-<-=.5] (1,1) -- (2,2);
    \draw[->-=.5] (1,1) -- (1,0);
    %\draw node at (1.25, 0.9) {};
    %\draw node at (0.35, 1.05) {\tiny $\mathcal L_{2/3^*}$};
    \draw node at (0.7, 1.75) {\tiny $\left[\mathcal L_{1/16}\right]$};
\draw node at (1.7, 1.25) {\tiny $\left[\mathcal L_{1/16}\right]$};
\draw node at (0.7, 0.25) {\tiny $\left[\mathcal L_{1/3}\right]$};
\draw node at (.8, 1) {\tiny $\psi_1$};
    \node at (1,1)[circle,fill,inner sep=1.5pt,red,opacity=1]{};
\end{tikzpicture}
\end{gathered}
\end{align}  
To summarize, we have the following fusion rules for $\widetilde{\mathcal C}$:
\begin{align}
&\left[\mathcal L_{i}\right]\cdot \left[\mathcal L_{j}\right]=\left[\mathcal L_{j}\right]\cdot\left[\mathcal L_{i}\right]\,,\quad
\left[\mathcal L_{0}\right]\cdot \left[\mathcal L_{i}\right]=\left[\mathcal L_{i}\right]\,,\notag\\
&\left[\mathcal L_{1/3}\right]\cdot \left[\mathcal L_{1/3}\right]=\mathbb C^{1|0|1|0}\left[\mathcal L_{0}\right]+\mathbb C^{1|0|1|0}\left[\mathcal L_{1/3}\right]\,,\quad
\left[\mathcal L_{1/3}\right]\cdot \left[\mathcal L_{1/16}\right]=\mathbb C^{1|0|1|0}\left[\mathcal L_{1/16}\right]\,,\notag\\
&\left[\mathcal L_{1/16}\right]\cdot \left[\mathcal L_{1/16}\right]=\mathbb C^{0|1|0|0}\left[\mathcal L_{0}\right]+\mathbb C^{0|1|0|1}\left[\mathcal L_{1/3}\right]\,.
\end{align}
\paragraph{Quantum dimensions.}
Let $A=[\mathcal L_{1/3}]$ and $B=[\mathcal L_{1/16}]$. To extract the fusion multiplicities relevant for Frobenius--Perron dimensions from the graded junction spaces above, one takes the rank over the endomorphism algebra of the output object. Here $A$ is a $\mathbb Z_2$ $q$-type object, so $\dim\End(A)=2$, whereas $[\mathcal L_0]$ and $B$ are $m$-type objects. The resulting fusion rules for the dimensions are therefore
\begin{equation}
 A\cdot A=2[\mathcal L_0]+A,\qquad
 A\cdot B=2B,\qquad
 B\cdot B=[\mathcal L_0]+A.
\end{equation}
Consequently, $d_A^2=2+d_A$, $d_A d_B=2d_B$, and $d_B^2=1+d_A$, which give
\begin{equation}
 d_{[\mathcal L_0]}=1,\qquad d_A=2,\qquad d_B=\sqrt{3}.
 \label{eq:Z4-quantum-dimensions}
\end{equation}
As a consistency check, these values give
\begin{equation}
 \dim(\widetilde{\mathcal C})
 =d_{[\mathcal L_0]}^2+\frac{d_A^2}{\dim\End(A)}+d_B^2
 =1+\frac{2^2}{2}+3=6
 =\frac{\dim(\mathcal C)}{4},
\end{equation}
where $\dim(\mathcal C)=24$ for the parent category.

Based on the fusion rule, we can solve the para-pentagon equations they need to satisfy. It turns out that there are 16 solutions to para-pentagons. One can compute their spin selection rules to further determine which one is for the parafermionic category $\widetilde{\mathcal C}$. Here, for brevity, we only spell out the two non-trivial F-moves with four legs of all $\left[\mathcal L_{1/3}\right]$ or $\left[\mathcal L_{1/16}\right]$. For $\mathcal F^{\left[\mathcal L_{1/3}\right],\left[\mathcal L_{1/3}\right],\left[\mathcal L_{1/3}\right]}_{\left[\mathcal L_{1/3}\right]}$, we found
\begin{align}
\begin{pmatrix}
\begin{gathered}
\begin{tikzpicture}[scale=.3]
\draw [line,blue] (-1,1) -- (-2,2);
\draw [line,blue] (-1,1) --(0,2);
\draw [line,blue] (0,0) --(2,2);
\draw [line,blue] (0,0)  -- (0,-1);
\draw [line,dotted] (0,0) -- (-1,1) ;
\node at (-1,1)[circle,fill,inner sep=.75pt,red,opacity=1]{};
\node at (0,0)[circle,fill,inner sep=.75pt,red,opacity=1]{};
\draw node at (-1.35, .7) {\tiny $0$};
\draw node at (-.4, -.4) {\tiny $0$};
\end{tikzpicture}
\end{gathered}
\\
\begin{gathered}
\begin{tikzpicture}[scale=.3]
\draw [line,blue] (-1,1) -- (-2,2);
\draw [line,blue] (-1,1) --(0,2);
\draw [line,blue] (0,0) --(2,2);
\draw [line,blue] (0,0)  -- (0,-1);
\draw [line,dotted] (0,0) -- (-1,1) ;
\node at (-1,1)[circle,fill,inner sep=.75pt,red,opacity=1]{};
\node at (0,0)[circle,fill,inner sep=.75pt,red,opacity=1]{};
\draw node at (-1.35, .7) {\tiny $0$};
\draw node at (-.4, -.3) {\tiny $2$};
\end{tikzpicture}
\end{gathered}
\\
\begin{gathered}
\begin{tikzpicture}[scale=.3]
\draw [line,blue] (-1,1) -- (-2,2);
\draw [line,blue] (-1,1) --(0,2);
\draw [line,blue] (0,0) --(2,2);
\draw [line,blue] (0,0)  -- (0,-1);
\draw [line,dotted] (0,0) -- (-1,1) ;
\node at (-1,1)[circle,fill,inner sep=.75pt,red,opacity=1]{};
\node at (0,0)[circle,fill,inner sep=.75pt,red,opacity=1]{};
\draw node at (-1.35, .7) {\tiny $2$};
\draw node at (-.4, -.3) {\tiny $0$};
\end{tikzpicture}
\end{gathered}
\\
\begin{gathered}
\begin{tikzpicture}[scale=.3]
\draw [line,blue] (-1,1) -- (-2,2);
\draw [line,blue] (-1,1) --(0,2);
\draw [line,blue] (0,0) --(2,2);
\draw [line,blue] (0,0)  -- (0,-1);
\draw [line,dotted] (0,0) -- (-1,1) ;
\node at (-1,1)[circle,fill,inner sep=.75pt,red,opacity=1]{};
\node at (0,0)[circle,fill,inner sep=.75pt,red,opacity=1]{};
\draw node at (-1.35, .7) {\tiny $2$};
\draw node at (-.4, -.3) {\tiny $2$};
\end{tikzpicture}
\end{gathered}
\\
\begin{gathered}
\begin{tikzpicture}[scale=.3]
\draw [line,blue] (-1,1) -- (-2,2);
\draw [line,blue] (-1,1) --(0,2);
\draw [line,blue] (0,0) --(2,2);
\draw [line,blue] (0,0)  -- (0,-1);
\draw [line,blue] (0,0) -- (-1,1) ;
\node at (-1,1)[circle,fill,inner sep=.75pt,red,opacity=1]{};
\node at (0,0)[circle,fill,inner sep=.75pt,red,opacity=1]{};
\draw node at (-1.35, .7) {\tiny $0$};
\draw node at (-.4, -.4) {\tiny $0$};
\end{tikzpicture}
\end{gathered}
\\
\begin{gathered}
\begin{tikzpicture}[scale=.3]
\draw [line,blue] (-1,1) -- (-2,2);
\draw [line,blue] (-1,1) --(0,2);
\draw [line,blue] (0,0) --(2,2);
\draw [line,blue] (0,0)  -- (0,-1);
\draw [line,blue] (0,0) -- (-1,1) ;
\node at (-1,1)[circle,fill,inner sep=.75pt,red,opacity=1]{};
\node at (0,0)[circle,fill,inner sep=.75pt,red,opacity=1]{};
\draw node at (-1.35, .7) {\tiny $0$};
\draw node at (-.4, -.3) {\tiny $2$};
\end{tikzpicture}
\end{gathered}
\\
\begin{gathered}
\begin{tikzpicture}[scale=.3]
\draw [line,blue] (-1,1) -- (-2,2);
\draw [line,blue] (-1,1) --(0,2);
\draw [line,blue] (0,0) --(2,2);
\draw [line,blue] (0,0)  -- (0,-1);
\draw [line,blue] (0,0) -- (-1,1) ;
\node at (-1,1)[circle,fill,inner sep=.75pt,red,opacity=1]{};
\node at (0,0)[circle,fill,inner sep=.75pt,red,opacity=1]{};
\draw node at (-1.35, .7) {\tiny $2$};
\draw node at (-.4, -.3) {\tiny $0$};
\end{tikzpicture}
\end{gathered}
\\
\begin{gathered}
\begin{tikzpicture}[scale=.3]
\draw [line,blue] (-1,1) -- (-2,2);
\draw [line,blue] (-1,1) --(0,2);
\draw [line,blue] (0,0) --(2,2);
\draw [line,blue] (0,0)  -- (0,-1);
\draw [line,blue] (0,0) -- (-1,1) ;
\node at (-1,1)[circle,fill,inner sep=.75pt,red,opacity=1]{};
\node at (0,0)[circle,fill,inner sep=.75pt,red,opacity=1]{};
\draw node at (-1.35, .7) {\tiny $2$};
\draw node at (-.4, -.3) {\tiny $2$};
\end{tikzpicture}
\end{gathered}
\end{pmatrix}
\quad=\quad
\frac{1}{2}\begin{pmatrix}
1 & 0 & 0 & -1 & 1 & 0 & 0 & -1 \\
0 & 1 & -1 & 0 & 0 & 1 & -1 & 0 \\
0 & 1 & -1 & 0 & 0 & -1 & 1 & 0 \\
1 & 0 & 0 & -1 & -1 & 0 & 0 & 1 \\
1 & 0 & 0 & 1 & 0 & 0 & 0 & 0 \\
0 & 1 & 1 & 0 & 0 & 0 & 0 & 0 \\
0 & -1 & -1 & 0 & 0 & 0 & 0 & 0 \\
-1 & 0 & 0 & -1 & 0 & 0 & 0 & 0 \\
\end{pmatrix}
\cdot
\begin{pmatrix}
\begin{gathered}
\begin{tikzpicture}[scale=.3]
\draw [line,blue] (0,0) -- (-2,2);
\draw [line,blue] (1,1) --(0,2);
\draw [line,blue] (1,1) --(2,2);
\draw [line,blue] (0,0)  -- (0,-1);
\draw [line,dotted] (0,0) -- (1,1) ;
\node at (1,1)[circle,fill,inner sep=.75pt,red,opacity=1]{};
\node at (0,0)[circle,fill,inner sep=.75pt,red,opacity=1]{};
\draw node at (1.4, .7) {\tiny $0$};
\draw node at (.4, -.3) {\tiny $0$};
\end{tikzpicture}
\end{gathered}
\\
\begin{gathered}
\begin{tikzpicture}[scale=.3]
\draw [line,blue] (0,0) -- (-2,2);
\draw [line,blue] (1,1) --(0,2);
\draw [line,blue] (1,1) --(2,2);
\draw [line,blue] (0,0)  -- (0,-1);
\draw [line,dotted] (0,0) -- (1,1) ;
\node at (1,1)[circle,fill,inner sep=.75pt,red,opacity=1]{};
\node at (0,0)[circle,fill,inner sep=.75pt,red,opacity=1]{};
\draw node at (1.4, .7) {\tiny $0$};
\draw node at (.4, -.3) {\tiny $2$};
\end{tikzpicture}
\end{gathered}
\\\begin{gathered}
\begin{tikzpicture}[scale=.3]
\draw [line,blue] (0,0) -- (-2,2);
\draw [line,blue] (1,1) --(0,2);
\draw [line,blue] (1,1) --(2,2);
\draw [line,blue] (0,0)  -- (0,-1);
\draw [line,dotted] (0,0) -- (1,1) ;
\node at (1,1)[circle,fill,inner sep=.75pt,red,opacity=1]{};
\node at (0,0)[circle,fill,inner sep=.75pt,red,opacity=1]{};
\draw node at (1.4, .7) {\tiny $2$};
\draw node at (.4, -.3) {\tiny $0$};
\end{tikzpicture}
\end{gathered}
\\
\begin{gathered}
\begin{tikzpicture}[scale=.3]
\draw [line,blue] (0,0) -- (-2,2);
\draw [line,blue] (1,1) --(0,2);
\draw [line,blue] (1,1) --(2,2);
\draw [line,blue] (0,0)  -- (0,-1);
\draw [line,dotted] (0,0) -- (1,1) ;
\node at (1,1)[circle,fill,inner sep=.75pt,red,opacity=1]{};
\node at (0,0)[circle,fill,inner sep=.75pt,red,opacity=1]{};
\draw node at (1.4, .7) {\tiny $2$};
\draw node at (.4, -.3) {\tiny $2$};
\end{tikzpicture}
\end{gathered}
\\
\begin{gathered}
\begin{tikzpicture}[scale=.3]
\draw [line,blue] (0,0) -- (-2,2);
\draw [line,blue] (1,1) --(0,2);
\draw [line,blue] (1,1) --(2,2);
\draw [line,blue] (0,0)  -- (0,-1);
\draw [line,blue] (0,0) -- (1,1) ;
\node at (1,1)[circle,fill,inner sep=.75pt,red,opacity=1]{};
\node at (0,0)[circle,fill,inner sep=.75pt,red,opacity=1]{};
\draw node at (1.4, .7) {\tiny $0$};
\draw node at (.4, -.3) {\tiny $0$};
\end{tikzpicture}
\end{gathered}
\\
\begin{gathered}
\begin{tikzpicture}[scale=.3]
\draw [line,blue] (0,0) -- (-2,2);
\draw [line,blue] (1,1) --(0,2);
\draw [line,blue] (1,1) --(2,2);
\draw [line,blue] (0,0)  -- (0,-1);
\draw [line,blue] (0,0) -- (1,1) ;
\node at (1,1)[circle,fill,inner sep=.75pt,red,opacity=1]{};
\node at (0,0)[circle,fill,inner sep=.75pt,red,opacity=1]{};
\draw node at (1.4, .7) {\tiny $0$};
\draw node at (.4, -.3) {\tiny $2$};
\end{tikzpicture}
\end{gathered}
\\
\begin{gathered}
\begin{tikzpicture}[scale=.3]
\draw [line,blue] (0,0) -- (-2,2);
\draw [line,blue] (1,1) --(0,2);
\draw [line,blue] (1,1) --(2,2);
\draw [line,blue] (0,0)  -- (0,-1);
\draw [line,blue] (0,0) -- (1,1) ;
\node at (1,1)[circle,fill,inner sep=.75pt,red,opacity=1]{};
\node at (0,0)[circle,fill,inner sep=.75pt,red,opacity=1]{};
\draw node at (1.4, .7) {\tiny $2$};
\draw node at (.4, -.3) {\tiny $0$};
\end{tikzpicture}
\end{gathered}
\\
\begin{gathered}
\begin{tikzpicture}[scale=.3]
\draw [line,blue] (0,0) -- (-2,2);
\draw [line,blue] (1,1) --(0,2);
\draw [line,blue] (1,1) --(2,2);
\draw [line,blue] (0,0)  -- (0,-1);
\draw [line,blue] (0,0) -- (1,1) ;
\node at (1,1)[circle,fill,inner sep=.75pt,red,opacity=1]{};
\node at (0,0)[circle,fill,inner sep=.75pt,red,opacity=1]{};
\draw node at (1.4, .7) {\tiny $2$};
\draw node at (.4, -.3) {\tiny $2$};
\end{tikzpicture}
\end{gathered}
\end{pmatrix}\,,
\end{align}
and for $\mathcal F^{\left[\mathcal L_{1/16}\right],\left[\mathcal L_{1/16}\right],\left[\mathcal L_{1/16}\right]}_{\left[\mathcal L_{1/16}\right]}$ we  have
\begin{align}
\begin{pmatrix}
\begin{gathered}
\begin{tikzpicture}[scale=.3]
\draw [line] (-1,1) -- (-2,2);
\draw [line] (-1,1) --(0,2);
\draw [line] (0,0) --(2,2);
\draw [line] (0,0)  -- (0,-1);
\draw [line,dotted] (0,0) -- (-1,1) ;
\node at (-1,1)[circle,fill,inner sep=.75pt,red,opacity=1]{};
\node at (0,0)[circle,fill,inner sep=.75pt,red,opacity=1]{};
\draw node at (-1.35, .7) {\tiny $1$};
\draw node at (-.4, -.4) {\tiny $0$};
\end{tikzpicture}
\end{gathered}
\\
\begin{gathered}
\begin{tikzpicture}[scale=.3]
\draw [line] (-1,1) -- (-2,2);
\draw [line] (-1,1) --(0,2);
\draw [line] (0,0) --(2,2);
\draw [line] (0,0)  -- (0,-1);
\draw [line,blue] (0,0) -- (-1,1) ;
\node at (-1,1)[circle,fill,inner sep=.75pt,red,opacity=1]{};
\node at (0,0)[circle,fill,inner sep=.75pt,red,opacity=1]{};
\draw node at (-1.35, .7) {\tiny $1$};
\draw node at (-.4, -.3) {\tiny $0$};
\end{tikzpicture}
\end{gathered}
\\
\begin{gathered}
\begin{tikzpicture}[scale=.3]
\draw [line] (-1,1) -- (-2,2);
\draw [line] (-1,1) --(0,2);
\draw [line] (0,0) --(2,2);
\draw [line] (0,0)  -- (0,-1);
\draw [line,blue] (0,0) -- (-1,1) ;
\node at (-1,1)[circle,fill,inner sep=.75pt,red,opacity=1]{};
\node at (0,0)[circle,fill,inner sep=.75pt,red,opacity=1]{};
\draw node at (-1.35, .7) {\tiny $3$};
\draw node at (-.4, -.3) {\tiny $2$};
\end{tikzpicture}
\end{gathered}
\\
\begin{gathered}
\begin{tikzpicture}[scale=.3]
\draw [line] (-1,1) -- (-2,2);
\draw [line] (-1,1) --(0,2);
\draw [line] (0,0) --(2,2);
\draw [line] (0,0)  -- (0,-1);
\draw [line,blue] (0,0) -- (-1,1) ;
\node at (-1,1)[circle,fill,inner sep=.75pt,red,opacity=1]{};
\node at (0,0)[circle,fill,inner sep=.75pt,red,opacity=1]{};
\draw node at (-1.35, .7) {\tiny $1$};
\draw node at (-.4, -.3) {\tiny $2$};
\end{tikzpicture}
\end{gathered}
\\
\begin{gathered}
\begin{tikzpicture}[scale=.3]
\draw [line] (-1,1) -- (-2,2);
\draw [line] (-1,1) --(0,2);
\draw [line] (0,0) --(2,2);
\draw [line] (0,0)  -- (0,-1);
\draw [line,blue] (0,0) -- (-1,1) ;
\node at (-1,1)[circle,fill,inner sep=.75pt,red,opacity=1]{};
\node at (0,0)[circle,fill,inner sep=.75pt,red,opacity=1]{};
\draw node at (-1.35, .7) {\tiny $3$};
\draw node at (-.4, -.4) {\tiny $0$};
\end{tikzpicture}
\end{gathered}
\end{pmatrix}
\quad=\quad
(-1)^{\frac{1}{4}}\begin{pmatrix}
\frac{1}{\sqrt{3}} & \sqrt{\frac{2}{3}} & \sqrt{\frac{2}{3}} & 0 & 0\\
\frac{1}{\sqrt{3}} & -\frac{1}{\sqrt{6}} & -\frac{1}{\sqrt{6}} & 0 & 0\\
\frac{1}{\sqrt{3}} & -\frac{1}{\sqrt{6}} & -\frac{1}{\sqrt{6}} & 0 & 0\\
0 & 0 & 0 & \frac{i}{\sqrt{2}} & \frac{i}{\sqrt{2}} \\
0 & 0 & 0 & \frac{i}{\sqrt{2}} & \frac{i}{\sqrt{2}}
\end{pmatrix}
\cdot
\begin{pmatrix}
\begin{gathered}
\begin{tikzpicture}[scale=.3]
\draw [line] (0,0) -- (-2,2);
\draw [line] (1,1) --(0,2);
\draw [line] (1,1) --(2,2);
\draw [line] (0,0)  -- (0,-1);
\draw [line,dotted] (0,0) -- (1,1) ;
\node at (1,1)[circle,fill,inner sep=.75pt,red,opacity=1]{};
\node at (0,0)[circle,fill,inner sep=.75pt,red,opacity=1]{};
\draw node at (1.4, .7) {\tiny $1$};
\draw node at (.4, -.3) {\tiny $0$};
\end{tikzpicture}
\end{gathered}
\\
\begin{gathered}
\begin{tikzpicture}[scale=.3]
\draw [line] (0,0) -- (-2,2);
\draw [line] (1,1) --(0,2);
\draw [line] (1,1) --(2,2);
\draw [line] (0,0)  -- (0,-1);
\draw [line,blue] (0,0) -- (1,1) ;
\node at (1,1)[circle,fill,inner sep=.75pt,red,opacity=1]{};
\node at (0,0)[circle,fill,inner sep=.75pt,red,opacity=1]{};
\draw node at (1.4, .7) {\tiny $1$};
\draw node at (.4, -.3) {\tiny $0$};
\end{tikzpicture}
\end{gathered}
\\\begin{gathered}
\begin{tikzpicture}[scale=.3]
\draw [line] (0,0) -- (-2,2);
\draw [line] (1,1) --(0,2);
\draw [line] (1,1) --(2,2);
\draw [line] (0,0)  -- (0,-1);
\draw [line,blue] (0,0) -- (1,1) ;
\node at (1,1)[circle,fill,inner sep=.75pt,red,opacity=1]{};
\node at (0,0)[circle,fill,inner sep=.75pt,red,opacity=1]{};
\draw node at (1.4, .7) {\tiny $3$};
\draw node at (.4, -.3) {\tiny $2$};
\end{tikzpicture}
\end{gathered}
\\
\begin{gathered}
\begin{tikzpicture}[scale=.3]
\draw [line] (0,0) -- (-2,2);
\draw [line] (1,1) --(0,2);
\draw [line] (1,1) --(2,2);
\draw [line] (0,0)  -- (0,-1);
\draw [line,blue] (0,0) -- (1,1) ;
\node at (1,1)[circle,fill,inner sep=.75pt,red,opacity=1]{};
\node at (0,0)[circle,fill,inner sep=.75pt,red,opacity=1]{};
\draw node at (1.4, .7) {\tiny $1$};
\draw node at (.4, -.3) {\tiny $2$};
\end{tikzpicture}
\end{gathered}
\\
\begin{gathered}
\begin{tikzpicture}[scale=.3]
\draw [line] (0,0) -- (-2,2);
\draw [line] (1,1) --(0,2);
\draw [line] (1,1) --(2,2);
\draw [line] (0,0)  -- (0,-1);
\draw [line,blue] (0,0) -- (1,1) ;
\node at (1,1)[circle,fill,inner sep=.75pt,red,opacity=1]{};
\node at (0,0)[circle,fill,inner sep=.75pt,red,opacity=1]{};
\draw node at (1.4, .7) {\tiny $3$};
\draw node at (.4, -.3) {\tiny $0$};
\end{tikzpicture}
\end{gathered}
\end{pmatrix}\,,
\end{align}
where the dotted, black and blue lines denote the $\left[\mathcal L_{0}\right]$, $\left[\mathcal L_{1/3}\right]$ and $\left[\mathcal L_{1/16}\right]$ TDLs respectively, and the red dots and numbers stand for the parafermionic defects $\psi_{i}$. One can verify that the ranks of $\mathcal F^{\left[\mathcal L_{1/3}\right],\left[\mathcal L_{1/3}\right],\left[\mathcal L_{1/3}\right]}_{\left[\mathcal L_{1/3}\right]}$ and $\mathcal F^{\left[\mathcal L_{1/16}\right],\left[\mathcal L_{1/16}\right],\left[\mathcal L_{1/16}\right]}_{\left[\mathcal L_{1/16}\right]}$ are $6$ and $3$. This is because the parafermionic defects can ``move'' along the $\mathbb Z_2$ q-type TDL $\left[\mathcal L_{1/3}\right]$.

\subsection{\texorpdfstring{A $\mathbb Z_4$ example with $q_2$ and $q_4$ objects}{A Z4 example with q2 and q4 objects}}
\label{sec:q2-q4-example}
We now construct an example containing both $\mathbb Z_2$ and $\mathbb Z_4$ $q$-type objects. Consider the bosonic CFT
\begin{align}
\mathcal T=\left[\frac{{\rm SU}(2)_4}{{\rm U}(1)}\right]_{\rm diagonal}\times {\rm Ising}\,.
\label{eq:q2q4-cft}
\end{align}
The first factor admits a finite system of TDLs forming the Tambara--Yamagami category ${\rm TY}_{\mathbb Z_4}^{1,+1}$, with bicharacter $\chi(r,s)=i^{rs}$ and Frobenius--Schur indicator $+1$ \cite{Thorngren:2021yso}. Its simple objects are $1,a,a^2,a^3,D$, with fusion rules
\begin{align}
a^4&=1\,, & a^r\otimes D&=D\otimes a^r=D\,, & D\otimes D&=\bigoplus_{r=0}^3 a^r\,.
\label{eq:q2q4-ty-fusion}
\end{align}
The duality line $D$ does not preserve the full chiral algebra, so this example includes TDLs beyond the Verlinde lines of the preceding subsections. We select the closed defect subsystem
\begin{align}
\C={\rm TY}_{\mathbb Z_4}^{1,+1}\boxtimes {\rm Ising}\,,
\label{eq:q2q4-parent}
\end{align}
where the Ising factor has simple objects $1,\psi,\sigma$ satisfying $\psi^2=1$, $\psi\otimes\sigma=\sigma$ and $\sigma\otimes\sigma=1\oplus\psi$.

We parafermionize the diagonal $\mathbb Z_4$ symmetry generated by
\begin{align}
f=a\boxtimes\psi\,,\qquad f^2=a^2\boxtimes1\ne1\,,\qquad f^4=1\,.
\label{eq:q2q4-generator}
\end{align}
To apply the UFC construction of Section~\ref{sec:pfcond2}, we choose a central lift $\widetilde a\in Z({\rm TY}_{\mathbb Z_4}^{1,+1})$ of exact order four and twist $-i$, and the usual central lift $\widetilde\psi\in Z({\rm Ising})$ of order two and twist $-1$.\footnote{In the notation for the invertible objects of the TY center in refs.~\cite{Gelaki:2009zgc,Naidu:2009fin}, one may take $\widetilde a=X_{1,\epsilon}$ with $\epsilon=e^{-\pi i/4}$. In general, $X_{r,\epsilon}$ obeys $\epsilon^2=\chi(r,r)^{-1}$, $\theta_{X_{r,\epsilon}}=\chi(r,r)^{-1}$ and $X_{r,\epsilon}\otimes X_{s,\epsilon'}=X_{r+s,\epsilon\epsilon'\chi(r,s)^{-1}}$. Thus $\widetilde a^2=X_{2,-1}$, $\widetilde a^4=X_{0,1}=1$ and $\theta_{\widetilde a}=-i$.} Their product $\widetilde f=\widetilde a\boxtimes\widetilde\psi\in Z(\C)$ satisfies
\begin{align}
\widetilde f^4=1\,,\qquad
\theta_{\widetilde f}=(-i)(-1)=i\,,\qquad
h_{\widetilde f}=\frac14\pmod1\,.
\label{eq:q2q4-spin}
\end{align}
Since $f$ already has exact order four, so does $\widetilde f$. This gives $N=4$ and $m=1$ in Eq.~\eqref{eq:parafermionic-line-definition}. The pointed subcategory generated by $f$ has trivial associator, so the diagonal symmetry is nonanomalous and the object algebra to be condensed is $A^{pf}=1\oplus f\oplus f^2\oplus f^3$.

The fifteen simple objects of $\C$ form five orbits under fusion with $f$. Their representatives, orbit lengths and stabilizers are
\begin{align}
\begin{array}{c|c|c|c}
\text{representative }X & |\mathbb Z_4\cdot X| & H_X & \text{type of }[X] \\\hline
1\boxtimes1 & 4 & \{0\} & m \\
1\boxtimes\psi & 4 & \{0\} & m \\
1\boxtimes\sigma & 4 & \{0\} & m \\
D\boxtimes1 & 2 & \{0,2\}\simeq\mathbb Z_2 & q_2 \\
D\boxtimes\sigma & 1 & \mathbb Z_4 & q_4
\end{array}\,.
\label{eq:q2q4-orbits}
\end{align}
In particular, $f\otimes(D\boxtimes1)=D\boxtimes\psi$, while $f^2\otimes(D\boxtimes1)=D\boxtimes1$. The object $D\boxtimes\sigma$ is fixed by $f$ itself. Equation~\eqref{eq:pf_Hom_cond} then gives the graded endomorphism spaces
\begin{align}
\End([D\boxtimes1])&\simeq\bC_0\oplus\bC_2\,,\notag\\
\End([D\boxtimes\sigma])&\simeq\bC_0\oplus\bC_1\oplus\bC_2\oplus\bC_3\,,
\label{eq:q2q4-endomorphisms}
\end{align}
where $\bC_j$ denotes a one dimensional homogeneous component of degree $j\in\mathbb Z_4$. Their algebra structures are the graded division algebras associated with the respective cyclic stabilizers, as explained in Appendix~\ref{app:graded-division-algebra}. Thus the selected subsystem gives three $m$-type objects, one $\mathbb Z_2$ $q$-type object and one $\mathbb Z_4$ $q$-type object after $\mathbb Z_4$ parafermionization.

%%%%%%%%%%%%%%%%%%%%%%%%%%%%%%%%55
\section{Example: Para-condensed ${\rm TY}(\mathbb Z_N)$ Category}
\label{5}
The preceding example used a duality line beyond the Verlinde sector. We now study the parafermionic condensation of a general ${\rm TY}(\mathbb Z_N)$ category, whose duality line supports $N$ types of parafermionic defect operators after condensation. These TDLs can be obtained by para-condensing the non-invertible duality line $\mathcal N$ of the parent bosonic WZW model $\mathfrak {su}(2)_N/\mathfrak u(1)$. At the conformal fixed point, these theories are self-dual under gauging their $\mathbb Z_N$ symmetries. Gauging the $\mathbb Z_N$ symmetry on a half-plane \cite{Thorngren:2021yso} therefore produces a non-invertible interface $\mathcal N$ between the gauged and ungauged regions, with fusion rule
\begin{align}
\mathcal N\cdot\mathcal N=\sum_{a\in \mathbb Z_N} a\,\quad\quad {\rm and} \quad \mathcal N\cdot a=a\cdot \mathcal N=\mathcal N\,, \quad {\rm for} \quad a\in \mathbb Z_N\,.
\label{eq:Zn_fusion}
\end{align} 
The $\mathcal N$-line together with the $\mathbb Z_N$ symmetry combined are the well-known Tambara-Yamagami category ${\rm TY}_{\mathbb Z_N}^{t,\,\kappa}$, where $t$, satisfying $1\leq t\leq N-1$ and ${\rm gcd}(t,N)=1$, defines a symmetric bi-character $\chi_t$ as
\begin{align}
\chi_{t}(a,b)=e^{\frac{2\pi i t a b}{N}}\equiv \omega_{N,t}^{ a b}\,, \quad {\rm for}\quad
a,b\in\mathbb Z_N\,,
\end{align}
and $\kappa=\pm 1$ is the Frobenius--Schur indicator. The $\mathbb Z_N$ TY category is completely determined by $t$ and $\kappa$. Its nontrivial F-symbols are
\begin{align}
\begin{cases}
\begin{gathered}
\begin{tikzpicture}[scale=0.5]
\draw [line, red] (-1,1) -- (-2,2);
\draw [line] (-1,1) --(0,2);
\draw [line, red] (0,0) --(2,2);
\draw [line] (0,0)  -- (0,-1);
\draw [line] (0,0) -- (-1,1) ;
\draw node at (-1.5,1.8) {\tiny $a$};
\draw node at (1.5,1.8) {\tiny $b$};
%\node at (-1,1)[circle,fill,inner sep=1.5pt,red,opacity=1]{};
%\node at (0,0)[circle,fill,inner sep=1.5pt,red,opacity=1]{};
%\draw node at (-1.3, .7) {\tiny $\psi_2$};
%\draw node at (-.3, -.3) {\tiny $\psi_1$};
\end{tikzpicture}
\end{gathered}
\quad=\quad \mathcal F^{a,\,\mathcal N,\,b}_{\mathcal N}
\begin{gathered}
\begin{tikzpicture}[scale=0.5]
\draw [line, red] (-2,2) -- (0,0);
\draw [line] (0,2) --(1,1);
\draw [line, red] (1,1) --(2,2);
\draw [line] (0,0)  -- (1,1);
\draw [line] (0,0) -- (0,-1) ;
\draw node at (-1.5,1.8) {\tiny $a$};
\draw node at (1.5,1.8) {\tiny $b$};
%\node at (1,1)[circle,fill,inner sep=1.5pt,red,opacity=1]{};
%\node at (0,0)[circle,fill,inner sep=1.5pt,red,opacity=1]{};
%\draw node at (1.4, .7) {\tiny $\psi_2$};
%\draw node at (.4, -.3) {\tiny $\psi_1$};
\end{tikzpicture}
\end{gathered}\,,\quad \mathcal F^{a,\,\mathcal N,\,b}_{\mathcal N}=\chi_t(a,b)\,,\\\\
\begin{gathered}
\begin{tikzpicture}[scale=0.5]
\draw [line] (-1,1) -- (-2,2);
\draw [line,red] (-1,1) --(0,2);
\draw [line] (0,0) --(2,2);
\draw [line,red] (0,0)  -- (0,-1);
\draw [line] (0,0) -- (-1,1) ;
\draw node at (-.5,1.8) {\tiny $a$};
\draw node at (0.3,-.8) {\tiny $b$};
%\node at (-1,1)[circle,fill,inner sep=1.5pt,red,opacity=1]{};
%\node at (0,0)[circle,fill,inner sep=1.5pt,red,opacity=1]{};
%\draw node at (-1.3, .7) {\tiny $\psi_2$};
%\draw node at (-.3, -.3) {\tiny $\psi_1$};
\end{tikzpicture}
\end{gathered}
\quad=\quad \mathcal F^{\mathcal N,\,a,\,\mathcal N}_{b}
\begin{gathered}
\begin{tikzpicture}[scale=0.5]
\draw [line] (-2,2) -- (0,0);
\draw [line,red] (0,2) --(1,1);
\draw [line] (1,1) --(2,2);
\draw [line] (0,0)  -- (1,1);
\draw [line,red] (0,0) -- (0,-1) ;
\draw node at (-.2,1.8) {\tiny $a$};
\draw node at (0.3,-.8) {\tiny $b$};
%\node at (1,1)[circle,fill,inner sep=1.5pt,red,opacity=1]{};
%\node at (0,0)[circle,fill,inner sep=1.5pt,red,opacity=1]{};
%\draw node at (1.4, .7) {\tiny $\psi_2$};
%\draw node at (.4, -.3) {\tiny $\psi_1$};
\end{tikzpicture}
\end{gathered}\,,
\quad \mathcal F^{\mathcal N,\,a,\,\mathcal N}_{b}=\chi_t(a,b)\,,\\\\
\begin{gathered}
\begin{tikzpicture}[scale=0.5]
\draw [line] (-1,1) -- (-2,2);
\draw [line] (-1,1) --(0,2);
\draw [line] (0,0) --(2,2);
\draw [line] (0,0)  -- (0,-1);
\draw [line,red] (0,0) -- (-1,1) ;
\draw node at (-.75,.4) {\tiny $a$}; 
%\node at (-1,1)[circle,fill,inner sep=1.5pt,red,opacity=1]{};
%\node at (0,0)[circle,fill,inner sep=1.5pt,red,opacity=1]{};
%\draw node at (-1.3, .7) {\tiny $\psi_2$};
%\draw node at (-.3, -.3) {\tiny $\psi_1$};
\end{tikzpicture}
\end{gathered}
\quad=\quad \sum_{b\in\mathbb Z_N}\left(\mathcal F^{\mathcal N,\,\mathcal N,\,\mathcal N}_{\mathcal N}\right)_{ab}
\begin{gathered}
\begin{tikzpicture}[scale=0.5]
\draw [line] (-2,2) -- (0,0);
\draw [line] (0,2) --(1,1);
\draw [line] (1,1) --(2,2);
\draw [line,red] (0,0)  -- (1,1);
\draw [line] (0,0) -- (0,-1) ;
\draw node at (.8,.4) {\tiny $b$}; 
%\node at (1,1)[circle,fill,inner sep=1.5pt,red,opacity=1]{};
%\node at (0,0)[circle,fill,inner sep=1.5pt,red,opacity=1]{};
%\draw node at (1.4, .7) {\tiny $\psi_2$};
%\draw node at (.4, -.3) {\tiny $\psi_1$};
\end{tikzpicture}
\end{gathered}\,,
\quad \left(\mathcal F^{\mathcal N,\,\mathcal N,\,\mathcal N}_{\mathcal N}\right)_{ab}=\frac{\kappa}{\sqrt{N}}\chi^{-1}_t(a,b)\,,
\end{cases}
\label{eq:TY_Fmove}
\end{align}
where the black and red lines in the graphs denote the TDLs for $\mathcal N$-line and $\mathbb Z_N$-lines respectively. For the bosonic $\mathfrak {su}(2)_N/\mathfrak u(1)$ model, the $\mathcal N$-line combined with $\mathbb Z_N$ symmetry specify one of ${\rm TY}_{\mathbb Z_N}^{t,\,\kappa}$'s. For example, in the case of $\mathbb Z_2$, $\mathfrak {su}(2)_2/\mathfrak u(1)\simeq {\rm Ising}$, so the TY category is given by ${\rm TY}_{\mathbb Z_N}^{1,\,+1}$.

After para-condensation, the $\mathbb Z_N$-TDLs are all condensed to the trivial line. The category thus only contains two objects\footnote{By abuse of notation $\mathcal N$ denotes the duality line in both the parent ${\rm TY}_{\mathbb Z_N}^{t,\,\kappa}$ category and its para-condensation.}
\begin{align}
\left\{\mathds 1\,,\ \mathcal N\right\}\,,
\end{align}
with the fusion rule
\begin{align}
\mathcal N\cdot\mathcal N=\mathbb C^{\overbrace{1|1|\cdots|1}^{N}}\mathds 1\,,\quad{\rm and}\quad \mathcal N\cdot\mathds 1=\mathds 1\cdot \mathcal N=\mathcal N
\label{eq:pfZn_fusion}
\end{align}
Apparently, the fusion \eqref{eq:pfZn_fusion} condensed from \eqref{eq:Zn_fusion} implies that there are $N$ different types of 1d parafermionic defects, denoted as ``color", that can live on the $\mathcal N$-line. 

In this subsection, we propose a classification of the $\mathbb Z_N$ para-condensed categories ${\rm TY}_{\mathbb Z_N}^{t,\,\kappa}$, denoted by pf-${\rm TY}_{\mathbb Z_N}^{t,\kappa,\beta}$, where the additional parameter $\beta$ is defined below. The fusion rule implies that only F-symbols with an even number of $\mathcal N$ lines are nonzero:
\begin{align}
    \left\{
    \mathcal F^{\mathds 1,\,\mathds 1,\,\mathds 1}_{\mathds 1}\,,\,
    \mathcal F^{\mathds 1,\,\mathds 1,\,\mathcal N}_{\mathcal N}\,,\,
    \mathcal F^{\mathds 1,\,\mathcal N,\,\mathds 1}_{\mathcal N}\,,\,
    \mathcal F^{\mathds 1,\,\mathcal N,\,\mathcal N}_{\mathds 1}\,,\,
    \mathcal F^{\mathcal N,\,\mathds 1,\,\mathds 1}_{\mathcal N}\,,\,
    \mathcal F^{\mathcal N,\,\mathds 1,\,\mathcal N}_{\mathds 1}\,,\,
    \mathcal F^{\mathcal N,\,\mathcal N,\,\mathds 1}_{\mathds 1}\,,\,
    \mathcal F^{\mathcal N,\,\mathcal N,\,\mathcal N}_{\mathcal N}\right\}\,.
\label{Fsymb}
\end{align}
Since, at trivalent junctions involving the $\mathcal N$-line, there could be $N$ different colors, these F-symbols are generically matrices. For example, for $\mathcal F^{\mathds 1,\,\mathds 1,\,\mathcal N}_{\mathcal N}$,
\begin{align}
\begin{gathered}
\begin{tikzpicture}[scale=0.5]
% \draw [lightgray] (-2,-2)  -- (-2,2) -- (2,2) -- (2,-2) -- (-2,-2) ;
\draw [line,dashed] (-1,1) -- (-2,2);
\draw [line,dashed] (-1,1) --(0,2);
\draw [line] (0,0) --(2,2);
\draw [line] (0,0)  -- (0,-2);
\draw [line,dashed] (0,0) -- (-1,1) ;
%\node at (-1,1)[circle,fill,inner sep=1.5pt,red,opacity=1]{};
\node at (0,0)[circle,fill,inner sep=1.5pt,red,opacity=1]{};
%\draw node at (-1.3, .7) {\tiny $\psi_2$};
%\draw node at (-.3, -.3) {\tiny $\psi_1$};
\end{tikzpicture}
\end{gathered}
\quad=\mathcal F^{\mathds 1,\,\mathds 1,\,\mathcal N}_{\mathcal N}\quad 
\begin{gathered}
\begin{tikzpicture}[scale=0.5]
% \draw [lightgray] (-2,-2)  -- (-2,2) -- (2,2) -- (2,-2) -- (-2,-2) ;
\draw [line,dashed] (-2,2) -- (0,0);
\draw [line,dashed] (0,2) --(1,1);
\draw [line] (1,1) --(2,2);
\draw [line] (0,0)  -- (1,1);
\draw [line] (0,0) -- (0,-2) ;
\node at (1,1)[circle,fill,inner sep=1.5pt,red,opacity=1]{};
\node at (0,0)[circle,fill,inner sep=1.5pt,red,opacity=1]{};
%\draw node at (1.4, .7) {\tiny $\psi_2$};
%\draw node at (.4, -.3) {\tiny $\psi_1$};
\end{tikzpicture}
\end{gathered}\,,
\end{align}
where the dashed and solid lines denote the trivial and $\mathcal N$-line, respectively, and the red dot implies there are $N$ different colors at each junction. Therefore the F-symbol $ \mathcal F^{\mathds 1,\,\mathds 1,\,\mathcal N}_{\mathcal N}$ is an $N\times N^2$ matrix. We denote the entries of the F-symbol as $\mathcal F^{\mathcal L ,\,\mathcal L,\,\mathcal L}_{\mathcal L}\left(a,b,c,d\right)$. Notice that, because of the parafermionic parity \eqref{eq:pf_parity},
\begin{align}
    \mathcal F^{\mathcal L ,\,\mathcal L,\,\mathcal L}_{\mathcal L}\left(a,b,c,d\right)=0\,,\quad\quad {\rm if}\quad\quad a+b\neq c+d\mod{N}\,.
\end{align}

\subsection{Half braiding structure}
For parafermionic condensation of \TY, 
we need to introduce a consistent braiding structure for some elements in \TY. In fact, there is no such a structure for the whole \TY category except for $N=2$. However a \emph{half braiding} structure does exist for elements in $\mathbb Z_N$ and $\mathcal N$-line.

We define the braiding between $\mathbb Z_N$-anyons $a$ and $b$, and $\mathbb Z_N$-anyon $a$ and $\mathcal N$ as
\begin{align}
\begin{gathered}
\begin{tikzpicture}[scale=0.5]
\draw [line,red] (0,2) -- (2,0);
\draw [line,red] (0,0) -- (.8,.8);
\draw [line,red] (1.2,1.2) -- (2,2);
\draw node at (0, -.3) {\tiny $a$};
\draw node at (2, -.3) {\tiny $b$};
\end{tikzpicture}
\end{gathered}
\quad\equiv\beta_a(b)\quad
\begin{gathered}
\begin{tikzpicture}[scale=0.5]
\draw [line,red] (0,2) -- (2,0);
\draw [line,red] (.4,0) -- (1.2,.8);
\draw [line,red] (1.6,2) -- (.8,1.2);
\draw node at (0, -.3) {\tiny $a$};
\draw node at (2, -.3) {\tiny $b$};
\draw node at (1.9,2) {\tiny $a$};
\end{tikzpicture}
\end{gathered}
\qquad{\rm and}\qquad
%%%%
\begin{gathered}
\begin{tikzpicture}[scale=0.5]
\draw [line] (0,2) -- (2,0);
\draw [line,red] (0,0) -- (.8,.8);
\draw [line,red] (1.2,1.2) -- (2,2);
\draw node at (0, -.3) {\tiny $a$};
\draw node at (2, -.3) {\tiny $\mathcal N$};
\end{tikzpicture}
\end{gathered}
\quad\equiv\beta_a(\mathcal N)\quad
\begin{gathered}
\begin{tikzpicture}[scale=0.5]
\draw [line] (0,2) -- (2,0);
\draw [line,red] (.4,0) -- (1.2,.8);
\draw [line,red] (1.6,2) -- (.8,1.2);
\draw node at (0, -.3) {\tiny $a$};
\draw node at (2, -.3) {\tiny $\mathcal N$};
\draw node at (1.9,2) {\tiny $a$};
\end{tikzpicture}
\end{gathered}\,,
\end{align}
where the braiding $\beta_a(b)$, for consistency, needs to be identical to the braiding structure of $\mathbb Z_N$-anyons, i.e.
\begin{align}
\label{eq:Zn_braiding}
\beta_a(b)=\omega_{N,t}^{ab}=e^{\frac{2\pi itab}{N}}\,.
\end{align}
On the other hand, for $\beta_a(\mathcal N)$, first note that
\begin{align}
\label{eq:recursion}
\begin{gathered}
\begin{tikzpicture}[scale=0.5]
\draw [line] (0,2) -- (2,0);
\draw [line,red] (0,0) -- (.8,.8);
\draw [line,red] (1.2,1.2) -- (2,2);
\draw node at (0, -.3) {\tiny $a$};
\draw node at (2, -.3) {\tiny $\mathcal N$};
\end{tikzpicture}
\end{gathered}
&\quad=\quad
\begin{gathered}
\begin{tikzpicture}[scale=0.5]
\draw [line] (0,2) -- (2,0);
\draw [line,red] (.5,0) -- (1.15,.65);
\draw [line,red] (1.4,.9) -- (2.2,1.7);
\draw [line,red] (0,0.5) -- (.65,1.15);
\draw [line,red] (.9,1.4) -- (1.6,2.1);
\draw node at (0.3, -.3) {\tiny $a-1$};
\draw node at (2, -.3) {\tiny $\mathcal N$};
\draw node at (-.2, .3) {\tiny $1$};
\end{tikzpicture}
\end{gathered}
\quad=\beta_1(\mathcal N)\beta_{a-1}(\mathcal N)\quad
\begin{gathered}
\begin{tikzpicture}[scale=0.5]
\draw [line] (0,2) -- (2,0);
\draw [line,red] (.5,0) -- (1.25,.75);
\draw [line,red] (1.05,.95) -- (1.8,1.7);
\draw [line,red] (0,0.5) -- (.75,1.25);
\draw [line,red] (.6,1.4) -- (1.3,2.1);
\draw node at (0.3, -.3) {\tiny $a-1$};
\draw node at (2, -.3) {\tiny $\mathcal N$};
\draw node at (-.2,.3) {\tiny $1$};
\draw node at (2.3, 1.9) {\tiny $a-1$};
\draw node at (1.45,2.25) {\tiny $1$};
\end{tikzpicture}
\end{gathered}
\notag\\
%%%%%
&\quad=\beta_1(\mathcal N)\beta_{a-1}(\mathcal N)\chi_t(-1,a-1)\quad
\begin{gathered}
\begin{tikzpicture}[scale=0.5]
\draw [line] (0,2) -- (2,0);
\draw [line,red] (.5,0) -- (1.25,.75);
%\draw [line,red] (1.05,.95) -- (1.8,1.7);
\draw [line,red] (1.55,1.95) -- (.8,1.2);
\draw [line,red] (1.05,.95) -- (0.3,0.2);
%\draw [line,red] (0,0.5) -- (.75,1.25);
\draw [line,red] (.6,1.4) -- (1.3,2.1);
\draw node at (0.3, -.3) {\tiny $a-1$};
\draw node at (2, -.3) {\tiny $\mathcal N$};
\draw node at (0,.3) {\tiny $1$};
\draw node at (2.3, 1.9) {\tiny $a-1$};
\draw node at (1.45,2.25) {\tiny $1$};
\end{tikzpicture}
\end{gathered}
\notag\\
%%%%
&\quad=\beta_1(\mathcal N)\beta_{a-1}(\mathcal N)\chi_t(-1,a-1)\quad
\begin{gathered}
\begin{tikzpicture}[scale=0.5]
\draw [line] (0,2) -- (2,0);
\draw [line,red] (.4,0) -- (1.2,.8);
\draw [line,red] (1.6,2) -- (.8,1.2);
\draw node at (0, -.3) {\tiny $a$};
\draw node at (2, -.3) {\tiny $\mathcal N$};
\draw node at (1.9,2) {\tiny $a$};
\end{tikzpicture}
\end{gathered}\,,
\end{align}
where we have used the following F-move
\begin{align}
\begin{gathered}
\begin{tikzpicture}[scale=0.5]
\draw [line] (1.5,0)--(1.5,3);
\draw [line, red] (1.5,2) -- (0,0);
\draw [line, red] (1.5,1) -- (3,3);
\draw node at (0.3, 0) {\tiny $a$};
\draw node at (2.7,3) {\tiny $b$};
\end{tikzpicture}
\end{gathered}
\quad=\quad
\begin{gathered}
\begin{tikzpicture}[scale=0.5]
\draw [line] (1.5,0)--(1.5,3);
%\draw [line, red] (1.5,2) -- (0,0);
\draw [line, red] (1.5,1) -- (3,3);
%%%
\draw [line,red] (1.5,2) -- (1.05,2.25);
\draw [red, line width=.8] (1.05, 2.25) arc (60:145:.25);
\draw [line, red] (.72,2.2) -- (0,0);
\draw node at (0.3, 0) {\tiny $a$};
\draw node at (2.7,3) {\tiny $b$};
\draw node at (1, 1.9) {\tiny $-a$};
\end{tikzpicture}
\end{gathered}
\quad=\chi_t(-a,b)\quad
\begin{gathered}
\begin{tikzpicture}[scale=0.5]
\draw [line] (1.5,0)--(1.5,3);
%\draw [line, red] (1.5,2) -- (0,0);
\draw [line, red] (1.5,2) -- (3,3);
%%%
\draw [line,red] (1.5,1) -- (1.05,1.25);
\draw [red, line width=.8] (1.05, 1.25) arc (60:145:.25);
\draw [line, red] (.725,1.2) -- (0,0);
\draw node at (0.3, 0) {\tiny $a$};
\draw node at (2.7,3) {\tiny $b$};
\draw node at (1, .9) {\tiny $-a$};
\end{tikzpicture}
\end{gathered}
\quad=\chi_t(-a,b)\quad
\begin{gathered}
\begin{tikzpicture}[scale=0.5]
\draw [line] (1.5,0)--(1.5,3);
\draw [line, red] (1.5,1) -- (0,0);
\draw [line, red] (1.5,2) -- (3,3);
\draw node at (0.3, .5) {\tiny $a$};
\draw node at (2.7,2.5) {\tiny $b$};
\end{tikzpicture}
\end{gathered}
\end{align}
in the third equality. Therefore we have, from \eqref{eq:recursion}, 
\begin{align}
    \beta_a(\mathcal N)=\beta_1(\mathcal N)^a\,\omega_{N,t}^{-\sum_{j=1}^{a-1} j}=\beta_1(\mathcal N)^a\,\omega_{N,t}^{-\frac{a(a-1)}{2}}\,,
\end{align}
i.e. the braiding $\beta_1(\mathcal N)$ determines all other braiding data between the $\mathcal N$- and $\mathbb Z_N$-anyons.

To find consistent braiding data $\beta_1(\mathcal N)$, the naturality conditions have to be imposed:
\begin{align}
\begin{gathered}
\begin{tikzpicture}[scale=0.5]
\draw [line, red] (2,2) -- (4,0);
\draw [line,red] (2,0) -- (2.8,.8);
\draw [line,red] (3.2,1.2) -- (4,2);
\draw [line] (2,2)--(0,2.5);
\draw [line] (2,2)--(1.5,4);
\draw node at (2, -.3) {\tiny $1$};
\draw node at (4, -.3) {\tiny $a$};
\draw node at (.2, 2) {\tiny $\mathcal N$};
\draw node at (1.2, 3.5) {\tiny $\mathcal N$};
\end{tikzpicture}
\end{gathered}
\quad=\quad
\begin{gathered}
\begin{tikzpicture}[scale=0.5]
\draw [line, red] (3,1) -- (4,0);
\draw [line,red] (0,0) -- (1.5,1.5);
\draw [line,red] (2.5,2.5) -- (4,4);
\draw [line, red] (1.8,1.8)-- (2.2, 2.2);
\draw [line] (3,1)--(0,2.5);
\draw [line] (3,1)--(1.5,4);
\draw node at (.3, -.3) {\tiny $1$};
\draw node at (4, -.3) {\tiny $a$};
\draw node at (.2, 2) {\tiny $\mathcal N$};
\draw node at (1.2, 3.5) {\tiny $\mathcal N$};
\end{tikzpicture}
\end{gathered}
\end{align}
The LHS of the above graph gives
\begin{align}
\begin{gathered}
\begin{tikzpicture}[scale=0.5]
\draw [line, red] (2,2) -- (4,0);
\draw [line,red] (2,0) -- (2.8,.8);
\draw [line,red] (3.2,1.2) -- (4,2);
\draw [line] (2,2)--(0,2.5);
\draw [line] (2,2)--(1.5,4);
\draw node at (2, -.3) {\tiny $1$};
\draw node at (4, -.3) {\tiny $a$};
\draw node at (.2, 2) {\tiny $\mathcal N$};
\draw node at (1.2, 3.5) {\tiny $\mathcal N$};
\end{tikzpicture}
\end{gathered}
\quad =\beta_1(a)\quad
\begin{gathered}
\begin{tikzpicture}[scale=0.5]
\draw [line, red] (2,2) -- (4,0);
\draw [line,red] (2.4,0) -- (3.2,.8);
\draw [line,red] (3.6,2) -- (2.8,1.2);
\draw [line] (2,2)--(0,2.5);
\draw [line] (2,2)--(1.5,4);
\draw node at (2, -.3) {\tiny $1$};
\draw node at (4, -.3) {\tiny $a$};
\draw node at (3.9,2) {\tiny $1$};
\draw node at (.2, 2) {\tiny $\mathcal N$};
\draw node at (1.2, 3.5) {\tiny $\mathcal N$};
\end{tikzpicture}
\end{gathered}
\,,
\end{align}
while the RHS gives
\begin{align}
\begin{gathered}
\begin{tikzpicture}[scale=0.5]
\draw [line, red] (3,1) -- (4,0);
\draw [line,red] (0,0) -- (1.5,1.5);
\draw [line,red] (2.5,2.5) -- (4,4);
\draw [line, red] (1.8,1.8)-- (2.2, 2.2);
\draw [line] (3,1)--(0,2.5);
\draw [line] (3,1)--(1.5,4);
\draw node at (.3, -.3) {\tiny $1$};
\draw node at (4, -.3) {\tiny $a$};
\draw node at (.2, 2) {\tiny $\mathcal N$};
\draw node at (1.2, 3.5) {\tiny $\mathcal N$};
\end{tikzpicture}
\end{gathered}
&\quad=\beta_1(\mathcal N)^2\quad
\begin{gathered}
\begin{tikzpicture}[scale=0.5]
\draw [line, red] (3,1) -- (4,0);
\draw [line] (3,1)--(0,2.5);
\draw [line] (3,1)--(1.5,4);
%%%%
\draw [line,red] (0.2,0) -- (1.82,1.62);
\draw [line,red] (2,3) -- (3,4);
\draw [line, red] (1.4,1.8)-- (2.2, 2.6);
%%%%
\draw node at (.3, -.3) {\tiny $1$};
\draw node at (2.7, 4) {\tiny $1$};
\draw node at (1.6, 2.4) {\tiny $1$};
\draw node at (4, -.3) {\tiny $a$};
\draw node at (.2, 2) {\tiny $\mathcal N$};
\draw node at (1.2, 3.5) {\tiny $\mathcal N$};
\end{tikzpicture}
\end{gathered}
\quad =\beta_1(\mathcal N)^2\quad
\begin{gathered}
\begin{tikzpicture}[scale=0.5]
\draw [line, red] (3,1) -- (4,0);
\draw [line] (3,1)--(0,2.5);
\draw [line] (3,1)--(1.5,4);
%%%%
\draw [line,red] (.2,0) -- (3.5, .5);
\draw [line,red] (2,3) -- (3,4);
\draw [line, red] (1.4,1.8)-- (2.2, 2.6);
%%%%
\draw node at (.3, -.3) {\tiny $1$};
\draw node at (2.7, 4) {\tiny $1$};
\draw node at (1.6, 2.4) {\tiny $1$};
\draw node at (4, -.3) {\tiny $a$};
\draw node at (.2, 2) {\tiny $\mathcal N$};
\draw node at (1.2, 3.5) {\tiny $\mathcal N$};
\end{tikzpicture}
\end{gathered}
\notag\\
&\quad = \beta_1(\mathcal N)^2\,\chi_t(1,a+1)\quad
\begin{gathered}
\begin{tikzpicture}[scale=0.5]
\draw [line, red] (3,1) -- (4,0);
\draw [line] (3,1)--(0,2.5);
\draw [line] (3,1)--(1.5,4);
%%%%
\draw [line,red] (0.2,0) -- (3.5, .5);
\draw [line,red] (2,3) -- (3,4);
\draw [red,line width=1] (2.2,2.6) arc (145:270:.5);
%%%%
\draw node at (.3, -.3) {\tiny $1$};
\draw node at (2.7, 4) {\tiny $1$};
\draw node at (1.6, 2.4) {\tiny $1$};
\draw node at (4, -.3) {\tiny $a$};
\draw node at (.2, 2) {\tiny $\mathcal N$};
\draw node at (1.2, 3.5) {\tiny $\mathcal N$};
\end{tikzpicture}
\end{gathered}
\notag\\
%%%%
&\quad = \beta_1(\mathcal N)^2\,\chi_t(1,a+1)\quad
\begin{gathered}
\begin{tikzpicture}[scale=0.5]
\draw [line, red] (2,2) -- (4,0);
\draw [line,red] (2.4,0) -- (3.2,.8);
\draw [line,red] (3.6,2) -- (2.8,1.2);
\draw [line] (2,2)--(0,2.5);
\draw [line] (2,2)--(1.5,4);
\draw node at (2, -.3) {\tiny $1$};
\draw node at (4, -.3) {\tiny $a$};
\draw node at (3.9,2) {\tiny $1$};
\draw node at (.2, 2) {\tiny $\mathcal N$};
\draw node at (1.2, 3.5) {\tiny $\mathcal N$};
\end{tikzpicture}
\end{gathered}\,.
\end{align}
Therefore, for consistency, we have
\begin{align}
\beta_1(a)=\beta_1(\mathcal N)^2\,\chi_t(1,a+1)\quad\Longrightarrow \quad \beta_1(\mathcal N)=\pm e^{-\frac{\pi i t}{N}}\,,
\label{eq:beta_1}
\end{align}
as the allowable values, where we have used \eqref{eq:Zn_braiding}.
Using \eqref{eq:recursion}, one arrives at
\begin{align}
    \beta_a(\mathcal N)=(\pm 1)^{a}\,\omega_{N,t}^{-\frac{a^2}{2}}\,.
\end{align}
We need to further require that
\begin{align}
    \beta_N(\mathcal N)=(\pm 1)^N(-1)^{tN}\equiv\beta_0(\mathcal N)=1\,.
\label{eq:beta_N}
\end{align}
Obviously, for even $N$, \eqref{eq:beta_N} will be automatically satisfied. Therefore $\beta_1(\mathcal N)$ can be assigned $\pm e^{-\frac{\pi i t}{N}}$ as \eqref{eq:beta_1}. On the other hand, for odd $N$, $\beta_1(\mathcal N)$ has to be assigned as $(-1)^t e^{-\frac{\pi i t}{N}}$. In sum, we have
\begin{equation}
\beta\equiv\beta_1(\mathcal N)=\left\{
\begin{aligned}
(-1)^t e^{-\frac{\pi i t}{N}}\qquad {\rm for}\quad {\rm odd}\quad N\\
\pm e^{\frac{-\pi i t}{N}}\qquad {\rm for}\quad {\rm even}\quad N
\end{aligned}
\right.
\label{eq:beta}
\end{equation}

With these preparations, we can compute the braiding coefficient $B(a,\mathcal N)$ introduced in \eqref{eq:braiding_coef}. Notice that
\begin{align}
   \begin{gathered}
\begin{tikzpicture}[scale=0.5]
\draw [line] (1.5,0)--(1.5,3);
\draw [line, red] (1.5,.5) -- (2,1);
\draw [red, line width=.8] (2,1) arc (-45:60:.5);
\draw [line, red] (1.3, 2.1) -- (0,3);
%\draw [line, red] (1.5,2) -- (0,0);
\draw node at (1.9, 2.8) {\tiny $\mathcal N$};
\draw node at (0.1, 2.6) {\tiny $a$};
\draw node at (1.8, .3) {\tiny $a$};
\end{tikzpicture}
\end{gathered}
&\quad=\quad
\begin{gathered}
\begin{tikzpicture}[scale=0.5]
\draw [line] (1.5,0)--(1.5,3);
\draw [line, red] (1.5,.5) -- (2.6,1.7);
\draw [red, line width=.8] (2.6,1.7) arc (-45:130:.2);
\draw [red, line] (1.7, 1.3) -- (2.37,2.03);
\draw [line, red] (1.3, .95) -- (1,.6);
\draw [red, line width=.8] (1,.6) arc (-45:-160:.27);
\draw [line, red] (.57,.65) -- (0,2);
\draw node at (1.9, 2.8) {\tiny $\mathcal N$};
\draw node at (0.1, 2.3) {\tiny $a$};
\draw node at (1.9, .3) {\tiny $a$};
\draw node at (1.9, 2) {\tiny $-a$};
\draw node at (.85,.95) {\tiny $-a$};
\end{tikzpicture}
\end{gathered}
\quad=\beta_{-a}(\mathcal N)\quad
\begin{gathered}
\begin{tikzpicture}[scale=0.5]
\draw [line] (1.5,0)--(1.5,3);
\draw [line, red] (1.5,.5) -- (2,1);
\draw [red, line width=.8] (2,1) arc (-45:105:.5);
%\draw [red, line width=.8] (2.6,1.7) arc (-45:130:.2);
%\draw [red, line] (1.7, 1.3) -- (2.37,2.03);
\draw [line, red] (1.5, 1.1) -- (1,.6);
\draw [red, line width=.8] (1,.6) arc (-45:-160:.27);
\draw [line, red] (.57,.65) -- (0,2);
\draw node at (1.9, 2.8) {\tiny $\mathcal N$};
\draw node at (0.1, 2.3) {\tiny $a$};
\draw node at (1.9, .3) {\tiny $a$};
\draw node at (1.9, 2.1) {\tiny $-a$};
\draw node at (.85,.95) {\tiny $-a$};
\end{tikzpicture}
\end{gathered}
\notag\\
&\quad=\beta_{-a}(\mathcal N)\,\chi^{-1}_t(a,-a)\quad
\begin{gathered}
\begin{tikzpicture}[scale=0.5]
\draw [line] (1.5,0)--(1.5,3);
\draw [red, line width=.8] (1.5,.5) arc (-90:90:.5);
%\draw [line, red] (1.5, 2) -- (1,1.5);
%\draw [red, line width=.8] (1,1.5) arc (-45:-160:.27);
\draw [line, red] (1.5, 2) -- (0.2,3);
\draw node at (1.9, 2.8) {\tiny $\mathcal N$};
\draw node at (0.8, 2.8) {\tiny $a$};
\draw node at (1.9, .4) {\tiny $a$};
%\draw node at (.85,1.85) {\tiny $-a$};
\end{tikzpicture}
\end{gathered}
\quad=\beta_{a}(\mathcal N)\,\chi_t(a,a)\quad
\begin{gathered}
\begin{tikzpicture}[scale=0.5]
\draw [line] (1.5,0)--(1.5,3);
%\draw [red, line width=.8] (1.5,.5) arc (-90:90:.5);
%\draw [line, red] (1.5, 2) -- (1,1.5);
%\draw [red, line width=.8] (1,1.5) arc (-45:-160:.27);
\draw [line, red] (1.5,.5) -- (0,3);
\draw node at (1.9, 2.8) {\tiny $\mathcal N$};
\draw node at (0.5, 2.8) {\tiny $a$};
\end{tikzpicture}
\end{gathered}
\,.
\end{align}
We thus have
\begin{align}
    B(a,\mathcal N)=\beta_a(\mathcal N)\,\chi_t(a,a)\,.
\end{align}

\subsection{Parafermionic condensation}
All F-symbols are constrained by the para-pentagon equations introduced in Section~\ref{3}. By Ocneanu rigidity, only finitely many gauge-equivalence classes of solutions exist. However, the number of para-pentagon equations grows as $\mathcal O(N^6)$, so a direct brute-force solution is practical only for $N=2$, the fermionic case. Instead, we use the data of ${\rm TY}_{\mathbb Z_N}^{t,\kappa}$ to simplify the F-symbols of its condensed category pf-${\rm TY}_{\mathbb Z_N}^{t,\kappa,\beta}$. This procedure requires the half-braiding data introduced below.

We separate all F-symbols \eqref{Fsymb} into three sectors:
\begin{align}
    \mathcal S_1&=\left\{ \mathcal F^{\mathds 1,\,\mathds 1,\,\mathds 1}_{\mathds 1}\right\}\,,\notag\\
    \mathcal S_2&=\left\{ \mathcal F^{\mathds 1,\,\mathds 1,\,\mathcal N}_{\mathcal N}\,,\,
    \mathcal F^{\mathds 1,\,\mathcal N,\,\mathds 1}_{\mathcal N}\,,\,
    \mathcal F^{\mathds 1,\,\mathcal N,\,\mathcal N}_{\mathds 1}\,,\,
    \mathcal F^{\mathcal N,\,\mathds 1,\,\mathds 1}_{\mathcal N}\,,\,
    \mathcal F^{\mathcal N,\,\mathds 1,\,\mathcal N}_{\mathds 1}\,,\,
    \mathcal F^{\mathcal N,\,\mathcal N,\,\mathds 1}_{\mathds 1}\right\}\,,\notag\\
    \mathcal S_3&=\left\{ \mathcal F^{\mathcal N,\,\mathcal N,\,\mathcal N}_{\mathcal N}\right\}\,.
\end{align}
The F-symbol $\mathcal F^{\mathds 1,\,\mathds 1,\,\mathds 1}_{\mathds 1}$ in $\mathcal S_1$ is fixed trivially by the para-pentagon formed by five trivial lines:
\begin{align}
    \mathcal F^{\mathds 1,\,\mathds 1,\,\mathds 1}_{\mathds 1}=1\,.
\end{align}
Now we will establish some linear relations among entries of each F-symbol in $\mathcal S_2$. Notice that the 1d parafermionic defects can ``move" along the q-type $\mathcal N$-line. We can thus merge two different colors of defects by lifting them to the parent \TY, performing F-moves, and then condensing back,
\begin{align}
\begin{gathered}

\end{gathered}
\,,
\label{eq:1-way_lift}
\end{align}
where we have used the F-move \eqref{eq:TY_Fmove} in \TY.

For a trivalent junction in \pfTY, we can also use \eqref{eq:1-way_lift} to define its lifting in \TY. Notice that there are three trivalent junctions involving $\mathcal N$-line from \eqref{eq:pfZn_fusion},
\begin{align}
\begin{gathered}
%
\end{gathered}
\label{eq:junctions}
\end{align}
There are ambiguities when one tries to lift the above junctions to \TY. For example we can lift the first junction in \eqref{eq:junctions} in two different ways as
\begin{align}
\begin{gathered}
%
\end{gathered}
\,,
\label{eq:lift_1}
\end{align}
where we have chosen $\mathds 1=\left[0\right]$ as the representative of $\mathbb Z_N$. However by the F-move in \eqref{eq:TY_Fmove}, we have
\begin{align}
\begin{gathered}
%
\end{gathered}
\,.
\end{align}
Therefore there is actually no ambiguity for the lifting of \eqref{eq:lift_1}. We now look at the second junction in \eqref{eq:junctions}, and have
\begin{align}
\begin{gathered}
%
\end{gathered}
\,.
\label{eq:lift_2}
\end{align}
The two ways of lifting can be related in the following F-move and braiding,
\begin{align}
\begin{gathered}
%
\end{gathered}
\,,
\end{align}
where the coefficient $B(a,0)$ denotes the braiding of anyon ``$a$" and trivial line, which is always unity. Therefore there is no ambiguity either in \eqref{eq:lift_2}. Finally let's come to the third junction in \eqref{eq:junctions}, and its two ways of lifting,
\begin{align}
\begin{gathered}
%
\end{gathered}
\,.
\label{eq:lift_3}
\end{align}
Similar to the previous calculation, we can establish the relation between the two liftings,
\begin{align}
\begin{gathered}
%
\end{gathered}
\,,
\label{eq:braiding_coef}
\end{align}
where $B(a,\mathcal N)$ is the braiding coefficient between anyon ``$a$" and $\mathcal N$-line.

Because of the ambiguity in \eqref{eq:lift_3}, let us assign the lifting of the trivalent junctions \eqref{eq:junctions} in the following way:
\begin{align}
\begin{cases}
&\quad   
\begin{gathered}
%
\end{gathered}\,.
\end{cases}
\label{eq:lift_cond}
\end{align}
Here we briefly remark that one can also use the second equation in \eqref{eq:lift_3} to define the lifting and condensation. The two ways are actually equivalent up to a gauge transformation by redefining the trivalent junction.

Now, using \eqref{eq:lift_cond}, we can proceed to establish various linear relations for the components of F-symbols in $\mathcal S_2$.\\

\noindent For $\mathcal F^{\mathds 1,\,\mathds 1,\,\mathcal N}_{\mathcal N}$,
\begin{align}
\begin{gathered}
%
\end{gathered}
\notag\\\notag\\
\Longrightarrow\quad\quad
&\mathcal F^{\mathcal N ,\,\mathcal N,\,\mathds 1 }_{\mathds 1}\left(a+b,0,a,b\right)=B(a,\mathcal N)\,\chi_t(a,b)\,\mathcal F^{\mathcal N,\,\mathds 1,\,\mathcal N}_{\mathds 1}\left(a+b,0,0,a+b\right)
\\\notag
\label{eq:F2211}
\end{align}

\subsection{Gauge fixings and solutions to \pfTY}
The condensation equations \eqref{eq:F1122}--\eqref{eq:F2211} substantially simplify the para-pentagon equations. The remaining gauge freedom comes from redefinitions of the trivalent junctions and must be fixed to solve the equations. For \pfTY, there are $3N-1$ gauge parameters, counted as in \cite{Chang:2022hud}. Of these, $3(N-1)$ are implicitly fixed by \eqref{eq:lift_cond}, because different choices of condensation correspond to redefinitions of the three types of junctions in each of the $N-1$ nonzero colors. Thus two gauge parameters remain, since $3N-1-3(N-1)=2$. For convenience, we choose
\begin{align}
    \mathcal F^{\mathds 1,\,\mathds 1,\, \mathcal N}_{\mathcal N}(0,0,0,0)
    =\mathcal F^{\mathcal N,\,\mathds 1,\,\mathds 1}_{\mathcal N}(0,0,0,0)=\frac{1}{\sqrt{N}}\,.
\end{align}
Solving pentagons with two external $\mathcal N$-lines, one can find all F-moves in $\mathcal S_2$ as,
\begin{align}
\begin{cases}
    &\mathcal F^{\mathds 1,\,\mathds 1,\,\mathcal N}_{\mathcal N}(0,a+b,a,b)=\frac{1}{\sqrt{N}}\\\\
    &\mathcal F^{\mathds 1,\,\mathcal N,\,\mathds 1}_{\mathcal N}(a,b,c,d)=\frac{1}{N}\,,\quad {\rm for}\quad a+b=c+d\,\mod{N}\\\\
    &\mathcal F^{\mathds 1,\, \mathcal N,\,\mathcal N}_{\mathds 1}(a,b,a+b,0)=\frac{1}{\sqrt{N}}\\\\
    &\mathcal F^{\mathcal N,\,\mathds 1,\,\mathds 1}_{\mathcal N}(a,b,0,a+b)=\frac{1}{\sqrt{N}}\\\\
    &\mathcal F^{\mathcal N,\,\mathds 1,\mathcal N}_{\mathds 1}(a,b,c,d)=\frac{1}{N}\,\beta_c(\mathcal N)\,\chi_t(c,a+b)\,,\quad {\rm for}\quad a+b=c+d\,\mod{N}\\\\
    &\mathcal F^{\mathcal N,\,\mathcal N,\mathds 1}_{\mathds 1}(a,b,0,a+b)=\frac{1}{\sqrt{N}}\,\beta_a(\mathcal N)\,\chi_t(a,a+b)\,.
\end{cases}
\end{align}
Finally, solving pentagons with four external $\mathcal N$-lines, we find the F-move $\mathcal F^{\mathcal N,\,\mathcal N,\,\mathcal N}_{\mathcal N}$ in $\mathcal S_4$,
\begin{align}
    \mathcal F^{\mathcal N,\,\mathcal N,\,\mathcal N}_{\mathcal N}(a,b,c,d)=\frac{\kappa}{\sqrt{N}}\,\beta_c(\mathcal N)\,\chi_t(a,c)\,,\quad {\rm for}\quad a+b=c+d\,\mod{N}
\end{align}
Therefore, we have shown that \pfTY, the parafermionic condensation of \TY category, is uniquely determined by the data $t$, $\kappa$ and half-braiding $\beta$. For example, in the case of $N=2$, i.e. the fermionic condensation of ${\rm TY}(\mathbb Z_2)$, we have $t=1$, $\kappa=\pm 1$ and $\beta=\pm i$ as in \eqref{eq:beta}. Therefore, overall there are 4 gauge independent solutions that have been found in \cite{Chang:2022hud}.
\\

\paragraph{Acknowledgments}

We would like to thank Yongchao L\"u who mention this interesting topic to us, and also Chi-Ming Chang, Zhuo Chen, Wei Cui, Zhihao Duan, Zheng-Cheng Gu, Yongchao L\"u, Jia Qiang, and Fengjun Xu for their helpful discussions and related projects. The work of J.C. is supported by the Fundamental Research Funds for the Central Universities (No.20720230010) of China. B.H. is supported by the Young Thousand Talents grant of China as well as by the NSFC grant 12250610187. Q.-R.W. is supported by the National Natural Science Foundation of China (Grant No. 12274250).
\clearpage
\appendix

\section{Bosonic anyon condensation}
\label{app:bosonic-condensation}

Bosonic anyon condensation was first discussed in the mathematical community~\cite{kirillov2002q,ostrik2003module} and was later independently discovered as a phenomenon that can occur in 3d topological orders~\cite{bais2009condensate,eliens2014diagrammatics,kong2014anyon,burnell2018anyon}. In 3d topological orders, spatial point-like topological excitations, called anyons, can fuse and braid with each other. The process of fusing and braiding anyons can be mathematically described using a unitary modular tensor category (UMTC) framework~\cite{Kitaev:2005hzj}. Anyon condensation is then a procedure in which a new UMTC is obtained from an old one by condensing some bosonic anyons. It is also recognized that the condensation of bosonic anyons in 3d topological order is intimately connected to the conformal extension of 2d CFTs~\cite{bais2009condensate}.

Apart from considering anyon condensation as a phase transition from one topological order to another, one can also view it as a gapped domain wall separating two distinct topological orders~\cite{kitaev_models_2012,fuchs_bicategories_2013}. As a special case, the gapped boundary of a topological order can be seen as the domain wall between the topological order and the vacuum state. In this way, we can understand both the domain wall and the boundary on equal terms.

In fact, we can differentiate between two types of bosonic anyon condensations, which will be discussed separately in the following. The first type is more commonly discussed in the condensed matter physics literature. It occurs in a UMTC $\B$, where both the fusion and braiding of the anyons are defined. The second type, which is more relevant for the TDLs studied in this paper, occurs in the unitary fusion category (UFC) $\C$ where there are no braiding operations in general. In the following, we will use the notation $\B$ for a UMTC to indicate that it is braided. On the contrary, a fusion category without braiding will be denoted as $\C$.

The constructions reviewed here will be used in Section~\ref{3} as the bosonic template for fermionic and parafermionic condensation. The key ingredients are the module category $\C_A$, the bimodule category $_A\C_A$, and the adjunction formula for Hom spaces. In the fermionic and parafermionic cases, the same ideas are generalized by replacing ordinary Hom spaces with $\Z_2$- and $\Z_N$-graded Hom spaces, respectively.

\subsection{Bosonic anyon condensation in UMTC $\B$}

In a quantum field theory, boson condensation or Bose-Einstein condensation occurs when a boson $b$ is condensed by changing its effective mass to negative. In the condensed phase, the boson $b$ can be created or annihilated arbitrarily in the new vacuum. For this new vacuum to be consistent, the self-braiding of $b$ must be trivial, meaning that $b$ should be a boson. Otherwise, creating, braiding, and annihilating a pair of $b$ will result in a nontrivial phase factor that is ambiguous for the vacuum.

The concept of boson condensation is generalized to 3d topological orders or anyon models described by a unitary modular tensor category (UMTC) $\B$~\cite{kirillov2002q,bais2009condensate,eliens2014diagrammatics,kong2014anyon,burnell2018anyon}. Given a subset $A$ of anyons in $\B$, we can perform anyon condensation by identifying $A$ as the new vacuum in the condensed phase. To ensure the consistency of this identification, certain technical conditions must be satisfied such that $A$ forms a connected commutative separable algebra in $\B$~\cite{kirillov2002q,kong2014anyon}. For example, if we intend to condense a single Abelian boson $b$ with $\Z_2$ fusion rule, the algebra $A$ could be represented as $1\oplus b$.

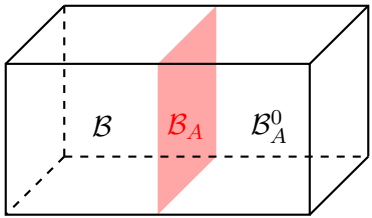
\begin{figure}[h]
\centering
\begin{tikzpicture}
% Highlight y-z plane
\fill[red!50, opacity=0.7] (0,0,0) -- (0,2,0) -- (0,2,2) -- (0,0,2) -- cycle;
% Cube
\draw[black, thick] (-2,2,0) -- (2,2,0) -- (2,0,0);
\draw[black, thick, dashed] (-2,0,0) -- (2,0,0);
\draw[black, thick, dashed] (-2,0,0) -- (-2,2,0);
\draw[black, thick] (-2,0,2) -- (-2,2,2) -- (2,2,2) -- (2,0,2) -- cycle;
\draw[black, thick, dashed] (-2,0,0) -- (-2,0,2);
\draw[black, thick] (-2,2,0) -- (-2,2,2);
\draw[black, thick] (2,0,0) -- (2,0,2);
\draw[black, thick] (2,2,0) -- (2,2,2);
% label
% \node at (-1.4,1.6,0) {$\C$};
\node at (-1.5,.4,0) {$\B$};
% \node at (.8,1.6,0) {$_A\C_A$};
\node at (.7,.4,0) {$\B_A^0$};
\node[red] at (-.4,.4,0) {$\B_A$};
% Axis labels
% \draw[->] (-3,0,0) -- (3,0,0) node[below right] {xx};
% \draw[->] (0,-1,0) -- (0,3,0) node[above] {yy};
% \draw[->] (0,0,-1) -- (0,0,3) node[right] {zz};
\node at (-.4,-1.3,0){(a)};
\end{tikzpicture}
\caption{Bosonic anyon condensation in 3d topological order of UMTC $\B$. $A$ is a connected commutative separable algebra in $\B$. The red domain wall that separates the topological orders before and after the condensation is described by the fusion category $\B_A$ of the right $A$-modules. Subsequently, the condensed phase is characterized by the UMTC $\B_A^0$ which consists of local (or dyslectic) right $A$-modules in $\B$.}
\label{fig:cond_B}
\end{figure}

When we perform an anyon condensation of $A$ by identifying $A$ as the new vacuum, we obtain an intermediate phase described by a fusion category $\B_A$ (without braidings in general). The fusion category $\B_A$ consists of $A$-modules and module morphisms. The total quantum dimension of $\B_A$ is related to that of $\B$ by
\begin{align}
\dim(\B_A):= \sum_{a\in \B_A}d_a^2 = \frac{\dim \B}{\dim A}.
\end{align}
Physically, this intermediate phase can be considered as a 2d \emph{gapped domain wall} between the original and final 3d phases of the condensation procedure. The objects in $\B_A$ correspond to the excitations localized at the domain wall.

The condensed phase resulting from anyon condensation by $A$ gives rise to a 3d topological order that can be described by a UMTC $\mathcal{B}_A^0$. Unlike the intermediate fusion category phase $\mathcal{B}_A$, $\mathcal{B}_A^0$ possesses braidings as part of its structure. To obtain $\mathcal{B}_A^0$, we select the local or dyslectic $A$-modules~\cite{pareigis1995braiding} from $\mathcal{B}_A$. The key principle is that anyons in $\mathcal{B}_A^0$ should have trivial full braidings with $A$ since $A$ is new vacuum. Any anyons that exhibit nontrivial full braidings with $A$ become confined in the condensed phase due to the quantum interference of Aharonov-Bohm effects. It can be shown that the total quantum dimension of $\mathcal{B}_A^0$ is given by 
\begin{align}
\dim(\B_A^0) = \frac{\dim (\B_A)}{\dim A}= \frac{\dim \B}{(\dim A)^2}.
\end{align}

\subsection{Bosonic anyon condensation in UFC $\C$}

Anyon condensation can be defined not only in braided tensor category $\B$ but also in fusion category $\C$. Strictly speaking, an algebra in a fusion category $\C$ is not an anyon, since $\C$ does not have braidings in general. We use the term ``anyon condensation in $\C$" in two related senses: as generalized gauging on a 2d boundary, and, when $A$ admits a commutative lift to $Z(\C)$, as the boundary avatar of ordinary anyon condensation in the bulk UMTC $Z(\C)$. In either case, we obtain another fusion category from a given fusion category $\C$ by condensing an algebra $A$ with some conditions.

\begin{figure}[h]
    \centering
\begin{tikzpicture}
% Highlight y-z plane
\fill[red!50, opacity=0.2] (0,0,0) -- (0,2,0) -- (0,2,2) -- (0,0,2) -- cycle;
% Cube
\draw[red, thick] (0,2,0) -- (0,2,2);
\draw[black, thick] (-2,2,0) -- (2,2,0) -- (2,0,0);
\draw[black, thick, dashed] (-2,0,0) -- (2,0,0);
\draw[black, thick, dashed] (-2,0,0) -- (-2,2,0);
\draw[black, thick] (-2,0,2) -- (-2,2,2) -- (2,2,2) -- (2,0,2) -- cycle;
\draw[black, thick, dashed] (-2,0,0) -- (-2,0,2);
\draw[black, thick] (-2,2,0) -- (-2,2,2);
\draw[black, thick] (2,0,0) -- (2,0,2);
\draw[black, thick] (2,2,0) -- (2,2,2);
% label
\node at (-1.4,1.6,0) {$\C$};
\node at (-1.5,.4,0) {$Z(\C)$};
\node at (.8,1.6,0) {$_A\C_A$};
\node at (.7,.4,0) {$Z(_A\C_A)$};
\node[red] at (-.35,1.6,0) {$\C_A$};
\node[red] at (-.4,.4,0) {$\simeq$};
% Axis labels
% \draw[->] (-3,0,0) -- (3,0,0) node[below right] {xx};
% \draw[->] (0,-1,0) -- (0,3,0) node[above] {yy};
% \draw[->] (0,0,-1) -- (0,0,3) node[right] {zz};
\node at (-.4,-1.3,0){(a)};
\end{tikzpicture}
\qquad
\begin{tikzpicture}
% Highlight y-z plane
\fill[red!50, opacity=0.7] (0,0,0) -- (0,2,0) -- (0,2,2) -- (0,0,2) -- cycle;
% Cube
\draw[red, thick] (0,2,0) -- (0,2,2);
\draw[black, thick] (-2,2,0) -- (2,2,0) -- (2,0,0);
\draw[black, thick, dashed] (-2,0,0) -- (2,0,0);
\draw[black, thick, dashed] (-2,0,0) -- (-2,2,0);
\draw[black, thick] (-2,0,2) -- (-2,2,2) -- (2,2,2) -- (2,0,2) -- cycle;
\draw[black, thick, dashed] (-2,0,0) -- (-2,0,2);
\draw[black, thick] (-2,2,0) -- (-2,2,2);
\draw[black, thick] (2,0,0) -- (2,0,2);
\draw[black, thick] (2,2,0) -- (2,2,2);
% label
\node at (-1.4,1.6,0) {$\C$};
\node at (-1.5,.4,0) {$Z(\C)$};
\node at (.8,1.6,0) {$\C_A$};
\node at (1.1,.4,0) {$Z(\C_A)=[Z(\C)]_A^0$};
\node[red] at (-.35,1.6,0) {$\C_A$};
% Axis labels
% \draw[->] (-3,0,0) -- (3,0,0) node[below right] {xx};
% \draw[->] (0,-1,0) -- (0,3,0) node[above] {yy};
% \draw[->] (0,0,-1) -- (0,0,3) node[right] {zz};
\node at (-.4,-1.3,0){(b)};
\end{tikzpicture}
\caption{Two types of bosonic anyon condensation in UFC $\C$ corresponding to a 2d boundary or UMTC $Z(\C)$ associated with a 3d bulk topological order. The 2d boundary can be understood from two different perspectives: as a 2d spatial manifold used to define the Hamiltonian, or as a 1+1d boundary through a Wick rotation.
(a) $A$ is an algebra in fusion category $\C$. The domain wall $\C_A$ has no monoidal/tensor structure in general. The original fusion category $\C$ and the bimodule category $_A\C_A$ are Morita equivalent, i.e., $Z(\C)\simeq Z(_A\C_A)$. So the red domain wall between them is invertible or an equivalence of bulk UMTC. A prototypical example is $\C=\mathrm{Vec}_G$, $A=\mathbb C[G]$, $\C_A=\mathrm{Vec}$ and $_A\C_A=\mathrm{Rep}(G)$, where the two sides are related by gauging/condensing $A$. 
(b) $A$ is a central commutative algebra, i.e., a commutative algebra in $Z(\C)$. The category of right-$A$ modules already has a monoidal/tensor structure. The bulk of $\C$ and $\C_A$ are related by $Z(\C_A)=[Z(\C)]_A^0$~\cite{schauenburg_monoidal_2001}. So the 3d red domain wall corresponds to anyon condensation of topological orders from UMTC $Z(\C)$ to $[Z(\C)]_A^0$.
}
\label{fig:relation}
\end{figure}
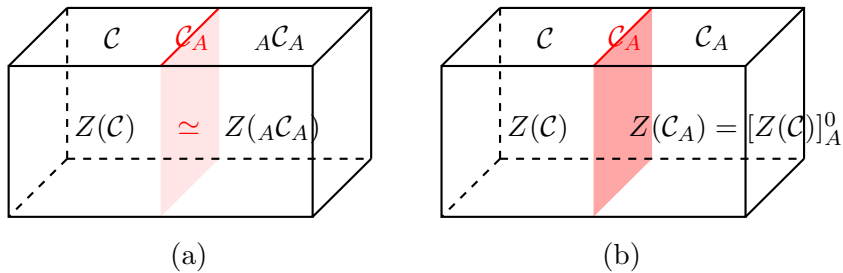

In 2d theory, anyon condensation of an algebra $A$ in $\C$ has a physical interpretation as a generalized notion of gauging~\cite{Bhardwaj:2017xup}. Suppose we have a 2d theory with fusion category symmetry $\C$, which means that the TDLs of the theory are labeled by objects in $\C$. Gauging an algebra $A$ in $\C$ involves labeling all lines of a fine enough mesh in 2d spacetime with $A$. In the subsequent discussion, it becomes clear that the resulting theory exhibits an emergent fusion category symmetry $_A\C_A$, depicted on the upper surface of Figure~\ref{fig:relation}(a).

In the context of 3D bulk theory, the physical interpretation of anyon condensation within a fusion category becomes most clear when considering the topological order of Levin-Wen string-net models~\cite{LW2005}. The string-net model is essentially a Hamiltonian version on a 2D spatial manifold that corresponds to the Turaev-Viro state sum construction used for calculating the 3D partition function~\cite{TV1992}. The input data for the Hamiltonian is a fusion category $\C$, and the output topological order is a 3d TQFT with UMTC $Z(\C)$ of Reshetikhin-Turaev type~\cite{reshetikhin_invariants_1991}. The process of condensation enables the creation of a gapped domain wall that separates two distinct Levin-Wen models, one before the condensation and one after. Depending on the details of the algebra $A$, the input category for the Hamiltonian of the condensed Levin-Wen model can be either $_A\C_A$ or $\C_A$ [see Figures~\ref{fig:relation}(a) and \ref{fig:relation}(b)]. In the following, we will discuss these two types of condensation in a fusion category $\C$ separately.\\

{\bf (a) $A$ is an algebra of $\C$ [see Figure~\ref{fig:relation}(a)]}

Let's begin by assuming that $A$ is an algebra within the fusion category $\C$, which characterizes the TDLs of a 2D theory, whether it's a gapped system or a gapless one like a CFT. Physically, condensing $A$ means gauging or orbifolding $A$ in the original 2d theory~\cite{Bhardwaj:2017xup}. To be more precise, gauging or condensing $A$ means summing over all configurations of TDLs labeled by $A$ on the 2d spacetime. Therefore, the final 2D theory is defined on a 2D spacetime with a fine enough mesh, where all lines are labeled by $A$.

We can inquire about the fusion category symmetry exhibited by the condensed theory. The TDLs of the condensed theory are lines that can absorb or emit $A$ lines from both the left and right sides. Mathematically, they form the fusion category $_A\C_A$ of $A$-bimodules. It is fusion because we can define the tensor structure $\otimes_A$ over $A$. On the domain wall separating the two theories, the $A$ lines can only merge from the right-hand side. Therefore, the 1d domain wall is labeled by objects in the category (not necessarily a fusion category) $\C_A$ of right $A$-modules. In fact, $\C_A$ is a ($\C$,$_A\C_A$)-bimodule category since TDLs in $\C$ and $_A\C_A$ can fuse to it from the left and right-hand sides, respectively.

Let us delve into more details about the $A$-module category $\C_A$~\cite{etingof_tensor_2015}. For any right $A$-module $M\in\C_A$, there exists an object $X\in\C$ and a surjection $X\otimes A\rightarrow M$. Consequently, any irreducible right $A$-module in $\C_A$ is a quotient of a module of the form $X\otimes A$. The object $X\otimes A$ possesses a natural right $A$-module structure given by the composition of the associator in $\C$ and the multiplication of $A$: $(X\otimes A)\otimes A\rightarrow X\otimes (A\otimes A)\rightarrow X\otimes A$. Therefore, we have a $\C$-module functor $-\otimes A: \C\rightarrow\C_A, X\mapsto X\otimes A$. 
There is another $\C$-module functor $\mathrm{Forget}: \C_A\rightarrow\C$ that forgets the $A$-module structure of an object. In fact, these two functors are adjoint to each other, meaning that the Hom spaces are related as follows:
\begin{align}\label{adjoint}
\Hom_{\C_A}(X\otimes A, M) \simeq \Hom_\C(X,\mathrm{Forget}(M)).
\end{align}
Sometimes, we abuse the notation by letting $M$ denote $\mathrm{Forget}(M)\in \C$. It is a consequence of the special case that $\Hom_{\C_A}(A, M) = \Hom_\C(1,M)$ for any $M$ in $\C_A$.

The adjoint relation of the two functors helps us understand the relations of objects in $\C_A$ in terms of those in $\C$. Let's take the example of $A=1\oplus b$ to illustrate this. We have:
\begin{align}\nonumber
\Hom_{\C_A}(X\otimes A, Y\otimes A) 
&= \Hom_\C(X,Y\otimes A)\\
&= \Hom_\C(X,Y)\oplus\Hom_\C(X,Y\otimes b).
\end{align}
We see that the Hom space of $\C_A$ after the $A=1\oplus b$ condensation is a direct sum of two Hom spaces from the original category $\C$. The physical interpretation is that we do not distinguish between $Y$ and $Y\otimes b$ after the condensation, as $b$ is considered to be in the new vacuum. If we set $X=Y$, we can also use it to calculate
\begin{align}
\End_{\C_A}(X\otimes A) := \Hom_{\C_A}(X\otimes A,X\otimes A) = \Hom_{\C}(X,X\otimes A).
\end{align}

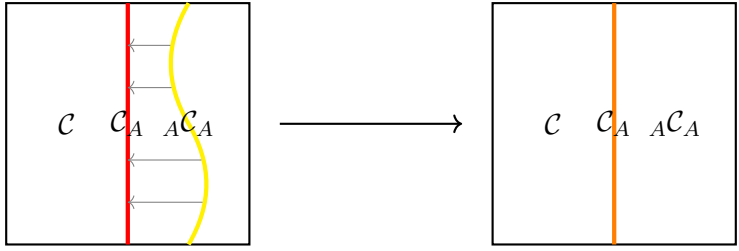
\begin{figure}[h]
\centering
\begin{tikzpicture}[scale=.8]
    % Draw the first square
    \draw[thick] (0,0) -- (4,0) -- (4,4) -- (0,4) -- cycle;
    % Draw the red vertical line in the first square
    \draw[red, ultra thick] (2,0) -- (2,4);
    \draw[yellow, ultra thick] (3,4) to [bend right=30] (3,2) to [bend left=30] (3,0);
    % Draw several small arrows originating from the yellow line
    \draw[->, gray] (2.7,3.3) -- (2,3.3);
    \draw[->, gray] (2.72,2.6) -- (2,2.6);
    \draw[->, gray] (3.21,1.4) -- (2,1.4);
    \draw[->, gray] (3.23,.7) -- (2,.7);
    %node
    \node[] at (1,2) {$\C$};
    \node[] at (2,2) {$\C_A$};
    \node[] at (3,2) {$_A\C_A$};
    % Draw the arrow
    \draw[->, thick] (4.5,2) -- (7.5,2);
    % Draw the second square
    \draw[thick] (8,0) -- (12,0) -- (12,4) -- (8,4) -- cycle;
    % Draw the red vertical line in the second square
    \draw[orange, ultra thick] (10,0) -- (10,4);
    \node[] at (9,2) {$\C$};
    \node[] at (10,2) {$\C_A$};
    \node[] at (11,2) {$_A\C_A$};
\end{tikzpicture}
\caption{
Physical interpretation of the fusion category equivalence $_A\C_A \simeq \mathrm{Fun}_\C(\C_A,\C_A)$. The fusion category $\C$ and $_A\C_A$ describe the TDLs of the left original theory and the right $A$-condensed theory in 2d, respectively. They are separated by a (red) domain wall in the middle. The domain wall is labeled by objects in the module category $\C_A$. When moving a (yellow) TDL $M\in {}_A\C_A$ from the right condensed theory to the middle (red) domain wall labeled by object in $\C_A$, the domain wall is changed to another (orange) one in $\C_A$. This process of fusing $M\in {}_A\C_A$ provides a functor from $\C_A$ to itself, establishing an equivalence between the TDLs in $_A\C_A$ and the boundary condition changing processes in $\mathrm{Fun}_\C(\C_A,\C_A)$ from $\C_A$ to itself.
}
\label{fig:ACA}
\end{figure}

As previously mentioned, the TDLs of the condensed theory are described by the fusion category symmetry $_A\C_A$. In terms of right module category, this bimodule category can also be understood as the right $A^\mathrm{op}\otimes A$-module, where $A^\mathrm{op}$ is $A$ with the opposite tensor product. Another way to understand this bimodule category is through module categories, which is represented by the monoidal equivalence~\cite{etingof_tensor_2015}
\begin{align}
_A\C_A \simeq \mathrm{Fun}_\C(\C_A,\C_A).
\end{align}
Here, $\mathrm{Fun}_\C(\C_A,\C_A)$ is the category of $\C$-module functors between $\C_A$ and $\C_A$ with certain technical conditions. The tensor structure of this category comes from the composition of functors. In this case, the equivalence between the two fusion categories is established by mapping an $A$-bimodule $M$ to $-\otimes_A M$, forming a functor from $\C_A$ to $\C_A$. This mapping has a physical interpretation as moving a TDL in the condensed theory $_A\C_A$ to the domain wall labeled by $\C_A$, resulting in another type of domain wall. The equivalence $_A\C_A \simeq \mathrm{Fun}_\C(\C_A,\C_A)$ indicates that all domain wall-changing processes can be obtained in this manner (see Figure~\ref{fig:ACA}).

Another way to relate the $A$-bimodule category and the left/right $A$-module category is through the equivalence~\cite{etingof_fusion_2010}
\begin{align}
_A\C_A \simeq (_A\C)\boxtimes_\C (\C_A).
\end{align}
Here, the right-hand side involves the relative Deligne tensor product of two module categories over $\C$. The physical meaning of this relation can be understood as follows. Consider the original 2d theory with the geometry of a stripe and TDLs labeled by $\C$. Let us condense $A$ to obtain a new theory with TDLs in $_A\C_A$ from both the left and right sides. As a result, there are two domain walls, $_A\C$ and $\C_A$, separating three 2d theories: $_A\C_A$, $\C$, and $_A\C_A$. By shrinking the $\C$ stripe or fusing the two domain walls, we obtain a 1d topological line that lives in the 2d theory $_A\C_A$. This line should also be labeled by objects in $_A\C_A$. The fusion process is mathematically represented by the relative Deligne tensor product $(_A\C)\boxtimes_\C (\C_A)$. Consequently, we find the equivalence of the two fusion categories $(_A\C)\boxtimes_\C (\C_A)$ and $_A\C_A$.

The standard example of such anyon condensation in a fusion category is the case of $\C=\mathrm{Vec}_G$, where $G$ is a finite group~\cite{Bhardwaj:2017xup}. The algebra to be condensed is the group ring $A=\mathbb C[G]$. The process of condensing the algebra $A$ can be understood as physically gauging the symmetry group $G$. This interpretation arises because summing over all possible flat connections of $G$ is equivalent to introducing additional $A$ lines into the space. In this context, the resulting gapped domain wall $\C_A$ is simply $\C_A=\mathrm{Vec}$, indicating that there is only one nontrivial simple excitation on the domain wall. The gauged phase corresponds to $_A\C_A=\mathrm{Rep}(G)$, which represents the category of $G$ representations. It's worth noting that it is widely recognized that $Z(\mathrm{Vec}_G) \simeq Z(\mathrm{Rep}(G))$. This equivalence establishes that the two fusion categories $\mathrm{Vec}_G$ and $\mathrm{Rep}(G)$ are Morita equivalent.\\

%%%%%%%%%%%%
{\bf (b) $A$ is a central commutative algebra of $\C$ [see Figure~\ref{fig:relation}(b)]}

Here by a central commutative algebra of $\C$, we mean an algebra $A$ in $\C$ whose underlying object can be lifted to the Drinfeld center $Z(\C)$ as a commutative algebra. Equivalently, $A$ is equipped with a half-braiding with every object of $\C$, and the multiplication of $A$ is compatible with this half-braiding so that $A$ is commutative after the lift to $Z(\C)$. In this way, although $\C$ itself is not braided, the algebra $A$ has enough braiding data to be treated as a condensable algebra in the bulk UMTC $Z(\C)$. This is precisely the situation that will be generalized to fermionic and parafermionic condensation in Section~\ref{3}, where the condensed algebra carries nontrivial half-braiding data inherited from $Z(\C)$.

Let's now assume that $A$ is a central commutative algebra within $\C$. In a similar manner as before, condensing $A$ gives rise to a domain wall consisting of right-$A$ modules, collectively forming a category denoted as $\C_A$. The key aspect here is that the requirement for $A$ to be a commutative algebra within $Z(\C)$ enables us to braid $A$ from the left side of an object $X\in\C_A$ to its right side, subsequently acting on $X$ from the right. As a result, a right-$A$ module automatically becomes a left-$A$ module as well.

Consequently, $\C_A$ can be understood as a subcategory of $_A\C_A$, where the monoidal/tensor structure is naturally present within $\C_A$. This observation implies that, following the condensation, the original fusion category $\C$ transforms into a theory exhibiting a fusion category (sub-)symmetry represented by $\C_A$.

\subsection{Relation between boundary condensation in $\C$ and bulk condensation in $Z(\C)$}

Previously, we approached the topic of bosonic anyon condensation within $\C$ from a 2d theory viewpoint. Alternatively, it is also possible to investigate anyon condensation from a 3d bulk theory perspective by employing Levin-Wen string-net models [as illustrated in Figure~\ref{fig:relation}].

Rather than contemplating a 1+1d CFT boundary, we can apply a Wick rotation to consider the boundary as a 2d spatial manifold. In this setup, the fusion category $\C$ governing the TDLs within the boundary serves as the input data for constructing the Levin-Wen Hamiltonians. The anyonic excitations present in the 3d topological order can then be described through the Drinfeld center $Z(\C)$.

The condensation process of $A$ in $\C$ on the 2d boundary can also be related to anyon condensation within the underlying UMTC $Z(\C)$ in the 3d bulk. In the subsequent discussions, we will delve into an exploration of the two distinct types of $A$ condensation separately.\\

{\bf (a) $A$ is an algebra of $\C$ [see Figure~\ref{fig:relation}(a)]}

For a given input fusion category $\C$, its objects are used to label the string types within the Levin-Wen string-net model. In the condensed model, the string types are labeled by objects from $_A\C_A$. In the context of the 3d bulk view, the anyonic excitations on the left and right sides correspond to Drinfeld centers $Z(\C)$ and $Z(_A\C_A)$, respectively. These two sides are separated by a gapped domain wall, depicted as the red line in Figure~\ref{fig:relation}(a).

A natural question emerges: what is the relationship between the anyons in the left and right Levin-Wen models? An important mathematical result establishes the equivalence between these two UMTCs:
\begin{align}
Z(\C) \simeq Z(_A\C_A),
\end{align}
known as Morita equivalence. Consequently, the red domain wall surface depicted in Figure~\ref{fig:relation} is invertible. It represents a self-duality wall within the same 3D topological order.\\

{\bf (b) $A$ is a central commutative algebra of $\C$ [see Figure~\ref{fig:relation}(b)]}

When considering the scenario where $A$ is a central commutative algebra within $\C$, the fusion category on the right-hand side can be selected as $\C_A$. We now have two Levin-Wen string-net models with input fusion categories $\C$ and $\C_A$. In this situation, the topological orders described by $Z(\C)$ and $Z(\C_A)$ on the two sides are not equivalent. Instead, the latter can be interconnected through the anyon condensation process within the initial UMTC $Z(\C)$. Schauenburg's theorem~\cite{schauenburg_monoidal_2001} reveals that
\begin{align}
Z(\C_A) \simeq [Z(\C)]_A^0.
\end{align}
Here, the $A$ on the right-hand side is treated as a commutative algebra within $Z(\C)$, thereby allowing us to perform anyon condensation within this UMTC.

Lattice models for bosonic anyon condensation have also been formulated~\cite{zhao_characteristic_2023,christian2023lattice}. The complete Hamiltonian, including the terms on both sides and on the intermediate domain wall, can be written as a commuting-projector Hamiltonian.

\section{Pre-metric group $\C(G,q)$}
\label{app:premetric-group}

\subsection{General definition}

In this appendix, we review the pre-metric group data that are used in the discussion of $\Z_N$ parafermions and their condensation. More precisely, a pre-metric group is a finite Abelian group $G$ equipped with a quadratic form $q:G\rightarrow \mathrm{U}(1)$. We denote by $\C(G,q)$ the corresponding pointed braided fusion category: its simple objects are labeled by elements of $G$, and the fusion rule is the group law of $G$~\cite{etingof_tensor_2015}. The quadratic form obeys $q(g)=q(g^{-1})$ and determines a bicharacter
\begin{align}\label{bicha}
b(g,h):= \frac{q(gh)}{q(g)q(h)}
\end{align}
which is multiplicative in both variables. Physically, $q(g)$ is the topological twist $\theta_g$ of the Abelian anyon $g$, while $b(g,h)$ is the mutual braiding or monodromy between $g$ and $h$~\cite{Kitaev:2005hzj}. When the bicharacter $b$ is non-degenerate, $(G,q)$ is called a metric group, and $\C(G,q)$ is a pointed modular category.

\subsection{The cyclic $\Z_N$ case and spin data}

We now specialize to $G=\Z_N=\{0,1,...,N-1\}$, written additively. This is the case relevant for the $\Z_N$ symmetry lines and $\Z_N$ parafermions in the main text. The pre-metric group $\C(\Z_N,k)=\C(\Z_N,q_k)$ is characterized by an integer parameter $k$ and the quadratic form
\begin{align}
q_k(n) = (\om_{N,k})^{n^2}
%= e^{2\pi i \frac{k}{\gcd(2,N)N}n^2}
=
\begin{cases}
e^{2\pi i \frac{k}{2N}n^2} (k=0,1,...,2N-1), & \text{if $N$ is even}\\
e^{2\pi i \frac{k}{N}n^2} (k=0,1,...,N-1), & \text{if $N$ is odd}
\end{cases},
\end{align}
where for simplicity we introduced the phase factor
\begin{align}
\om_{N,k}:=e^{2\pi i \frac{k}{\gcd(2,N)N}}.
\end{align}
Thus $k$ is defined modulo $N$ for odd $N$ and modulo $2N$ for even $N$. In particular, the generator of $\Z_N$ has topological twist
\begin{align}
q_k(1) = \om_{N,k},
\end{align}
and hence topological spin
\begin{align}\label{h1}
h_1=\frac{k}{\gcd(2,N)N}.
\end{align}
Thus, if the conformal spin of the generator of a $\Z_N$ subcategory is known in a CFT, Eq.~(\ref{h1}) determines the value of $k$ and hence the pre-metric group $\C(\Z_N,q_k)$. This is the first role of the pre-metric group in this paper: it packages the fractional spin data of the $\Z_N$ generator.

The $\Z_N$ parafermionic lines considered in Section~\ref{sec:pfcond} form the subclass for which
\begin{align}
h_1=\frac{m}{N}\mod 1,\qquad m\in\mathbb Z_N^\times.
\end{align}
Writing $d=\gcd(2,N)$, comparison with Eq.~\eqref{h1} gives $k=dm$ modulo $dN$. Hence
\begin{align}
q_k(a)=e^{2\pi i m a^2/N},\qquad a\in\mathbb Z_N.
\end{align}
For $N=3$, the choices $m=1$ and $m=2$ give topological spins $1/3$ and $2/3$ modulo one. Both choices define $\Z_3$ parafermionic lines in the convention of this paper.

\subsection{Associator and anomaly condition}

The next question is whether the corresponding $\Z_N$ symmetry is non-anomalous. This is not determined only by the twist $q_k$; one must also examine the associator, or equivalently the $F$ symbols, of the underlying pointed fusion category. In a convenient skeletal gauge, the relevant $F$ symbols are
\begin{align}\label{F_ZN}
F^{a,b,c} = 
\begin{cases}
(-1)^{ka\lfloor \frac{b+c}{N}\rfloor}, & \text{if $N$ is even}\\
1, & \text{if $N$ is odd}
\end{cases}.
\end{align}
The obstruction to gauging or condensing the $\Z_N$ lines is precisely the nontriviality of this associator. Therefore
\begin{align}
\Z_N \text{ anomalous} &\Longleftrightarrow \text{$N$ is even and $k$ is odd},\\\label{non-ano}
\Z_N \text{ non-anomalous} &\Longleftrightarrow \text{$N$ is odd or $k$ is even}.
\end{align}
This is the condition used in Section~\ref{3} when deciding whether the $\Z_N$ line generated by a parafermion can be condensed. In the non-anomalous cases, the $F$ symbols can be trivialized by a gauge choice, and the $\Z_N$ line can be treated as an ordinary condensable symmetry line.

\subsection{Braiding and modularity}

Finally, the same pre-metric data determine the braiding and modularity of the $\Z_N$ subcategory. The quadratic form is the twist of the braided tensor category,
\begin{align}
\theta_n = R^{n,n} = q_k(n) = (\om_{N,k})^{n^2}% = e^{2\pi i \frac{k}{\gcd(2,N)N}n^2}
,
\end{align}
and a compatible choice of braiding data is
\begin{align}
R^{m,n} = (\om_{N,k})^{mn}. %= e^{2\pi i \frac{k}{\gcd(2,N)N}mn}
\end{align}
These data make $\C(\Z_N,q_k)$ into a pointed pre-modular category.

The modular matrices of this pointed category can be computed from the general relation
\begin{eqnarray}
    S_{\alpha \beta} & = & \frac{1}{D}\sum_{\gamma} N^{\gamma}_{\alpha \beta} \frac{\theta_{\gamma}}{\theta_{\alpha}\theta_{\beta}}d_{\gamma} = S_{\beta \alpha}, \quad D = \sqrt{\sum_{\alpha} d_{\alpha}^2}, \\
    T_{\alpha \beta} & = & \theta_{\alpha} \delta_{\alpha \beta}.
\end{eqnarray}
The final results are
\begin{align}
S_{m,n} &= \frac{1}{\sqrt{N}} \frac{q_k(m+n)}{q_k(m)q_k(n)}
= \frac{1}{\sqrt{N}}(\om_{N,k})^{2mn}
%= \frac{1}{\sqrt{N}}e^{2\pi i \frac{k}{\gcd(2,N)N}2mn}
=
\begin{cases}
\frac{1}{\sqrt{N}} e^{2\pi i \frac{k}{N}mn} , & \text{if $N$ is even}\\
\frac{1}{\sqrt{N}} e^{2\pi i \frac{k}{N}2mn} , & \text{if $N$ is odd}
\end{cases},
\\
T_{m,n} &= q_k(n)\delta_{mn}.
\end{align}
Equivalently, the bicharacter in Eq.~(\ref{bicha}) is the monodromy matrix,
\begin{align}
b(m,n) = \frac{q(m+n)}{q(m)q(n)} = M_{m,n} = \frac{S_{mn}S_{00}}{S_{0m}S_{0n}}
= (\om_{N,k})^{2mn}.
\end{align}

The pre-metric group is metric precisely when this monodromy pairing is non-degenerate. Equivalently, the associated pointed category is modular exactly when
\begin{align}\label{mod}
\text{$\C(\Z_N,k)$ is modular $\Longleftrightarrow$ $q_k$ is non-degenerate $\Longleftrightarrow$ $\gcd(k,N)=1$.}
\end{align}

\subsection{Use in the main text}

Let us summarize how this appendix is used in the main text. First, a parafermionic line of type $m\in\mathbb Z_N^\times$ has spin $h_1=m/N$ and corresponds to $k=\gcd(2,N)m$ in the pre-metric parametrization. Second, Eq.~(\ref{non-ano}) tells us whether the corresponding $\Z_N$ symmetry can be orbifolded, gauged, or condensed. Third, Eq.~(\ref{mod}) tells us when the $\Z_N$ pointed subcategory is itself modular. In that case it decouples from its M\"uger centralizer, so the ambient modular category can be decomposed as a Deligne product of $\C(\Z_N,q_k)$ and the complementary modular subcategory. This is precisely the situation used in Section~\ref{sec:pfcond}, where parafermionic condensation removes a modular $\Z_N$ sector from the bulk UMTC.

\section{$\Z_N$-graded division algebras and $q_M$-type simple objects in para-fusion categories}
\label{app:graded-division-algebra}

In an ordinary fusion category over $\bC$, Schur's lemma implies that the endomorphism algebra of a simple object is a division algebra; hence it is simply $\bC$~\cite{etingof_tensor_2015}. In a para-fusion category, however, the Hom spaces carry a $\Z_N$ grading. Thus, if $X$ is simple, its endomorphism space
\begin{align}
\End(X)=\Hom_\C(X,X)=\bigoplus_{i\in\Z_N}\End(X)_i
\end{align}
should be regarded as a $\Z_N$-graded division algebra: every nonzero homogeneous endomorphism is invertible. This is the parafermionic generalization of the familiar distinction between $m$-type and $q$-type objects in a super-fusion category~\cite{Aasen:2017ubm}.

The case $N=2$ gives the standard super-fusion example. The endomorphism algebra of a simple object is either $\bC$, corresponding to an $m$-type object, or the two-dimensional complex Clifford algebra $\mathbb{C}l_1=\mathbb C^{1|1}$, corresponding to a $q$-type object. For general $\Z_N$ para-fusion categories, the same question becomes the classification of $\Z_N$-graded division algebras.

The classification theorem is as follows~\cite{bahturin2001group, bahturin2005finite}:
%~\cite{Bahturin2001,Bahturin2005}:
\begin{thm}
Let $R =\oplus_{g\in\Z_N}R_g$ be a finite-dimensional $\Z_N$-graded algebra over $\bC$. Then $R$ is a graded division algebra if and only if $R$ is isomorphic to the group algebra $\bC[\Z_M]=\oplus_{h\in \Z_M}\bC_h$, where $\Z_M$ is a finite subgroup of $\Z_N$, i.e., $M$ is a divisor of $N$.
\end{thm}

Here $\Z_M$ should be understood as the subgroup of $\Z_N$ of order $M$, so $M\mid N$. Consequently, the possible endomorphism algebras in a $\Z_N$ para-fusion category are labeled by the divisors of $N$. There are $\tau(N)$ possibilities, where $\tau(N)$ is the divisor function. For example, for $1\leq N\leq 10$, one has
\begin{align}
\tau(N)=1,2,2,3,2,4,2,4,3,4.
\end{align}
The case $M=1$ gives $\bC[\Z_1]=\bC$ and hence the ordinary $m$-type object. The cases $M>1$ give genuinely parafermionic analogs of $q$-type objects.

We therefore use the following terminology. If
\begin{align}
\End(X)\simeq \bC[\Z_M]
\end{align}
as a $\Z_N$-graded algebra, then $X$ is called a $q_M$-type object. This convention includes the trivial case $M=1$, where a $q_1$-type object is just an $m$-type object. For $N=2$, the two divisors $M=1,2$ reproduce the usual $m$- and $q$-type simple objects of a super-fusion category.

This terminology is used in Section~\ref{sec:pfcond2} after parafermion condensation. If $f$ is the condensed parafermion and $[X]$ is the image of a simple object $X$, then the nonzero homogeneous endomorphisms of $[X]$ are controlled by the stabilizer
\begin{align}
H_X=\{i\in \Z_N\mid X\otimes f^i\simeq X\}.
\end{align}
When $H_X\simeq \Z_M$, the endomorphism algebra of $[X]$ is isomorphic to $\bC[\Z_M]$ after choosing isomorphisms $X\simeq X\otimes f^i$ for the stabilizer elements. Hence $[X]$ is a $q_M$-type simple object. In this way, the graded division algebra classification explains why para-fusion categories can contain several different layers of $q_M$-type objects, rather than only the single $q$-type familiar from the $\Z_2$ super-fusion case.

\bibliographystyle{JHEP}

\newpage
\bibliography{refs.bib}

\end{document}